\documentclass[12pt,a4paper]{article}
\usepackage{amsmath,amssymb}
\usepackage{graphicx,rotating}
\pdfoutput=1 

\usepackage{color}
\usepackage{bm}

\usepackage[%breaklinks=true,
colorlinks=true,backref,pagebackref]{hyperref}

\newcommand{\fref}[1]{Fig.~\ref{#1}}
\def\a{\alpha}
\def\b{\beta}
\def\d{\delta}

\def\eps{\epsilon}

\def\vt{\vartheta}

\begin{document}
\newcommand{\bi}[1]{{\mathbf #1} }
\newcommand{\bGamma}{{\mathbf\Gamma}}
\newcommand{\brho}{{\mathbf\rho}}
%\unitlength 1mm

\title{Schwarzschild and Kerr Solutions\\
  of Einstein's Field Equation\\--- an introduction ---}
\author{Christian Heinicke$^{1,\dagger}$ and Friedrich W.\
  Hehl$^{1,2,\star}$ \\\small $^{1}\!$Inst.\ Theor.\ Physics, Univ.\
  of Cologne, 50923 K\"oln, Germany\\\small$^{2}\!$Dept.\ Physics \&
  Astron., Univ.\ of
  Missouri, Columbia, MO 65211, USA\\\small
  $^{\dagger}$christian.heinicke@t-online.de\qquad
  $^{\star}$hehl@thp.uni-koeln.de}
\date{7 March 2015}

\maketitle

\begin{abstract} {Starting {}from Newton's gravitational theory, we give
    a general introduction into the spherically symmetric solution of
    Einstein's vacuum field equation, the Schwarzschild(-Droste)
    solution, and into one specific stationary axially symmetric
    solution, the Kerr solution. The Schwarzschild solution is unique
    and its metric can be interpreted as the exterior gravitational
    field of a spherically symmetric mass. The Kerr solution is only
    unique if the multipole moments of its mass and its angular
    momentum take on prescribed values.  Its metric can be interpreted
    as the exterior gravitational field of a suitably rotating mass
    distribution. Both
  solutions describe objects exhibiting an {\it event horizon}, a
  frontier of no return. The corresponding notion of a black hole is
  explained to some extent. Eventually, we present some
  generalizations of the Kerr solution.}

\bigskip

{\hfill{\it\footnotesize file schwarzkerr29arxiv.tex, 7
March 2015}}
\end{abstract}\bigskip
\vfill

\noindent\begin{footnotesize}Invited review article. To appear in
Wei-Tou Ni (editor) ``One Hundred
  Years of General Relativity: Cosmology and Gravity,'' World
  Scientific, Singapore (2015).
Also published in  
Int.\ J.\ Mod.\ Phys.\ D {\bf 24} (2015) 1530006 (78 pages),
DOI: 10.1142/S0218271815300062.
\end{footnotesize}

\newpage

\tableofcontents

\pagebreak
%%%%%%%%%%%%%%%%%%%%%%%%%%%%%%%%%%%%%%%%%%%%%%%%%%%%%%%%%%%%%%%%%%%%%%%%%
\section{Prelude$^1$}
%%%%%%%%%%%%%%%%%%%%%%%%%%%%%%%%%%%%%%%%%%%%%%%%%%%%%%%%%%%%%%%%%%%%%%%%%

{\it In
Sec.1.1, we provide some background material on Newton's
  theory of gravity and, in Sec.1.2, on the flat and gravity-free
  Minkowski space of special relativity theory.\footnotetext[1]{Parts of 
Secs.1 \& 2 are adapted {}from 
our presentation\cite{HeinickeHehl} in Falcke et al.\cite{Falcke}.} Both theories were
  superseded by Einstein's gravitational theory, general
  relativity. In Sec.1.3, we supply some machinery for formulating
  Einstein's field equation without and with the cosmological constant.}

\addtocounter{footnote}{1}

%%%%%%%%%%%%%%%%%%%%%%%%%%%%%%%%%%%%%%%%%%%%%%%%%%%%%%%%%%%%%%%%%%%%%%%%%
\subsection{Newtonian gravity}
%%%%%%%%%%%%%%%%%%%%%%%%%%%%%%%%%%%%%%%%%%%%%%%%%%%%%%%%%%%%%%%%%%%%%%%%%

{\em Newton's gravitational theory is described---in particular tidal
  gravitational forces---and applied to a spherically symmetric body
  (a \hspace{-3pt}``star'').}\bigskip

\null\hfill\begin{minipage}{12cm} \small\it Gravity exists in all
  bodies universally and is proportional to the quantity of matter in
  each [\,\dots\,] If two globes gravitate towards each other, and
  their matter is homogeneous on all sides in regions that are equally
  distant {}from their centers, then the weight of either globe towards
  the other will be inversely as the square of the distance between
  the centers.  \null\hfill\rm Isaac Newton\cite{newton} (1687)
\end{minipage}

\bigskip

\begin{figure}%\fref{chap1:attraction1}.
\begin{center}
\includegraphics[width=6cm]{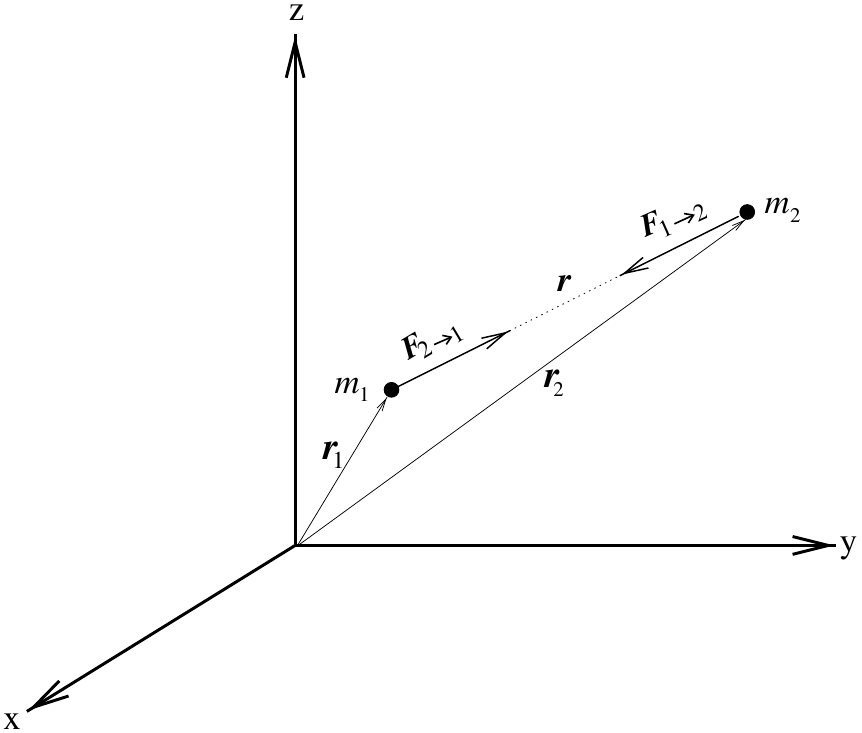}
\end{center}
\caption{Two mass points $m_1$ and $m_2$ attracting each other in
3-dimensional space, Cartesian coordinates $x$, $y$, $z$.}\label{chap1:masses}
\end{figure}

The gravitational force of a point--like mass $m_2$ on a similar one
of mass $m_1$ is given by Newton's attraction law,
\begin{equation}
\label{chap1:att.law}
\bi{F}_{2\to1}=-G \, \frac{m_1 \, m_2}{\left|\bi{r}\right|^2} \,
\frac{\bi{r}}{\left|\bi{r}\right|}\,,
\end{equation}
where $G$ is Newton's gravitational constant (CODATA 2010),   
\[
G\stackrel{\rm SI}{=} 6.67384(80) \times 10^{-11} \, 
\frac{\left({\rm m}/{\rm
s}\right)^4}{N}\,.
\]
The vector $\bi{r}:=\bi{r}_1-\bi{r}_2$ points {}from $m_2$ to $m_1$, see
the \fref{chap1:masses}.

According to {\em actio = reactio} (Newton's 3rd law), we have
$\bi{F}_{2\to1}=-\bi{F}_{1\to2}$.  Thus a complete symmetry exists of
the gravitational interaction of the two masses onto each other.  Let
us now distinguish the mass $m_2$ as field--generating active
gravitational mass and $m_1$ as (point--like) passive test--mass.
Accordingly, we introduce a hypothetical {\em gravitational field} as
describing the force per unit mass ($m_2\hookrightarrow M\,,\,\, m_1
\hookrightarrow m$):
\begin{equation}
\label{chap1:gravfield}
\bi{f} := \frac{\bi F}{m} = -\frac{GM}{\left|\bi{r}\right|^2} \, 
\frac{\bi{r}}{\left|\bi{r}\right|}\,.
\end{equation}
\begin{figure}
\begin{center}
\includegraphics[width=6cm]{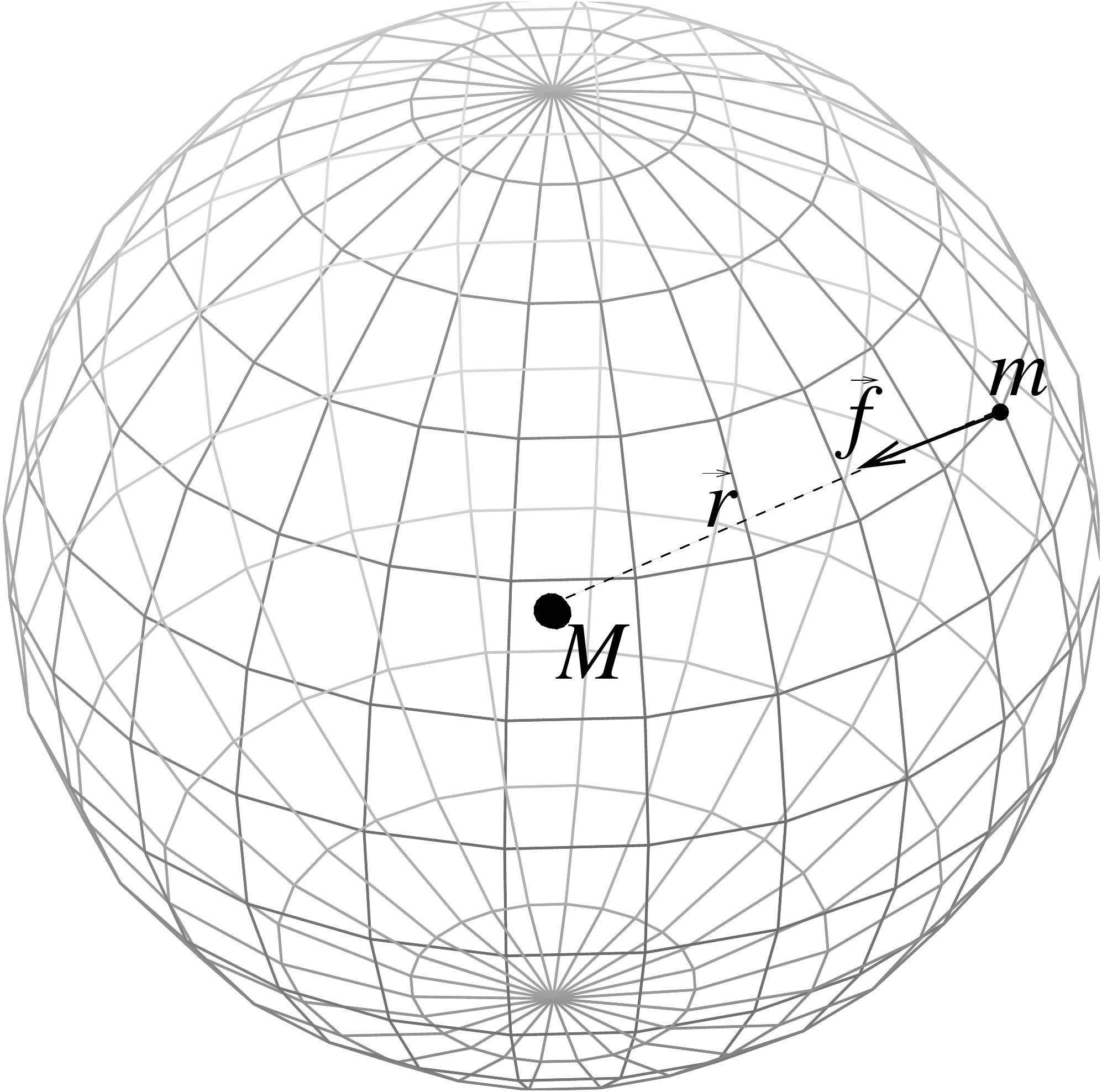}
\end{center}
\caption{The ``source'' $M$ attracts the test mass $m$.}
\label{fig2}
\end{figure}
With this definition, the force acting on the test--mass $m$ is equal
to {\em field strength} $\times$ {\em gravitational charge} (mass) or
$\bi{F}_{M\to m}= m \, \bi{f}$, in analogy to electrodynamics.  The
active gravitational mass $M$ is thought to emanate a gravitational
field which is always directed to the center of $M$ and has the same
magnitude on every sphere with $M$ as center, see \fref{fig2}.  
Let us now investigate
the properties of the gravitational field
(\ref{chap1:gravfield}). Obviously, there exists a potential
\begin{equation}
\label{potential}
\phi = - G\, \frac{M}{\left| \bi{r}\right|} \,,
\qquad \bi{f} = - \bm{\nabla} \phi\,.
\end{equation}      
Accordingly, the gravitational field is curl-free: $\bm{\nabla}
\bm{\times}\bi{f} = 0$.

By assumption it is clear that the source of the gravitational field is
the mass $M$. We find, indeed,
\begin{equation}
\label{chap1:divf}
\bm{\nabla}\bm{\cdot} \bi{f} = - 4\pi \, G M\, \delta^3(\bi{r})\,,
\end{equation}
where $\delta^3(\bi{r})$ is the 3-dimensional (3d) delta function.  By
means of the {\em Laplace operator} $\Delta := \bm{\nabla
  \!\cdot\!\nabla }$, we infer for the gravitational potential
\begin{equation}
\label{chap1:poisson1}
\Delta \, \phi = 4\pi \, G \, M \, \delta^3(\bi{r})\,.
\end{equation}
The term $M \, \delta^3(\bi{r})$ may be viewed as the mass density of
a point mass. Eq.(\ref{chap1:poisson1}) is a 2nd order linear partial
differential equation for $\phi$. Thus the gravitational potential
generated by several point masses is simply the linear superposition
of the respective single potentials.  Hence we can generalize the {\em
  Poisson equation} (\ref{chap1:poisson1}) straightforwardly to a
continuous matter distribution $\rho(\bi{r})$:
\begin{equation}
\label{chap1:poisson}
\Delta \, \phi = 4\pi \, G \, \rho\,.
\end{equation}
This equation interrelates the source $\rho$ of the gravitational field with
the gravitational potential $\phi$ and thus completes the quasi-field
theoretical description of Newton's gravitational theory. 

We speak here of {\em quasi}--field theoretical because the field
$\phi$ as such represents a convenient concept. However, it has no
{\em dynamical} properties, no genuine degrees of freedom. The
Newtonian gravitational theory is an {\em action at a distance} theory
(also called {\it mass-interaction theory}). When we remove the
source, the field vanishes instantaneously. Newton himself was very
unhappy about this consequence. Therefore, he emphasized the
preliminary and purely descriptive character of his theory. But before
we liberate the gravitational field {}from this constraint by equipping
it with its own degrees of freedom within the framework of general
relativity theory, we turn to some properties of the Newtonian theory.

A very peculiar fact characteristic to the gravitational field is that
the acceleration of a freely falling test-body does not depend on the
mass of this body but only on its position within the gravitational
field. This comes about because of the equality (in suitable units) of
the gravitational and the inertial mass:
\begin{equation}
\label{eqom}
\stackrel{\mbox{\tiny inertial}}{m} \, \ddot{\bi{r}}
= \bi{F} = \stackrel{\mbox{\tiny grav}}{m} \, \bi{f}\,.
\end{equation}
This equality has been well tested since Galileo's time by means of
pendulum and other experiments with an ever increasing accuracy, see 
Will~\cite{will}.

\begin{figure}
\begin{center}
\bigskip
\includegraphics[width=6cm]{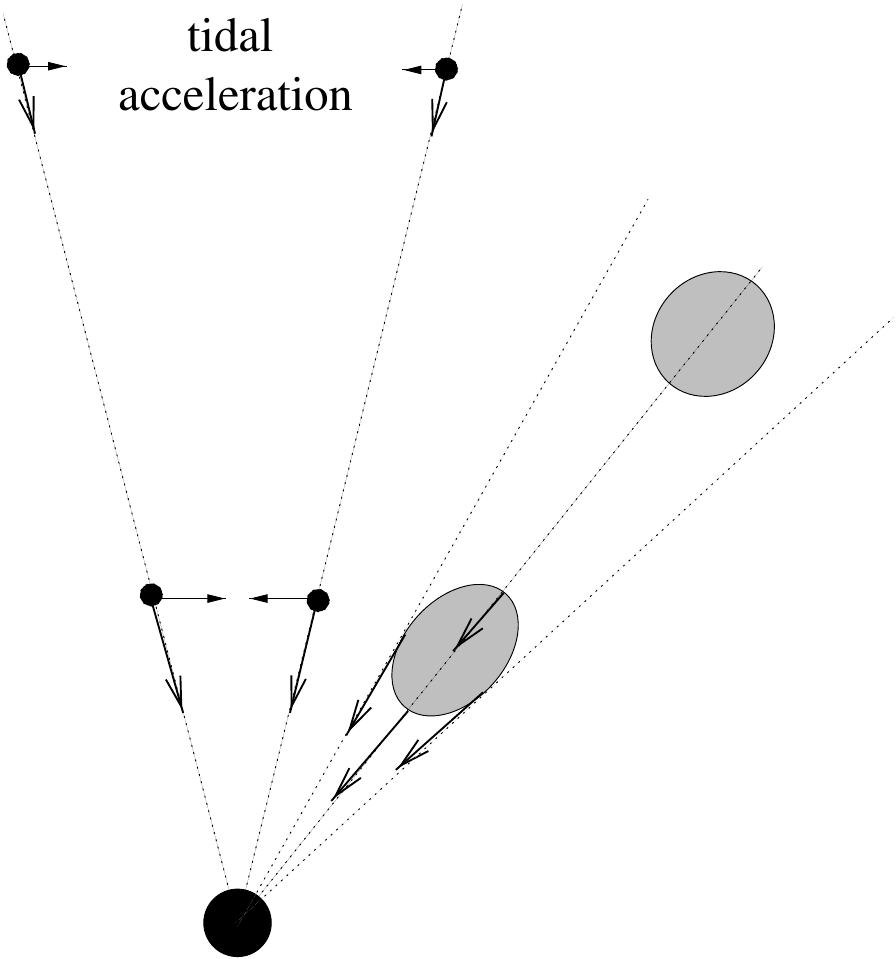}
\end{center}
\caption{Tidal forces emerging between two freely falling particles
and deforming a spherical body.}
\label{chap1:tidal}
\end{figure}

In order to allow for a more detailed description of the structure of
a gravitational field, we introduce the concept of {\em tidal force}.
This can be best illustrated by means of \fref{chap1:tidal}.  In a
spherically symmetric gravitational field, for example, two
test-masses will fall radially towards the center and thereby get
closer and closer. Similarly, a spherical drop of water is deformed to
an ellipsoidal shape because the gravitational force at its bottom is
bigger than at its top, which has a greater distance to the source.
If the distance between two freely falling test masses is relatively
small, we can derive an explicit expression for their relative
acceleration by means of a Taylor expansion.  Consider two mass points
with position vectors $\bi{r}$ and $\bi{r}+\delta\bi{r}\,,\text{ with
} \left|\delta\bi{r}\right|\ll1$.  Then the relative acceleration
reads
\begin{equation}\label{newtondeviation}
  \delta \bi{a}=
  \left[ \bi{f}(\bi{r}+\delta\bi{r})-\bi{f}(\bi{r})\right]
  =\delta\bi{r} \bm{\cdot} (\bm{\nabla f})\,.
\end{equation}
%where ${\rm Grad}$ denotes the vector gradient.  
We may rewrite this according to (the sign is conventional,
$\partial/\partial x^a =:\partial_a$,\,
$x^1\!=\!x,\,x^2\!=\!y,\,x^3\!=\!z$)
\begin{equation}\label{TidalForce}K_{ab} := - 
\left(\bm{\nabla}\bi{f}\right)_{ab}=-\partial_a \, f_b\,,\qquad
a,b=1,2,3\,.
\end{equation}
We call $K_{ab}$ the {\em tidal force} matrix. The vanishing curl of the
gravitational field is equivalent to its symmetry, $K_{ab}=K_{ba}$. 
Furthermore, $K_{ab}=\partial_a \, \partial_b \, \phi$. 
Thus, the Poisson equation becomes, 
\begin{equation}\label{tidaltrace}
\sum_{a=1}^3 \, K_{aa} = {\rm trace} \, K = 4\pi \, G \, \rho\,.
\end{equation}
Accordingly, in vacuum $K_{ab}$ is trace-free.

Let us now investigate the gravitational potential of a homogeneous
{\em star} with constant mass density $\rho_\odot$ and total mass
$M_\odot=(4/3) \, \pi \, R_\odot^3 \, \rho_\odot$.  For our Sun, the
radius is $R_\odot=6.9598 \times 10^{8}{\,\rm m}$ and the total mass
is $M=1.989 \times 10^{30}{\,\rm kg}$.

Outside the sun (in the idealized picture we are using here), 
we have vacuum. Accordingly, $\rho(\bi{r})=0$ for
$\left|\bi{r}\right|>R_\odot$. Then the Poisson equation reduces to
the {\em Laplace equation}
\begin{equation}
\label{chap1:laplace}
\Delta \, \phi = 0 \,, \quad\text{ for } r> R_\odot\,.
\end{equation}
In 3d polar coordinates, the $r$-dependent part of the Laplacian has the
form $(1/r^2) \, \partial_r \, (r^2 \, \partial_r)$.
Thus (\ref{chap1:laplace}) has the solution 
\begin{equation}
\phi = \frac{\alpha}{r} + \beta\,,
\end{equation}
where $\alpha$ and $\beta$ are integration constants. Requiring that the
potential tends to zero as $r$ goes to infinity, we get $\beta=0$.
The integration constant $\alpha$ will be determined {}from the
requirement that the force should change smoothly as we cross the star's
surface, that is, the interior and exterior potential and their first 
derivatives have to be matched continuously at $r=R_\odot$.

\unitlength 1mm
\begin{figure}[h]
\label{chap1:newtonpotsun}
\begin{picture}(100,75)
\put(7,10){\includegraphics[width=10cm,height=6cm]{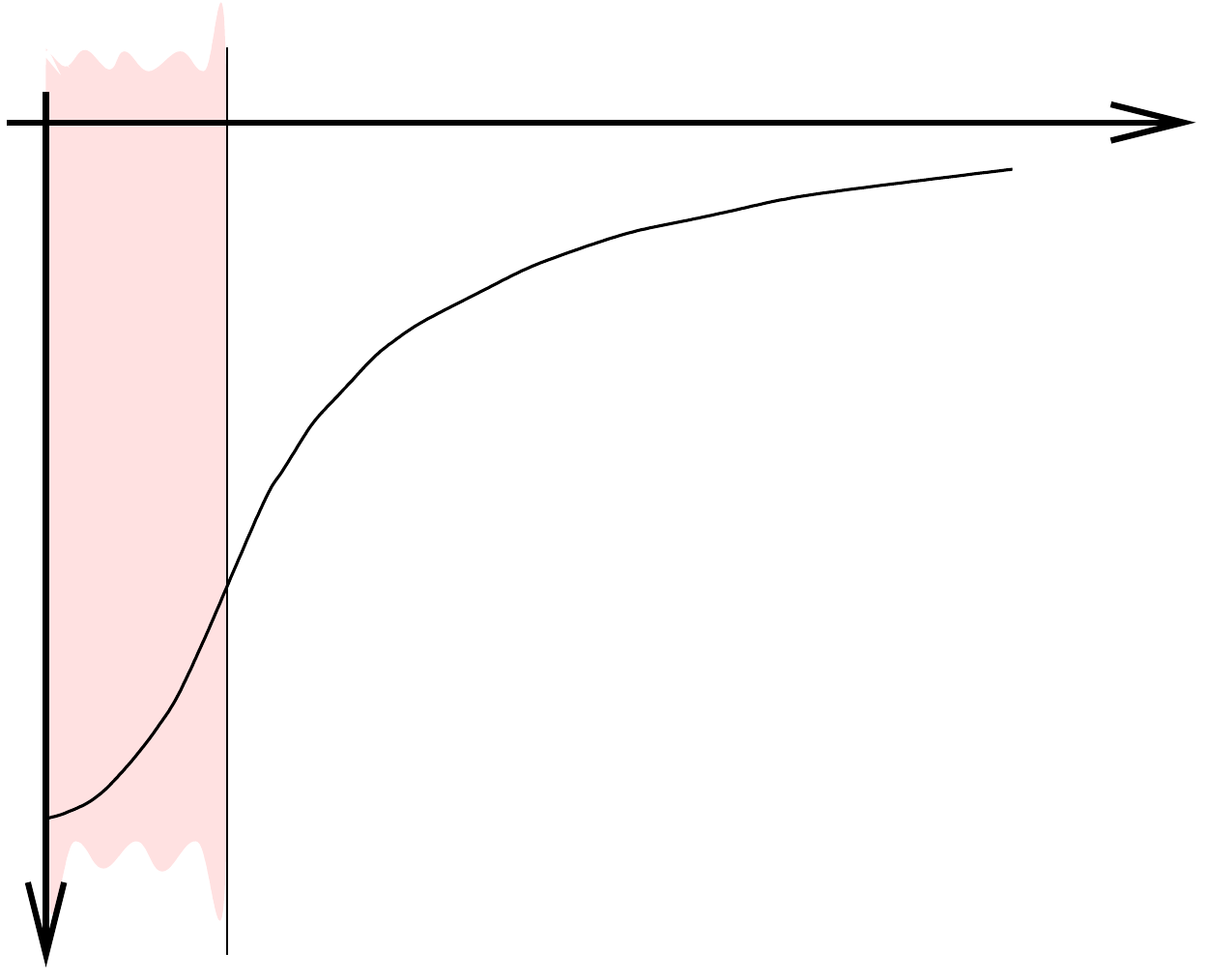}}
\put(12.5,63){\mbox{interior}}
\put(24,68){\mbox{$R_0$}}
\put(4,40){\mbox{$\phi$}}
\put(15,28){\mbox{$\sim r^2$}}
\put(4.5,19){\mbox{$\phi_0 \, -$}}
\put(45,45){\mbox{\large$\sim \frac{1}{r}$}}
\put(107.5,61.5){\mbox{$\infty$}}
\put(80,65){\mbox{\large$r$}}
\put(35,63){\mbox{exterior $\longrightarrow$}}
\end{picture}
\vspace{-20pt}
\caption{Newtonian potential of a homogeneous star.}
\end{figure}
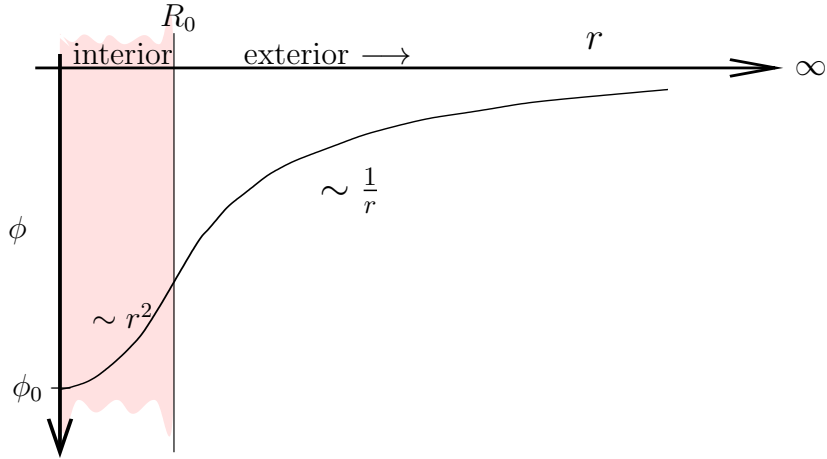
Inside the star we have to solve
\begin{equation}
\Delta \, \phi = 4\pi \, G \, \rho_\odot \,, \quad
\text{ for } r \leq R_\odot\,.
\end{equation}
We find
\begin{equation}
\phi=\frac{2}{3} \, \pi G \rho_\odot \, r^2 +\frac{C_1}{r} + C_2\,,
\end{equation} 
with integration constants $C_1$ and $C_2$. We demand that the potential 
in the center $r=0$ has a finite value, say $\phi_0$. This requires $C_1$=0.
Thus
\begin{equation}
\phi=\frac{2}{3}\pi \, G \, \rho_\odot \, r^2+\phi_0
=\frac{G \, M(r)}{2r}+\phi_0\,,
\end{equation}
where we introduced the {\em mass function} 
$M(r)=(4/3) \, \pi r^3  \rho_\odot$ which
measures the total mass inside a sphere of radius $r$.

Continuous matching of $\phi$ and its first derivatives 
at $r=R_\odot$ finally yields:
\begin{equation}
\label{chap1:potentialsun}
\phi(\bi{r})=\left\{\begin{array}{ccl}
-G \, \frac{\displaystyle M_\odot}{\displaystyle\left|\bi{r}\right|}&
\mbox{ for }
& \left|\bi{r}\right| \geq R_\odot\,,\vspace{5pt}\\
G \, \frac{\displaystyle M_\odot}{\displaystyle2R^3_\odot} \, 
\left|\bi{r}\right|^2
-\frac{\displaystyle 3G\,M_\odot}{\displaystyle 2R_\odot}
& \mbox{ for }&  \left|\bi{r}\right| < R_\odot\,.
\end{array}\right.
\end{equation}
The slope of this curve indicates the magnitude of the gravitational force,
the curvature (2nd derivative) the magnitude of the tidal force (or
acceleration).\\

\newpage

%%%%%%%%%%%%%%%%%%%%%%%%%%%%%%%%%%%%%%%%%%%%%%%%%%%%%%%%%%%%%%%%%%%%%%%%%%%%
\subsection{Minkowski space}
%%%%%%%%%%%%%%%%%%%%%%%%%%%%%%%%%%%%%%%%%%%%%%%%%%%%%%%%%%%%%%%%%%%%%%%%%%%%

{\it When, in a physical experiment, gravity can be safely neglected,
  we seem to live in the flat Minkowski space of special relativity
  theory. We introduce the metric of the Minkowksi space and rewrite
  it in terms of so-called null coordinates, that is, we use light
  rays for a parametrization of Minkowski space.}\bigskip

\null\hfill\begin{minipage}{10cm} \small\it Henceforth space by
  itself, and time by itself, are doomed to fade away into mere
  shadows, and only a kind of union of the two will preserve an
  independent reality.  \vspace{-13pt}

\null\hfill\rm Hermann Minkowski (1908)
\end{minipage}
  
\bigskip

It was Minkowski who welded space and time together into spacetime,
thereby abandoning the observer-independent meaning of spatial and
temporal distances. Instead, the spatio-temporal distance, the line
element,
\[
ds^2 = -c^2\,dt^2+dx^2+dy^2+dz^2
\]
is distinguished as the invariant measure of spacetime. The Poincar\'{e}
(or inhomogeneous Lorentz) transformations form the invariance group of
this spacetime metric. The principle of the constancy of the speed of
light is embodied in the equation $ds^2=0$. Suppressing one spatial
dimension, the solutions of this equation can be regarded as a double
cone. This light cone visualizes the paths of all possible light rays
arriving at or emitted {}from the cone's apex. Picturing the light cone
structure, and thereby the causal properties of spacetime, will be our
method for analyzing the meaning of the Schwarzschild and the Kerr
solution.

\subsubsection*{Null coordinates}

We first introduce so-called null coordinates. The Minkowski metric
(with $c=1$), in spherical polar coordinates reads
\begin{equation}
  ds^2 = -dt^2+dr^2+r^2\,\left(d\theta^2+\sin^2\theta \,d\phi^2\right)
  = -dt^2 + dr^2 + r^2 \, d\Omega^2 \,.
\end{equation}
We define {\em advanced} and {\em retarded null coordinates} according to 
\begin{equation}\label{vu}
  v := t+r\,,\quad u:= t-r \,,
\end{equation}
\begin{figure}[h]
%\bigskip
\begin{center}
\includegraphics[width=12cm]{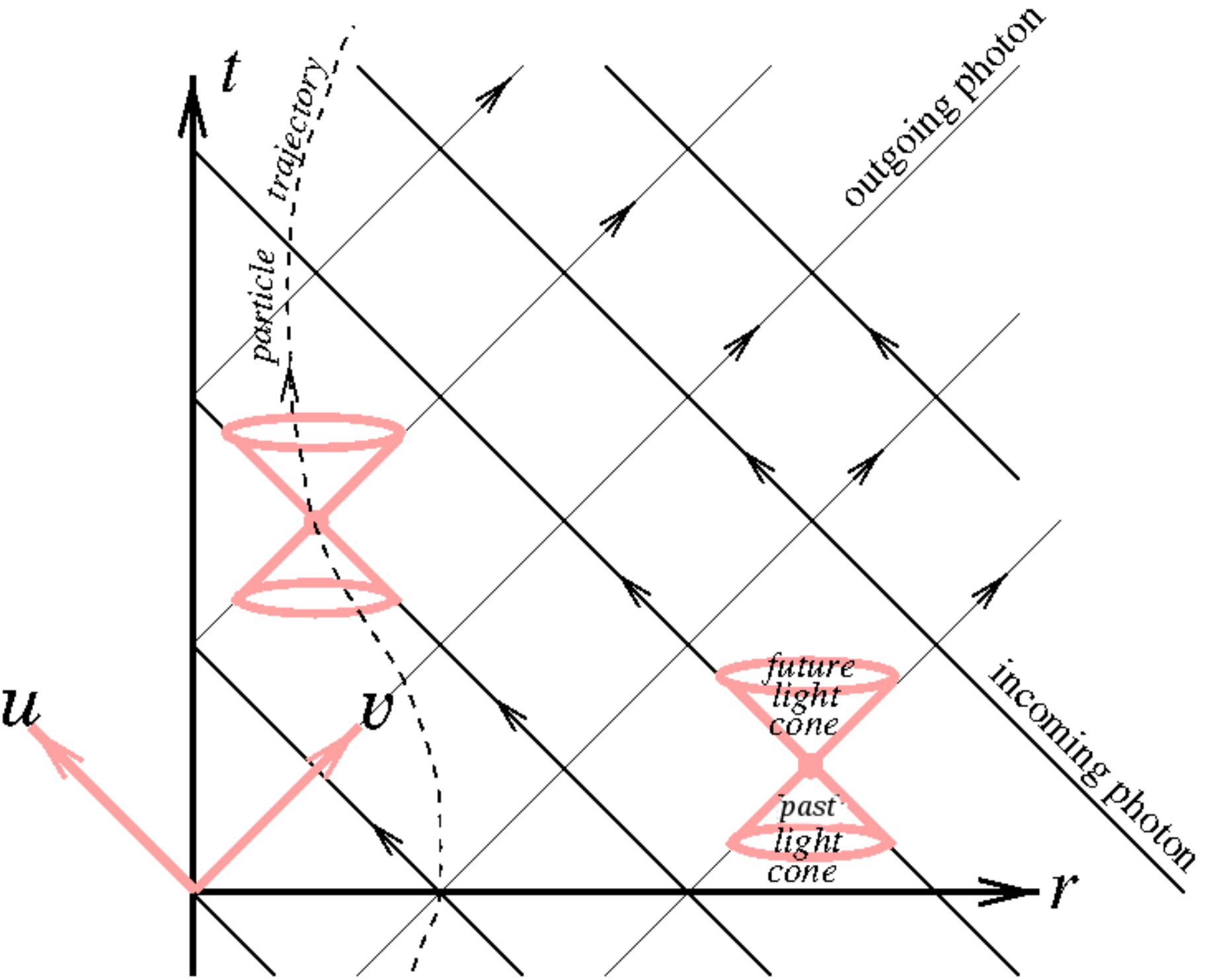}
\end{center}
\caption{Minkowski spacetime in null coordinates\label{chap1:mindianul}}
\end{figure}
and find
\begin{equation}
ds^2 = -dv \, du +\frac{1}{4} \, ( v-u )^2 \, d\Omega^2 \,.
\end{equation}
In \fref{chap1:mindianul} we show the Minkowski spacetime in terms of
the new coordinates.  Incoming photons, that is, point-like particles
with velocity $\dot r = -c=-1$, move on paths with $v=const.$
Correspondingly, we have for outgoing photons $u=const.$ The special
relativistic wave-equation is solved by any function $f(u)$ and
$f(v)$. The surfaces $f(u)=const.$ and $f(v)=const.$ represent the
wavefronts which evolve with the velocity of light.  The trajectory of
every material particle with $\dot{r}<c=1$ has to remain inside the region
defined by the surface $r=t$. In an $(r,t)$-diagram this surface is
represented by a cone, the so-called {\em light cone}. Any point in
the {\em future light cone} $r=t$ can be reached by a particle or
signal with a velocity less than $c$. A given spacetime point $P$ can
be reached by a particle or signal {}from the spacetime region enclosed
by the {\em past light cone} $r=-t$.

\newpage
\subsubsection*{ Penrose diagram}

We can map, following Penrose, the infinitely distant points of spacetime 
into finite regions by means of a conformal transformation which leaves 
the light cones intact. Then we can display the whole infinite Minkowski 
spacetime on a (finite) piece of paper. Accordingly, introduce the new 
coordinates
\begin{equation}
\tilde v := \arctan \, v \,, \quad \tilde u := \arctan \, u\,, \qquad
\text{ for } -\pi/2 \leq (\tilde v \,, \tilde u) \leq + \pi/2 \,.
\end{equation}
Then the metric reads 
\begin{equation}
ds^2 = \frac{1}{\cos^2 \, \tilde{v}} \,\, \frac{1}{\cos^2 \, \tilde u} \,\,
\left[ - d\tilde v \, d\tilde u + \frac{1}{4} \, \sin^2 \, (\tilde v -
\tilde u) \, d\Omega^2 \right]\,.
\end{equation}
We can go back to time- and space-like coordinates by means of the
transformation
\begin{equation}
\tilde t := \tilde v + \tilde u \,,\qquad \tilde r := \tilde v - \tilde u
\,,
\end{equation}
see (\ref{vu}). Then the metric reads,
\begin{equation}
ds^2 = \frac{-d\tilde t ^2 + d \tilde r^2 + \sin^2 \tilde r \,\,
d\Omega^2}{4 \, \cos^2 \, \frac{\tilde t + \tilde r}{2} \,\, \cos^2 \,
\frac{\tilde t - \tilde r}{2} }\,,
\end{equation}
that is, up to the function in the denominator, it appears as a flat
metric. Such a metric is called conformally flat (it is conformal to a
static Einstein cosmos).
The back-transformation to our good old Minkowski coordinates reads
\begin{eqnarray}\label{connulcoo1}
t & = & \frac{1}{2} \, \left(\tan \, \frac{\tilde t + \tilde r }{2} + \tan
\, \frac{\tilde t - \tilde r}{2} \right)\,,  \\ \label{connulcoo2}
r & = & \frac{1}{2} \, \left( \tan \, \frac{\tilde t + \tilde r}{2} - \tan
\, \frac{\tilde t -\tilde r}{2} \right) \,.
\end{eqnarray}\small
\begin{figure}
\bigskip
\begin{center}
\begin{picture}(120,80)
\put(27,0){\includegraphics[height=8cm]{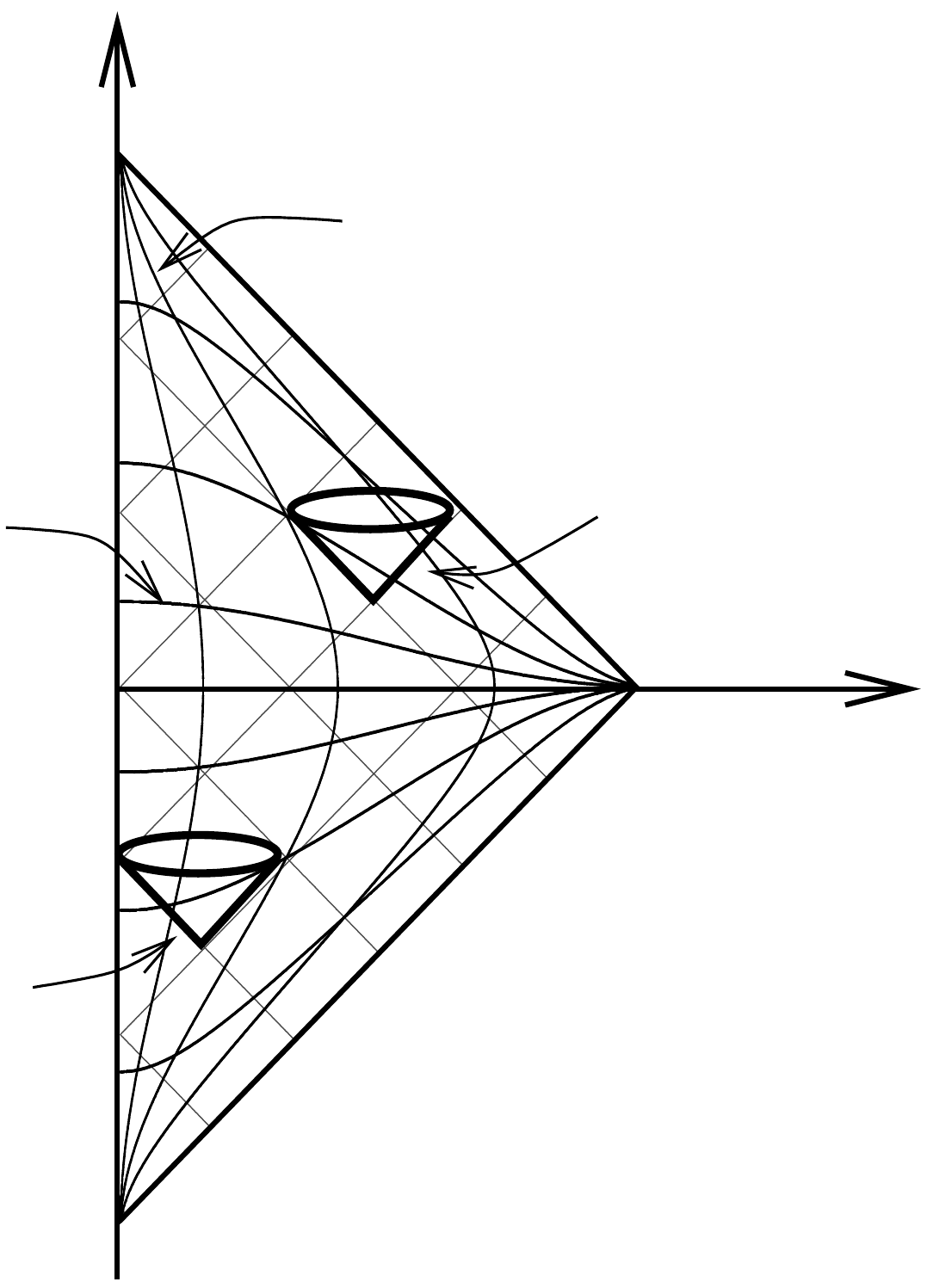}}
\put(36.5,2){\mbox{$-\pi$}}
\put(27.5,2){\mbox{$I^-$}}
\put(37,68){\mbox{$+\pi$}}
\put(28,67){\mbox{$I^+$}}
\put(67,32){\mbox{$+\pi$}}
\put(67,39){\mbox{$I_0$}}
\put(52,56){\mbox{$\cal I^+$}}
\put(52,17){\mbox{$\cal I^-$}}
\put(36.5,78){\mbox{$\tilde{t}$}}
\put(86,36){\mbox{$\tilde{r}$}}
\put(10,46){\mbox{${r}=const$}}
\put(50,65){\mbox{${t}=const$}}
\put(59,49){\mbox{light cone}}
\put(11,18){\mbox{light cone}}
\put(32,22){
\mbox{\begin{rotate}{90}{coordinate singularity}\end{rotate}}}
\end{picture}
\caption{Penrose diagram of Minkowski spacetime.\label{chap1:minpen}}
\end{center}
\end{figure}
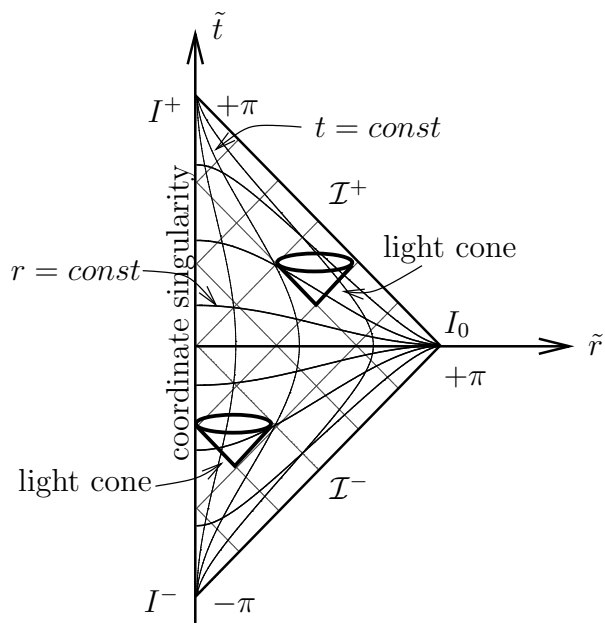\normalsize
Our new coordinates $\tilde{t},\tilde{r}$ extend only over a finite
range of values, as can be seen {}from (\ref{connulcoo1}) and
(\ref{connulcoo2}).  Thus, in the Penrose diagram of a Minkowski
spacetime, see \fref{chap1:minpen}, we can depict the whole Minkowski
spacetime, with a coordinate singularity along $\tilde{r}=0$.
All trajectories of uniformly moving particles (with velocity smaller than
$c$) emerge form one single point, past infinity $I^-$, and all will 
eventually arrive at the one single point $I^+$, namely at future infinity. 
All incoming photons have their origin on the segment ${\cal I}^-$ 
(script $I^-$ or ``scri minus''), light-like past-infinity, and will run 
into the coordinate singularity on the $\tilde{t}$-axis . 
All outgoing photons arise {}from the coordinate singularity 
and cease on the line ${\cal I^+}$, light-like future infinity (``scri
plus''). The entire spacelike infinity is mapped into the single point
$I^0$. For later reference we collect these notions in a
table:\medskip

\begin{center}
\renewcommand{\baselinestretch}{1.4}\small\normalsize
\framebox{\bf Table 1. The different infinities in Penrose diagrams}\\
\begin{tabular}{|l|l|l|}\hline
$I^-$ & timelike past infinity & origin of all particles\\\hline
$I^+$ & timelike future infinity & destination of all
particles\\\hline
$I_0$ & spacelike infinity & inaccessible for all particles\\\hline
$\cal I^-$ & lightlike past infinity & origin of all light rays\\\hline
$\cal I^+$ & lightlike future infinity & destination of all light
rays\\\hline
\end{tabular}
\renewcommand{\baselinestretch}{1}\small\normalsize
\end{center}\medskip

Now, we have a really compact picture of the the Minkowski space. Next, we
would like to proceed along similar lines in order to obtain an analogous 
picture for the Schwarzschild spacetime.

\newpage
\subsection{Einstein's field equation}

{\em We display our notations and conventions for the differential
  geometric tools used to formulate Einstein's field equation.}

\bigskip

We assume that our readers know at least the rudiments of general
relativity (GR) as represented, for instance, in Einstein's {\it
  Meaning of Relativity}\cite{einstein1}, which we still recommend as
a gentle introduction into GR. More advanced readers may then want to
turn to Rindler\cite{rindler} and/or to Landau-Lifshitz\cite{LL}.

We assume a 4d Riemannian spacetime with (Minkowski-)Lorentz signature
$(-++\,+)$, see Misner, Thorne, and Wheeler\cite{misner}. Thus, the
metric field, in arbitrary holonomic coordinates $x^\mu$, with
$\mu=0,1,2,3$, reads
\begin{equation}\label{metric}
\mathbf{g}\equiv ds^2=g_{\mu\nu}\, dx^\mu \otimes dx^\nu\,.
\end{equation}
By partial differentiation of the metric, we can calculate the
Christoffel symbols (Levi-Civita connection)
\begin{equation}\label{Christ}
  \Gamma^\mu{}_{\a\b}:=\frac{1}{2}g^{\mu\gamma}\left(\partial_\a g_{\b\gamma}
    +\partial_\b g_{\gamma\a}-\partial_\gamma g_{\a\b} \right)\,.
\end{equation}
This empowers us to determine the geodesics (curves of extremal
length) of the Riemannian spacetime:
\begin{equation}\label{geodesics}
\frac{D^2 x^\a}{D\tau^2}:= \frac{d^2
  x^\a}{d\tau^2}+\Gamma^\a{}_{\mu\nu}
\frac{dx^\mu}{d\tau}\frac{dx^\nu}{d\tau}=0\,.
\end{equation}
This equation can be read as a vanishing of the 4d covariant
acceleration. If we define the 4-velocity $ u^\a:={dx^\a}/{d\tau}$,
then the geodesics can be rewritten as
\begin{equation}\label{geodesics*}
\frac{D u^\a}{D\tau}= \frac{d
  u^\a}{d\tau}+\Gamma^\a{}_{\mu\nu}u^\mu u^\nu =0\,.
\end{equation}

In a neighborhood of any given point in spacetime we can
introduce Riemann\-ian normal coordinates, which are such that the
Christoffels vanish at that point. In
order to find a tensorial measure of the gravitational field, we have
to go one differentiation order higher. By partial differentiation of
the Christoffels, we find the Riemann curvature tensor\footnote{Always
  symmetrizing of indices is denoted by parentheses,
  $(\a\b):=\{\a\b+\b\a\}/2!$, antisymmetrization by brackets
  $[\a\b]:=\{\a\b-\a\b\}/2!$, with corresponding generalizations
  $(\a\b\gamma):=\{+\a\b\gamma+\b\gamma\a +\gamma\a\b+\cdots\}/3!$,
  etc.; indices standing between two vertical strokes $|\; |$ are
  excluded {}from the (anti)symmetrization process, see
  Schouten\cite{Schouten:1954}.}
\begin{equation}\label{curv}
R^\mu{}_{\nu\a\b}:=
2\left(\partial_{[\a}\Gamma^\mu{}_{|\nu|\b]}+\Gamma^\mu{}_{\sigma[\a} 
\Gamma^\sigma{}_{|\nu|\b]}\right)\,.
\end{equation}
The curvature is doubly antisymmetric, its two index pairs commute,
and its totally antisymmetric piece vanishes:
\begin{equation}\label{symmetries}
  R_{(\mu\nu)\a\b}=0\,,\,R_{\mu\nu(\a\b)}=0\,;\quad R_{\mu\nu\a\b}=R_{\a\b\mu\nu}\,;\quad
  R_{[\mu\nu\a\b]}=0\,.
\end{equation}
If we define collective indices $A,B,..=1,...,6$ for the antisymmetric
index pairs according to the rule $\{01,02,03;$ $23,31,12\}$
$\longrightarrow\{1,2,3;4,5,6 \}$, then the algebraic symmetries of
(\ref{symmetries}) can be rephrased as
\begin{equation}\label{symmetries*}
R_{AB}=R_{BA}\,,\qquad \text{trace}(R_{AB})=0.
\end{equation}
Thus, in 4d the curvature can be represented as a trace-free symmetric
$6\times 6$-matrix. Hence it has 20 independent components.

With the curvature tensor, we found a tensorial measure for the
gravitational field. Freely falling particles move along geodesics of
Riemannian spacetime. What about the tidal accelerations between two
freely falling particles? Let the ``infinitesimal'' vector $n^\a$
describe the distance between two particles moving on adjacent
geodesics. A standard calculation\cite{misner}, linear to the order of
$n$, yields the {\it geodesic deviation equation}
\begin{equation}\label{chap1:geodev}
  \frac{D^2 n^\a}{D\tau^2}=u^\b u^\gamma R^\a{}_{\b\gamma\d}\, n^\d\,.
\end{equation}
This equation describes the relative acceleration of neighboring particles,
similar as (\ref{newtondeviation}) and (\ref{TidalForce}) 
in the Newtonian case. The role of the tidal matrix
$K_{ab}$ is taken over by  ${\cal K}^\alpha{}_\delta :=
u^\b u^\gamma R^\a{}_{\b\gamma\d}$.

By contraction of the curvature, we can define the 2nd rank Ricci
tensor $R_{\mu\nu}$ and the curvature scalar $R$, respectively:
\begin{equation}\label{Ricci}
  R_{\mu\nu}:=R^\a{}_{\mu\a\nu}\,,\qquad R:=g^{\mu\nu}R_{\mu\nu}\,.
\end{equation}
For convenience, we can also introduce the Einstein tensor
$G_{\mu\nu}:=R_{\mu\nu}-\frac 12 g_{\mu\nu}R$.  The curvature with its
20 independent components can be irreducibly decomposed into smaller
pieces according to $20=10+9+1$. The Weyl curvature tensor
$C_{\a\b\gamma\d}$ is trace-free and has $10$ independent components,
whereas the trace-free Ricci tensor has $9$ components and the curvature
scalar just $1$.

Now we have all the
tools for displaying Einstein's field equation. With $G$ as Newton's
gravitational constant and $c$ as velocity of light, we define
Einstein's gravitational constant $\kappa:=8\pi G/c^4$. Then, the {\it
  Einstein field equation} with cosmological constant $\Lambda$ reads
\begin{equation}\label{FieldEq}
 R_{\mu\nu}-\frac 12 g_{\mu\nu}R+\Lambda g_{\mu\nu}=\kappa\, T_{\mu\nu}\,.
\end{equation}
The source on the right-hand side is the energy-momentum tensor of
matter. The {\it vacuum} field equation, without cosmological
constant, simply reduces to $R_{\mu\nu}=0$. Mostly this equation will
keep us busy in this article. A vanishing Ricci tensor implies that
only the Weyl curvature $C_{\a\b\gamma\d}\ne 0$. Accordingly, the
vacuum field in GR (without $\Lambda$) is represented by the Weyl
tensor.

Eq.(\ref{FieldEq}) represents a generalization of the Poisson equation
(\ref{tidaltrace}). There, the contraction of the tidal matrix is
proportional to the mass density; in GR, the contraction of the
curvature tensor is proportional to the energy-momentum tensor.

The physical mass is denoted by $M$. Usually, we use the {\em mass
parameter}, $m := \frac{GM}{c^2}$. The Schwarzschild radius reads
$r_S:=2m=\frac{2GM}{c^2}$. Usually we put $c=1$ and $G=1$. We make explicitly
use of $G$ and $c$ as soon as we stress analogies to Newtonian gravity or
allude to observational data.

\vfill
\null
\pagebreak

%%%%%%%%%%%%%%%%%%%%%%%%%%%%%%%%%%%%%%%%%%%%%%%%%%%%%%%%%%%%%%%%%%%%%%%%%
\section{The Schwarzschild metric (1916)}
%%%%%%%%%%%%%%%%%%%%%%%%%%%%%%%%%%%%%%%%%%%%%%%%%%%%%%%%%%%%%%%%%%%%%%%%%

{\it Spatial spherical symmetry is assumed and a corresponding exact
  solution for Einstein's theory searched for. After a historical
  outline (Sec.2.1), we apply the equivalence principle to a freely
  falling particle and try to implement that on top of the Minkowskian
  line element. In this way, we heuristically arrive at the
  Schwarzschild metric (Sec.2.2). In Sec.2.3, we display the
  Schwarzschild metric in six different classical coordinate systems.
  We outline the concept of a Schwarzschild black hole in Sec.2.4.
  In Secs.2.5 and 2.6, we construct the Penrose diagram for the
  Schwarzschild(-Kruskal) spacetime. We add electric charge to
  the Schwarzschild solution in Sec.2.7. The interior Schwarzschild metric, 
 with matter, is addressed in Sec.2.8.}

\bigskip

\null\hfill\begin{minipage}{10cm} \small{\it It is quite a wonderful
  thing that {}from such an abstract idea the explanation of the Mercury
  anomaly emerges so inevitably.}\vspace{-7pt}

%Es ist eine ganz wunderbare Sache, dass von einer so abstrakten Idee aus die
%Erkl\"arung der Merkuranomalie so zwingend herauskommt.\rm\\
\null\hfill{Karl Schwarzschild\cite{schwarzschild} (1915)}
\end{minipage}

\subsection{Historical remarks}

{\it The genesis of the Schwarzschild solution (1915/16) is
  described. In particular, we show that Droste, a bit later than
  Schwarzschild, arrived at the Schwarzschild metric independently. He
  put the Schwarzschild solution into that form in which we use it
  today.}\medskip

The first exact solution of Einstein's field equation was born in
hospital.  Unfortunately, the circumstances were more tragic than
joyful. The astronomer Karl Schwarzschild joined the German army right
at the beginning of World War I and served in Belgium, France, and
Russia. At the end of the year 1915, he was admitted to hospital with
an acute skin disease. There, not far {}from the Russian front, enduring
the distant gunfire, he found time to ``stroll through the land of
ideas" of Einstein's theory, as he puts it in a letter to Einstein\footnote{
The letters {}from and to Einstein can be found in Einstein's Collected
Works\cite{EinsteinCollWorks}, see also Schwarzschild's Collected
Works\cite{schwarzschild}.} dated 22 December 1915. According to this letter,
Schwarzschild started out
{}from the approximate solution in Einstein's ``perihelion paper",
published November 25th. Since presumably letters {}from
Berlin to the Russian front took a few days,
Schwarzschild\cite{schwarzschild1} found the solution within about a
fortnight. Fortunately, the premature field equation of the 
``perihelion paper" is correct in the vacuum case treated by Schwarzschild.

In February 1916, Schwarzschild\cite{schwarzschild2}
submitted the spherically symmetric solution with matter---the ``interior
Schwarzschild solution"---now based on Einstein's
final field equation.  In March 1916, he was sent home
were he passed away on 11 May 1916.

The field equation used by Schwarzschild requires $\det g = -1$.
To fulfill this condition, he uses modified polar coordinates  
(Schwarzschild's original notation used),
\[
x_1 = \frac{r^3}{2}\,,\quad x_2 = -\cos\theta\,,\quad x_3 = \phi\,,\quad
x_4 = t\,.
\]
The spherically symmetric ansatz then reads\small
\[
ds^2 = f_4\,dx_4^2-f_1\,dx_1^2 -
f_2\,\frac{dx_2^2}{1-x_2^2}-f_3\,dx_3^2\,(1-x_2^2)\,,
\]\normalsize
where $f_1$ to $f_4$ are functions of $x_1$ only.
The solution turns out to be
\[
f_1=\frac{1}{R^4}\,\frac{1}{1-\alpha/R}\,,\quad 
f_2=f_3=R^2\,,\quad
f_4=1-\alpha/R\,,\qquad R=(r^3+\alpha^3)^{1/3}\,.
\]
In this article, as well as in his letter to Einstein, he eventually
returns to the usual spherical polar coordinates,
\[
ds^2=(1-\alpha/R)\, dt^2 -\frac{dR^2}{1-\alpha/R} - R^2\, (d\theta^2 +
\sin^2\theta \, d\phi^2)\,,\quad R=(r^3+\alpha^3)^{1/3}\,.
\]
This looks like the Schwarzschild metric we are familiar with. One
should note, however, that the singularity at $R=\alpha$ is (as we
know today) a coordinate singularity, it corresponds to $r=0$. In the
early discussion the meaning of such a singularity was rather obscure.
Flamm\cite{Flamm:1916} in his 1916 article on embedding constant time
slices of the Schwarzschild metric into Euclidean space mentions ``the
oddity that a point mass has an finite circumference of $2\pi\alpha$".

In 1917, Weyl\cite{Weyl:1917} talks of the ``inside" and ``outside" of
the point mass and states that ``in nature, evidently, only that piece
of the solution is realized which does not touch the singular sphere."
In Hilbert's\cite{Hilbert:1917} opinion, the singularity $R=\alpha$
indicates the illusiveness of the concept of a pointlike mass; a point
mass is just the limiting case of a spherically symmetric mass
distribution. Illuminating the interior of ``Schwarzschild's sphere''
took quite a while and it was the discovery of new coordinates which
brought first elucidations.  Lanczos\cite{Lanczos:1922}, in 1922,
clearly speaks out that singularities of the metric components do not
necessarily have physical significance since they may vanish in
appropriate coordinates. However, it took another 38 years to find a
maximally extended fully regular coordinate system for the
Schwarzschild metric. We will become acquainted with these
Kruskal/Szekeres coordinates in Sec.\ref{kruskalszekeres}.

Schwarzschild's solution, published in the widely read minutes of the
Prussian Academy, communicated by Einstein himself, nearly instantly
triggered further investigations of the gravitational field of a point
mass. Already in March 1916, Reissner\cite{Reissner:1916}, a civil
engineer by education, published a generalization of the Schwarzschild
metric, including an electrical charge; this was later completed by
Weyl\cite{Weyl:1917} and by Nordstr\"om\cite{Nordstrom:1918}. Today it
is called Reissner-Nordstr\"om solution.

Nevertheless, one should not ignore the Dutch twin of Schwarzschild's
solution.  On 27 May 1916, Droste\cite{Droste:1916} communicated his
results on ``the field of a single centre in Einstein's theory of
gravitation, and the motion of a particle in that field'' to the Dutch
Academy of Sciences. He presents a very clear and easy to read
derivation of the metric and gives a quite comprehensive analysis of
the motion of a point particle. Since 1913, he had been working on
general relativity under the supervision of Lorentz at Leiden
University.  {\it Published in Dutch,} Droste's results are fairly
unknown today. Einstein, probably informed by his close friend
Ehrenfest, rather appreciated Droste's work, praising the graceful
mathematical style. Weyl\cite{Weyl:1917} also cites Droste, but in
Hilbert's\cite{Hilbert:1917} second communication the reference is not
found. Einstein, Hilbert, and Weyl always allude to ``Schwarzschild's
solution''.

After Droste took his PhD in 1916, he worked as school teacher and
eventually became professor for mathematics in Leiden. He never
resumed his work on Einstein's theory and his name faded {}from the
relativistic memoirs. In Leiden, people like Lorentz, de Sitter,
Nordstr\"om, or Fokker learned about the gravitational field of a
point mass primarily {}from Droste's work.  Thus, the name
``Schwarzschild--Droste solution'' would be quite justified {}from a
historical point of view.

The importance of the Schwarzschild metric is made evident by the
Birk\-hoff\cite{Birkhoff} theorem\footnote{The ``Birkhoff'' theorem was
  discovered by Jebsen\cite{Jebsen}, Birkhoff\cite{Birkhoff}, and
  Alexandrow\cite{Alexandrow}. For more details on Jebsen, see
  Johansen \& Ravndal\cite{Johansen}. The objections of Ehlers \&
  Krasi\'nski\cite{Ehlers} appear to us as nitpicking.}: For
vanishing cosmological constant, the unique spherically symmetric
vacuum spacetime is the Schwarzschild solution, which can be expressed
most conveniently in Schwarzschild coordinates, see Table 3, entry 1.
Thus, a spherically symmetric body is static (outside the horizon). In
particular, it cannot emit gravitational radiation. Moreover, the
asymptotic Minkowskian behavior of the Schwarzschild solution is
dictated by the solution itself, it is {\it not} imposed {}from the
outside.

%%%%%%%%%%%%%%%%%%%%%%%%%%%%%%%%%%%%%%%%%%%%%%%%%%%%%%%%%%%%%%%%%%%%%%%%%%%%
\subsection{Approaching the Schwarzschild metric}
%%%%%%%%%%%%%%%%%%%%%%%%%%%%%%%%%%%%%%%%%%%%%%%%%%%%%%%%%%%%%%%%%%%%%%%%%%%%

{\it We start {}from an ansatz for the metric of an accelerated motion
  in the radial direction and combine it, in the sense of the equivalence
  principle, with the free-fall velocity of a particle in a Newtonian
  gravitational field. In this way, we find a curved metric that,
  after a coordinate transformation, turns out to be the Schwarzschild
  metric.}\medskip

Einstein, in his 1907 {\em Jahrbuch} article\cite{Einstein:1907},
suggests the generalization of the relativity principle to arbitrarily
accelerated reference frames.

A plausible notion of a (local) rest frame in general relativity is a frame
where the coordinate time is equal to the proper time (for an observer
spatially at rest, of course). For a purely radial motion, the following
metric would be an obvious ansatz, see also Visser\cite{Visser:2003tt}:
\[
ds^2 = -dt^2 + [dr+f(r)\,dt]^2+
r^2\,d\Omega^2\,,\qquad\text{with}\qquad d\Omega^2
:=d\theta^2+\sin^2\theta d\phi^2\,.
\]
For $d\phi=0$, $d\theta=0$, and ${dr}/{dt}=-f(r)$, we have $ds^2=-dt^2$.
Thereby, $-f(r)$ is identified as a kind of ``radial infall velocity''.
Note also that constant time-slices, $dt=0$, are Euclidean.

In Newtonian gravity, a particle falling {}from infinity towards the
origin picks up a velocity
\begin{equation}\label{falling}
  \frac{dr}{dt}=v=-\sqrt{2\Phi(r)}=-\sqrt{\frac{2GM}{r}} 
  \quad \Longleftrightarrow\quad
  \frac{1}{2}m v^2(r) = m\Phi(r) = m\frac{GM}{r}\,.
\end{equation}
Here, $\Phi$ is the absolute value of the Newtonian potential of a
spherical body with mass $M$.

Hence, in some Newtonian limit, we demand $f(r)\rightarrow \sqrt{2\Phi}$.
This leads to the metric
\begin{equation}\label{gullstrand_ansatz}
ds^2 = -dt^2 +\big(dr + \sqrt{2\psi}\, dt \big)^2 + r^2\,d\Omega^2\,,
\end{equation}
where we allow for an arbitrary potential $\psi=\psi(r)$.  This metric
generates curvature. The calculations can be conveniently done even by
hand. The Ricci tensor reads
\[
R_0{}^0=R_1{}^1=\frac{1}{r}\,\partial_r\partial_r(r\, \psi)=0\,,\qquad
R_2{}^2=R_3{}^3=\frac{2\partial_r(r\,\psi)}{r^2}=0\,.\qquad
\]
The equations $R_0{}^0=0=R_1{}^1$ are mere integrability
conditions of the $R_2{}^2=0=R_3{}^3$ relations. Hence, $r\psi$ is determined by
its first order approximation alone and reads
\[
\psi = \frac{\alpha}{r}\,,
\]
with $\alpha$ as an unknown constant so far. By construction, we have
\[
\frac{dr}{dt}=-\sqrt{2\psi}=-\sqrt{\frac{2\alpha}{r}} 
\stackrel{!}{=}
-\sqrt{\frac{2GM}{r}}\quad \Longrightarrow \quad
\alpha = GM=:m\,.
\]

The metric (\ref{gullstrand_ansatz}), expanding the parenthesis and
collecting the terms in front of $d t^2$, reads
\begin{equation}\label{GPmetric}
ds^2 = - \left(1-\frac{2GM}{r}\right)\,dt^2
 + 2 \,\sqrt{\frac{2GM}{r}}\, dt
dr + dr^2 +r^2 d\Omega ^2\,.
\end{equation}%\setcounter{footnote}{0}
Using different methods, this metric was derived by
Gullstrand\cite{Gullstrand} in May 1921.  Gullstrand claimed to have
found a new spherically symmetric solution of Einstein's field
equation. In his opinion\footnote{Gullstrand, who was a member of the
  Nobel committee, was responsible that Einstein did not get his Nobel
  prize for relativity theory. He thought that GR is untenable.}, this
showed the ambiguity of Einstein's field equation.
However, the metric is of the form
\[
ds^2 = -A\, dt^2 + 2B dt\,dr+dr^2 + r^2\,d\Omega^2\,,\qquad
A:=1-\frac{2GM}{r}\,,\quad B:=\sqrt{\frac{2GM}{r}}\,,
\]
and can be diagonalized by completing the square via
\begin{eqnarray*}
ds^2&=&
-A\,\left(dt -\frac{B}{A}\,dr\right)^2 +\left(1+\frac{B^2}{A}\right)\,dr^2 
+r^2\,d\Omega^2\,.\\
\end{eqnarray*}
Introducing a new time coordinate,
\[
dt_{\rm S} := dt  -\frac{B}{A}\,dr
\]
or, explicitly,
%\[
$
t_{\rm S} = t - \left(2r\sqrt{\frac{2GM}{r}}-{4GM}\, {\rm Artanh}
\sqrt{\frac{2GM}{r}}\right)\,,
$
%\]
we arrive at ($A$ and $B$ re-substituted)
\[
ds^2=-\left(1-\frac{2GM}{r}\right)\,dt^2_S
+\left(1-\frac{2GM}{r}\right)^{-1}\,dr^2+r^2\,d\Omega^2\,.
\]
In contrast to what Gullstrand was aiming at, he ``just'' rederived
the Schwarz\-schild metric.

Later, applying a coordinate transformation to the Schwarzschild
metric, Painlev\'{e}\cite{Painleve} obtained the metric
(\ref{GPmetric}) independently and presented his result in October
1921. His aim was to demonstrate the vacuity of $ds^2$ by showing that
an exact solution does not determine the physical geometry and is
therefore meaningless. In a letter (Dec.\ 7$^{th}$ 1921) to
Painlev\'{e}, Einstein stresses on the contrary {\it the
meaninglessness of the coordinates!} In the words of Einstein himself
(our
translation): ``\dots merely results obtained by eliminating the coordinate 
dependence can claim an objective meaning.''

In the subsequent section, we
will meet the Schwarzschild metric in many different coordinate
systems. All of them have their merits and their shortcomings.

Using Gullstrand-Painlev\'{e} coordinates for the Schwarzschild metric
does not change the physics, of course. However, as a coordinate
system it is what Gustav Mie\cite{Mie:1920} calls a {\em sensible}
coordinate system. In contrast to many other coordinate systems, the
physics looks quite like we are used to. As an example, we analyze the
motion of a radial infalling particle in Schwarzschild and
{\it Gullstrand-Painlev\'{e} coordinates.}

The equations of motion for point particles in general relativity are
obtained via the geodesic equation (\ref{geodesics}). It can be shown
that this equation is equivalent to the solution of the variational
principle $\delta \,\int_{x^\alpha} ds^2 = \delta \,\int \dot x^\alpha
\, \dot x^\beta \,g_{\alpha\beta}\, d\tau^2$. We choose the proper time
$\tau$ for the parametrization of the curve, the dot denotes the
derivative with respect to $\tau$. In the present context, we are only
interested in the velocity of particles along ingoing geodesics
(``freely falling particles"). For time-like geodesics we have
$-1=\frac{ds^2}{d\tau^2}$. This allows the algebraic determination of
$\dot r$ provided we know $\dot t$. Since we consider static metrics
here, $t$ is a cyclic variable and
%$\frac{d}{d\tau}\,\left(\frac{\partial}{\partial\dot t}\,
%\frac{ds^2}{d\tau^2}\right) = 0$ or 
$\left(\frac{\partial}{\partial\dot t}\,
\frac{ds^2}{d\tau^2}\right) = K = const$. The constant is determined by the
boundary condition $\dot r = 0$ for $r\to\infty$. The calculation yields:

%\pagebreak
\begin{center}
{\bf Table 2.}
Velocities in different coordinate 
systems\footnote{The velocities of outgoing particles
are valid only for the boundary condition specified. The {\em coordinate} 
velocity for outgoing particles in GP coordinates does not fit in our table 
and is thus suppressed.}
\medskip

\small
\begin{tabular}{|c|c|c|c|}\hline
&&Schwarzschild&Gullstrand-Painlev\'{e}\\\hline
&\parbox{1.4cm}{\centerline{coordinate}\centerline{velocity}
\centerline{$\frac{dr}{dt}$}}&
$\pm\left(1-\frac{2GM}{r}\right) \sqrt{\frac{2GM}{r}}$&
$-\sqrt{\frac{2GM}{r}}$\\
particles&&&\\
&\parbox{1.2cm}{\centerline{proper}\centerline{velocity}
\centerline{$\frac{dr}{d\tau}$}}&$\pm\sqrt{\frac{2GM}{r}}$&$\pm\sqrt{\frac{2GM}{r}}$\\
&&&\\\hline
light rays &\parbox{1.2cm}{\centerline{coordinate}\centerline{velocity}
\centerline{$\frac{dr}{dt}$}}&$\pm(1-\frac{2GM}{r})$
&$\pm 1 -\sqrt{\frac{2GM}{r}}$
\\\hline
\end{tabular}
\end{center}\normalsize\medskip

The difference between the coordinate systems appears in the first line of
Table~2: In Gullstrand-Painlev\'e coordinates,
the coordinate velocity of a freely infalling particle increases
smoothly towards the center. Nothing special happens at $r=2GM$. 
{}From a given position, the particle will plunge into the center in a finite
time. Even numerically this looks quite Newtonian. In contrast, the 
velocity with respect to Schwarzschild coordinates approaches zero 
as the particle approaches $r=2GM$. Hence, the particle apparently will not be 
able to go further than $r=2GM$. 

For the Gullstrand-Painlev\'{e} metric for incoming light the radial
coordinate velocity is always larger in magnitude than $-1$, at $r=2GM$
it is $-2$, for outgoing rays it vanishes at $r=2GM$ and is negative
for $r<2GM$.

Taking the mere numerical values is misleading. Contemplate for
incoming light
\[
\frac{\left(\frac{dr}{dt}\right)_{\text{particle}}}
{\left(\frac{dr}{dt}\right)_{\text{light}}}
= \frac{1}{1+\sqrt{\frac{r}{2GM}}}\leq 1\,.
\]
So the particle is always slower than light, however it approaches the
velocity of light when approaching $r=0$.

The Gullstrand-Painlev\'{e} form of the metric is regular at the
surface $r=2GM$. This shows that it is not any kind of
barrier, but this observation was not made until much later, see
Eisenstaedt \cite{Eisenstaedt}.

%\newpage
%%%%%%%%%%%%%%%%%%%%%%%%%%%%%%%%%%%%%%%%%%%%%%%%%%%%%%%%%%%%%%%%%%%%%%%%%%%
\subsection{Six classical representations of the Schwarzschild metric}
%%%%%%%%%%%%%%%%%%%%%%%%%%%%%%%%%%%%%%%%%%%%%%%%%%%%%%%%%%%%%%%%%%%%%%%%%%%

{\em As we mentioned, a coordinate system should be chosen according to its
convenience for describing a certain situation. In the following table
(Table 3), we collect six widely used forms of the Schwarzschild metric.}

%{\bf Table 3. The Schwarzschild metric in various
%  coordinates}\bigskip

\begin{sidewaystable*}

\renewcommand{\baselinestretch}{1.3}\small\normalsize
%\tbl{The Schwarzschild metric in various coordinates}
{\begin{tabular}{||p{10.5cm}|p{4.5cm}|p{4.5cm}||} \hline\hline\large{\bf
      Table 3.}  Schwarzschild metric in various coordinates &
    coordinate transformation & characteristic
    properties\\
    \hline\normalsize
%%%%%%%%%%%%%%%%%%%%%%%%%%%%%%%%%%%%%%%%%%%%%%%%%%%%%%%%%%%%%%%%%%%%%%%%%%%%%
\vbox{\null%\vspace{2mm}\normalsize
{\bf Schwarzschild}\hfill $(t,r,\theta,\phi)$ 
%\vspace{0.5cm}
$
ds^2 = -\left(1-\frac{2m}{r} \right) \, dt^2 +
\frac{1}{1-\frac{2m}{r}} \, dr^2 + r^2 \, d\Omega^2 \,.
$}
&
\vbox{\qquad\hrule width 4cm}
&
\vbox{area of spheres $r=\text{const}$} 
\vbox{is the ``Euclidean'' $4\pi r^2$ }
\\\hline
%%%%%%%%%%%%%%%%%%%%%%%%%%%%%%%%%%%%%%%%%%%%%%%%%%%%%%%%%%%%%%%%%%%%%%%%%%%%%%
\vbox{\null%\vspace{2mm}
{\bf Isotropic}\hfill $(t,\bar r, \theta,\phi)$ 
%\vspace{0.5cm}
$%\label{IsotropicCoo}  
ds^2 = -\left(\frac{1-{m}/{2\bar{r}}}{1+{m}/{2\bar{r}}}\right)^{\!2} \,
dt^2  +\left(1+\frac{m}{2\bar{r}}\right)^{\hspace{0.5pt}4} \, \left(d\bar{r}^2 + \bar{r}^2 \,
d\Omega^2\right)
$}
&
\vbox{\null\vspace{1mm}
$
r = \bar{r} \, \left(1+\frac{m}{2\bar{r}}\right)^2 \,.
$}
&
\vbox{constant-curvature} 
\vbox{time slices}
\\\hline
%%%%%%%%%%%%%%%%%%%%%%%%%%%%%%%%%%%%%%%%%%%%%%%%%%%%%%%%%%%%%%%%%%%%%%%%%%%%%
\vbox{\null\vspace{2mm}
{\bf Eddington-Finkelstein}\hfill $(u,r,\theta,\phi)$
%\vspace{0.5cm}
$%\label{chap1:metoutedd}
ds^2 = - \left( 1-\frac{2m}{r}\right) \, dv^2 + 2 dv \, dr + r^2 \,
d\Omega^2 \,.  
$}
&
\vbox{$
v = t+r+2m\,\ln\left|\frac{r}{2m}-1\right|
$}
&
\vbox{
ingoing light rays:
$
dv= 0
$ 
}
\\\hline
%%%%%%%%%%%%%%%%%%%%%%%%%%%%%%%%%%%%%%%%%%%%%%%%%%%%%%%%%%%%%%%%%%%%%%%%%%%%%
\vbox{\null%\vspace{2mm}
{\bf Kerr-Schild}\hfill $(\bar t,x,y,z)$

$
ds^2= \left(\eta_{\alpha\beta} +2m \ell_\alpha
\ell_\beta\right)\,dx^\alpha \, dx^\beta
\mbox{ ; }
\ell_\alpha=\frac{1}{\sqrt{r}}\,\left(1,\frac{x}{\sqrt{r}},
\frac{y}{\sqrt{r}}\frac{z}{\sqrt{r}}\right)
$ }
&
\vbox{
$
\bar t = v-r
$

$
r^2=x^2+y^2+z^2
$}
&
``Cartesian'' coordinates 
\\\hline
%%%%%%%%%%%%%%%%%%%%%%%%%%%%%%%%%%%%%%%%%%%%%%%%%%%%%%%%%%%%%%%%%%%%%%%%%%%%%
\vbox{\null\vspace{2mm}
{\bf Lema\^{\i}tre}\hfill $(T,R,\theta,\phi)$
%\vspace{0.5cm}
$
ds^2= -dT^2 +\frac{2m}{r}\,dR^2 +r^2\,d\Omega^2
\,,\quad r=\left[\frac{2\sqrt{2m}}{3}\,(R-T)\right]^{\!\frac{2}{3}}
$}
&
\vbox{
$dT = dt + \sqrt{\frac{2m}{r}}\,\frac{1}{1-\frac{2m}{r}}\,dr$

$dR = dt + \sqrt{\frac{r}{2m}}\,\frac{1}{1-\frac{2m}{r}}\,dr$
}
&
\vbox{
infalling particles:
$
dR=0
$ 
}
\\\hline
%%%%%%%%%%%%%%%%%%%%%%%%%%%%%%%%%%%%%%%%%%%%%%%%%%%%%%%%%%%%%%%%%%%%%%%%%%%%%
\vbox{\null%\vspace{2mm}
{\bf Gullstrand-Painlev\'{e}}\hfill $(\tilde t,r,\theta,\phi)$
%\vspace{0.5cm}

$
ds^2 = - \left(1-\frac{2m}{r}\right)\,d\tilde t\,{}^2 
+ 2 \,\sqrt{\frac{2m}{r}}\, d\tilde t\,
dr + dr^2 +r^2 d\Omega ^2
$}
&
$dt= d\tilde t - \frac{dr}{\sqrt{\frac{r}{2m}}-\sqrt{\frac{2m}{r}}}$
%\tilde t = t - 2r\sqrt{\frac{2m}{r}}+\frac{2m}{c^3}\, {\rm Artanh}
%\sqrt{\frac{2m}{r}}
%$%}
&
\vbox{
infalling particles:

${d r}= -\sqrt{\frac{2m}{r}} \,d\tilde t$
}
\\\hline
\end{tabular}}\renewcommand{\baselinestretch}{1}\small\normalsize
\end{sidewaystable*}
\bigskip

%Before we analyze further the Schwarzschild metric, we would like to
%convey first some intuitive pictures of its properties.
\newpage
%%%%%%%%%%%%%%%%%%%%%%%%%%%%%%%%%%%%%%%%%%%%%%%%%%%%%%%%%%%%%%%%%%%%%%%%%%%%%
\subsection{The concept of a Schwarzschild black hole}
%%%%%%%%%%%%%%%%%%%%%%%%%%%%%%%%%%%%%%%%%%%%%%%%%%%%%%%%%%%%%%%%%%%%%%%%%%%%%

{\em We first draw a simple picture of a black hole. The event horizon and
the stationary limit emerge as characteristic features. These are
subsequently defined in a more mathematical way.}

In 1783 John Michell communicated his thoughts {\em on the means of
  discovering the Distance, magnitude, etc.\ of the fixed stars, in
  consequence of the diminuation of the velocity of their light \dots}
\cite{Michell:1784} to the Royal Society in London. In the context of
Newton's particle theory of light, he calculated that sufficiently
massive stars exhibit a gravitational attraction to such vast an
amount that even light could not escape.  A few years later (1796)
Pierre-Simon Laplace published similar ideas.

In modern notation, we may reconstruct the arguments as follows.  We
throw a mass $m$ {}from the surface of the Earth, assuming that there
were no air, in upward direction with an initial velocity $v$. It will
always fall back, unless its initial velocity reaches a sufficiently
high value $v_{\text{escape}}$ providing the mass with such a kinetic
energy that it can overpower the gravitational attraction of the
Earth. Energy conservation yields then immediately the formula
\[
%c \stackrel{!}{=} 
v_{\text{escape}} = \sqrt{\frac{2GM_\oplus}{R_\oplus}}\,,
\]
where $G$ is Newton's gravitational constant and $M$ and $R_\oplus$ the mass
and the radius of the spherically conceived Earth,
respectively. 

For the Earth we find $v_{\text{escape}}\approx 11.2\, {\rm km/s}$. If
we now compress the Earth appreciably (thought experiment!) until the
escape velocity coincides with the speed of light
$v_{\text{escape}}=c$, its compressed ``Schwarzschild'' radius
becomes $r_\oplus={2GM_\oplus}/{c^2}\approx 1\,\text{cm}$. For the
Sun, with its mass $M_\odot$, we 
have\footnote{For the sake of clarity, we display here the speed of light
$c$ explicitly.}
$$r_\odot=\frac{2GM_\odot}{c^2}\approx 3\,{\rm km}\,.$$
At any smaller radius the light will be confined to the corresponding
body. This is an intuitive picture of a spherically symmetric
invisible ``black hole''.\footnote{The phrase ``black hole'' is commonly
associated with Wheeler (1968). It appears definitely earlier in the
literature: In the January 1964 edition of the {\em Science News Letter} the
journalist Ann Ewing entitled her report at the meeting of the
American Association for the Advancement of Science in Cleveland {\em
``Black Holes'' in space}. And if you have a look into an arbitrary English
language dictionary published before ca.\ 1970, you will learn that ``black
hole'' refers to a notorious dungeon in Calcutta (now Kolkata)
in the 18th century,
apparently a place of no return \dots}

It is very intriguing to see how far-sighted Michell anticipated the
status of today's observational black hole physics:

{\em\small If there should really exist in nature bodies, whose density is
  not less than that of the sun, and whose diameters are more than 500
  times the diameter of the sun, since their light could not arrive at
  us; [\dots] we could have no information {}from sight; yet, if any
  other luminous bodies should happen to revolve about them we might
  still perhaps {}from the motions of these revolving bodies infer the
  existence of the central ones with some degree of probability \dots}

This could be a verdict on the current observations of the black hole
Sgr\! A$\!^*$ (``Sagittarius A-star'') in the center of our Milky
Way---and this is {\it not} a thought experiment---for a popular
account, see Sanders\cite{Sanders:2014}.  Sgr\! A$\!^*$ has a mass of
about $4\times 10^6\,M_\odot$. Thus its Schwarzschild radius is far
{}from being minute, it is about $3\times 4\times 10^6\,\text{km}$ or
about 17 solar radii.

\begin{figure}[h]
\includegraphics[height=4.6cm]{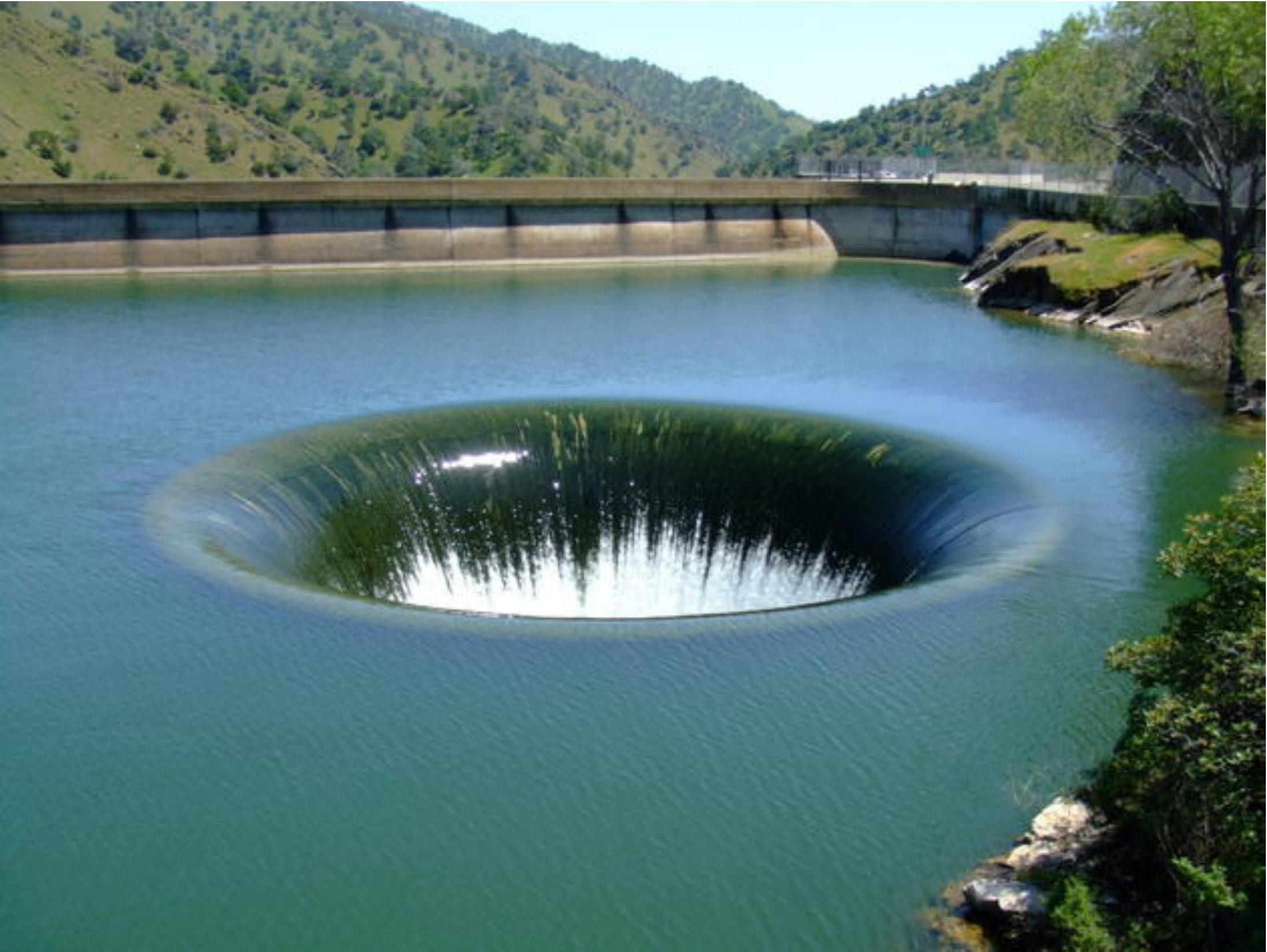}\qquad
\includegraphics[height=4.6cm]{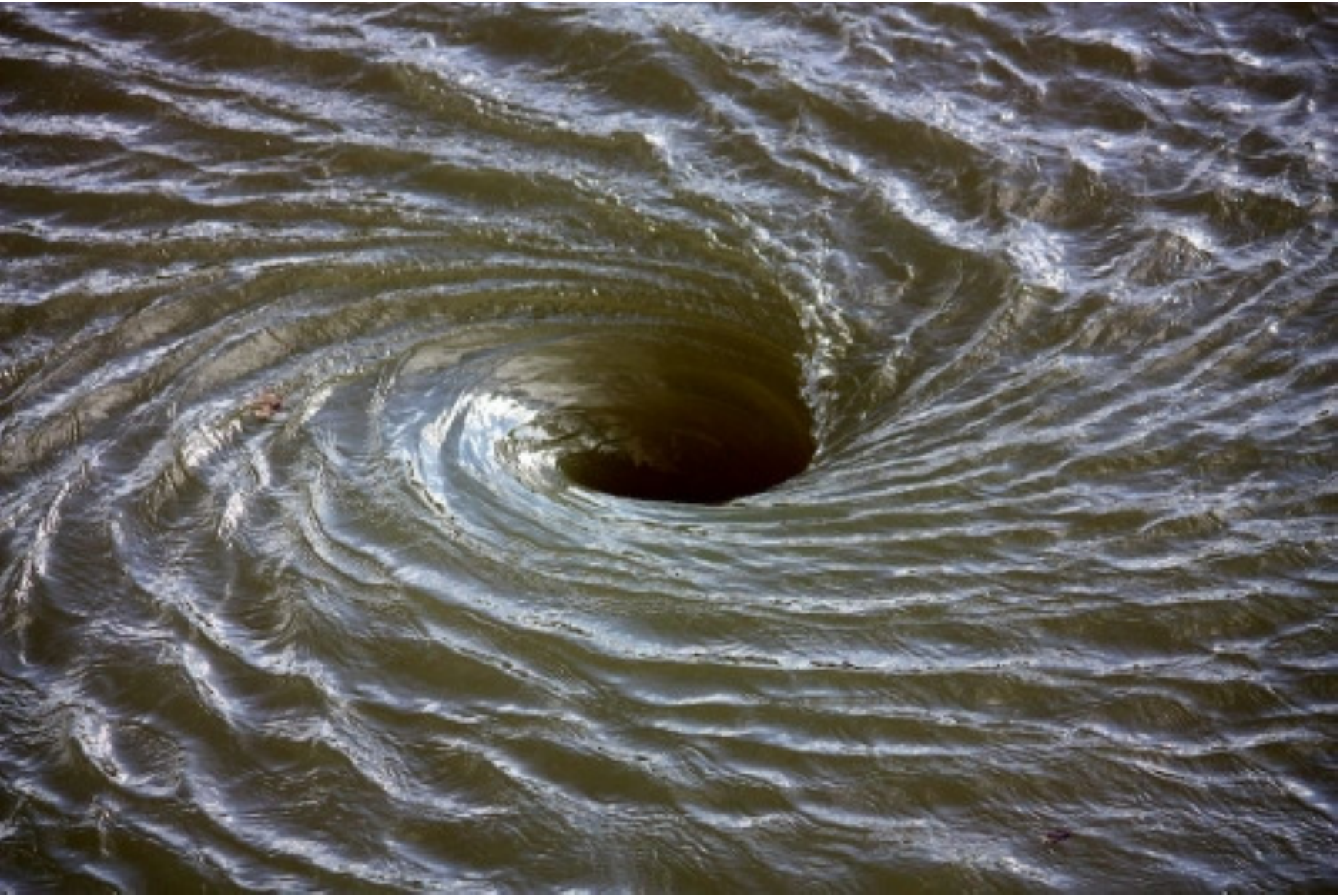}
  
\caption{Not quite seriously: ``Schwarzschild'' (left) versus ``Kerr''
    (right)\label{chap2:lakes}}\medskip
\end{figure}

A cautionary remark has to be made, though, see
Penrose\cite{Penrose:2005}.  In Newtonian gravity $c$ has no absolute
meaning like in special relativity. It is conceivable that the speed
of light in strong Newtonian gravitational fields could be larger than
$c$. Consequently, the Michell type argument becomes only pertinent if
$c$ is the maximal speed for all phenomena like in the Minkowski space
of special relativity, or, if gravity is involved, in the Riemannian
space of GR.

Let us follow the way of visualizing the black hole concept by means
of everyday physics a bit further: We explore the Schwarzschild and,
later in Sec.\ref{kerrconcept}, the Kerr spacetime by boat. Schwarzschild spacetime
is mimicked by a hole in a lake in which the surrounding water plunges
simply radially without whirling around (Monticello Dam,
California). The water flowing towards the hole will drag our boat to
the center. Our boat may move around quite freely as long as the
current is weak.  

However, at some distance {}from the hole, the current becomes so strong
that our boat, engines working at their maximum power, merely can keep
its position. This is the {\it stationary limit}. In the case of our
circularly symmetric water hole the stationary limit forms a ring. Bad
for the boat: The stationary limit is also the ring of no return. At
best, the boat remains at its position, it never will escape. Any
millimeter across the stationary limit will doom the boat, it will be
inevitably sucked into the throat.  Accordingly, the stationary limit
coincides in this spherically symmetric case with the so-called {\it
  event horizon}.

\subsubsection*{Event horizon}

In 1958, Finkelstein\cite{finkel:58} characterized the surface $r=2m$
as a ``semi-permeable membrane'' in spacetime, that is, a surface
which can be crossed only in one direction. As soon as our boat has
passed the event horizon, it can never come back. This property can be
formulated in an invariant way: The light cones at each point of the
surface have to nestle tangentially to the membrane. In 1964,
Penrose\cite{penrose:64} termed the null cone which divides observable
{}from unobservable regions an {\em event horizon}.  Mathematically
speaking, the event horizon is characterized by having {\it tangent
  vectors} which are {\it light-like} or null at all
points. Therefore, the event horizon is a {\it null hypersurface.}
This is what is meant by a {\it trapped surface}\cite{Carroll:2004},
see \fref{nullhypersurf} and \fref{chap1:eddi}, left image: a compact,
spacelike, 2-dimensional submanifold with the property that out\-going
future-directed light rays converge in both directions everywhere on
the sub\-manifold. All these characterizations quite intuitively show
up in the Penrose-Kruskal diagram to be discussed later.

In view of the preceding paragraph, we define
a black hole as a region of spacetime separated {}from infinity by an
event horizon, see Carroll\cite{Carroll:2004} and Brill\cite{Brill:2014}.

Observational evidence in favor of black holes was reviewed by Narayan and
McClintock\cite{Narayan:2013gca}.

\subsubsection*{Killing horizon}

\begin{figure}
\begin{center}
\includegraphics[width=14cm]{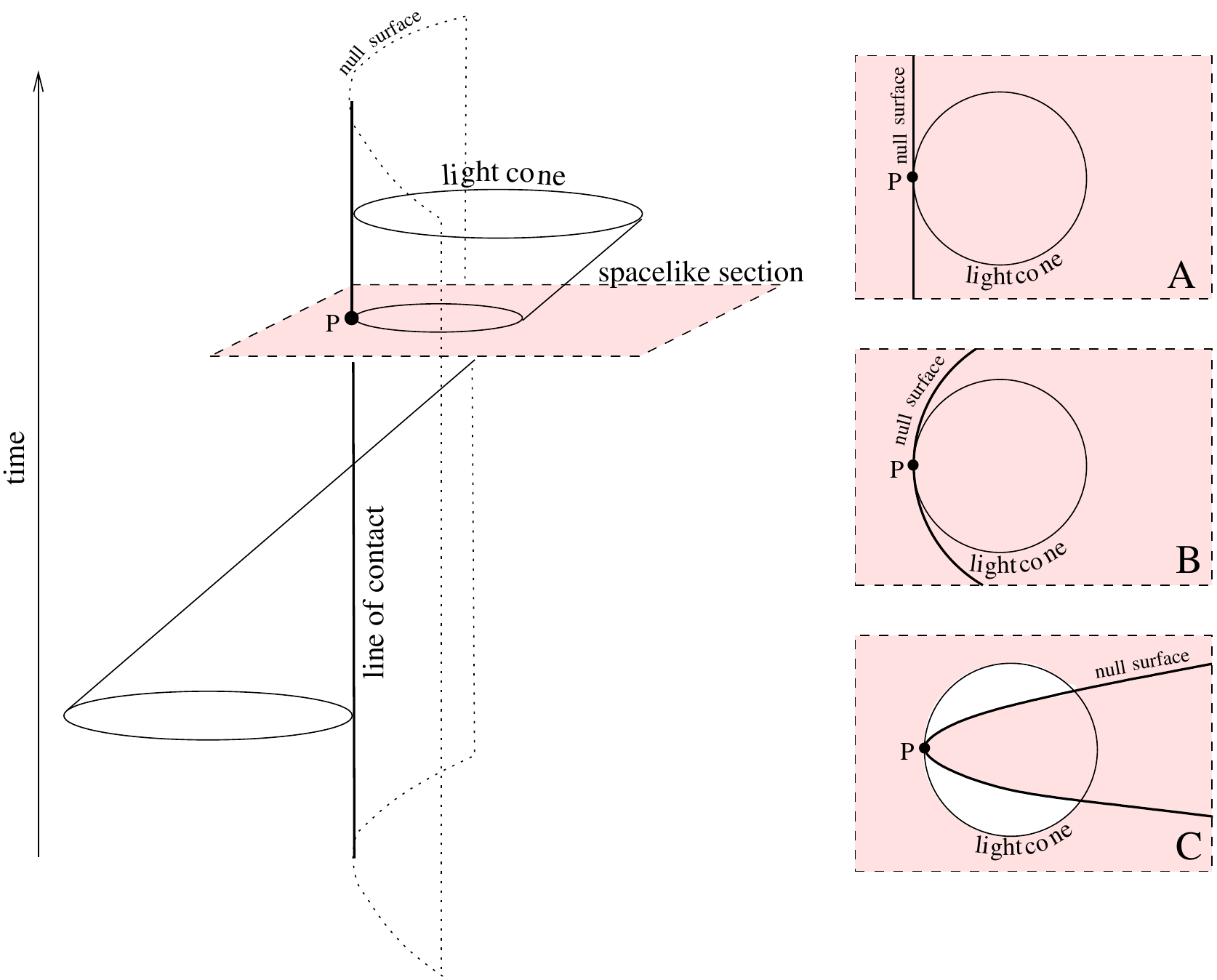}
\end{center}
\caption{\label{nullhypersurf} A null hypersurface is not necessarily
  an event horizon: Imagine a light cone that touches a hypersurface
  along the line of contact. Thus, the light cone is tangent as well
  as normal (in a spacetime sense) to the surface. Consequently, all
  such surfaces are null hypersurfaces. In the cases A and B, the
  light cone is entirely trapped inside the surface.  Case A suggests
  that the surface does not close in a finite region, therefore the
  enclosed volume is not compact. Case B represents a (part of a)
  circle, which encloses all tangential light cones, and this forms an
  (black hole) event horizon. In case C, the light cone intersects the
  hypersurface. The white domain is outside the null surface but
  inside the light cone and, thus, reachable {}from within the enclosed
  domain.}
\end{figure}

The stationary limit surface is rendered more precise in the notion of a {\em
Killing horizon}. A particle at rest (with respect to the infinity of an
asymptotically flat, stationary spacetime) is to be required to follow the
trajectories of the timelike Killing vector\footnote{
Using the definitions
of the covariant derivative and of the Christoffel symbols, we can derive the
following equation for an arbitrary vector $K$,
\begin{equation}
K^\alpha\partial_\alpha\,g_{\mu\nu} = 2 \nabla_{(\mu}K_{\nu)} -2
g_{\alpha(\mu}\partial_{\nu)}K^\alpha\,.
\end{equation}
Assuming $K^\alpha$ and $g_{\mu\nu}$ to be constant in time, demands
$\nabla_{(\mu}K_{\nu)}=0$. Hence $K$ has to be a Killing vector.  In
this coordinate system, we have $K^\alpha K_\alpha = g_{00}$.
Although $K$ acts as time translation, it is not necessarily
timelike!}.  However, if we have a Killing vector $K$ describing a
stationary spacetime, then at some points $K$ may become lightlike,
that is $K^\mu K_\mu =0$. If all these points build up a hypersurface
$\Sigma$, then this null hypersurface is called a Killing
horizon. Apparently, this notion is of a local character, in contrast
to the definition of an event horizon, the definition of which refers
to events in the future, it is of a nonlocal character, see
\fref{nullhypersurf}.

As we will see for the Schwarzschild black hole, see
\fref{chap1:eddi}, outside the black hole the Killing vector is
timelike, that is, $K^\mu K_\mu <0$, on the Killing horizon it becomes
null $K^\mu K_\mu =0$ (by definition of the horizon), and inside it
becomes spacelike $K^\mu K_\mu >0$.

In the Schwarzschild case it will turn out that the event horizon and
Killing horizon coincide, in the Kerr case they separate.

\subsubsection*{Surface gravity}

{}From the definition of the Killing horizon it can be
shown\cite{Carroll:2004} that the quantity
\begin{equation}
\kappa^2 := - \frac{1}{2}(\nabla_\mu K_{\nu})(\nabla^\mu K^{\nu})
\mid_\Sigma 
\end{equation}
is constant on the Killing horizon and positive. 
The quantity $\kappa$ is called {\em surface
gravity}. In simple cases, it has the interpretation of an acceleration or
gravitational force per unit mass on the horizon.
In the Schwarzschild spacetime it takes the value $\kappa=1/4m$, which is 
the acceleration of a particle with unit mass as seen {}from infinity, compare
with the Newtonian ``field strength'' (\ref{chap1:gravfield}) for $r=2m$:
\begin{equation}
f = \frac{GM}{r^2}=\frac{m}{(2m)^2}=\frac{1}{4m}\,.
\end{equation}
In general, there is no such simple interpretation. 

\subsubsection*{Infinite redshift}

Another property associated with the surface $K^\mu\,K_\nu =0$ 
is the infinite
redshift. In view of the
relation for the general relativistic time delay,
\[
\tau_0(\vec x_B) = \frac{\sqrt{g_{tt}(\vec x_B)}}{\sqrt{g_{tt}(\vec     
x_A)}}\,\tau_0(\vec x_A)\,.
\] 
$g_{tt}\to 0$ can be interpreted as follows. Consider $\tau_0(\vec
x_B)$ the time measured by a clock B resting well away {}from the Killing
horizon, whereas clock A with $\tau_0(\vec x_A)$ is nearly at the 
Killing horizon. If $g_{tt}(\vec x_A) \,\rightarrow 0$ we get 
$\tau_0(\vec x_B)\,\rightarrow\infty$.  {}From the point of view of clock B, clock
A's last signal, right before A hits the Killing horizon, 
will not reach B in a
finite time, that is, never. To put it a little bit different: Signals
sent with respect to A with constant frequency arrive increasingly
delayed at B; for B the frequency approaches zero. This is called
infinite redshift.

Let us work out these ideas for the Schwarzschild solution and let us
take ``photons'' in spacetime instead of boats on a lake.

%%%%%%%%%%%%%%%%%%%%%%%%%%%%%%%%%%%%%%%%%%%%%%%%%%%%%%%%%%%%%%%%%%%%%%%%%%%%%
\subsection{Using light rays as coordinate lines}
%%%%%%%%%%%%%%%%%%%%%%%%%%%%%%%%%%%%%%%%%%%%%%%%%%%%%%%%%%%%%%%%%%%%%%%%%%%%%
\label{kruskalszekeres}
{\em Schwarzschild coordinates exhibit a coordinate singularity at $r=2m$.
This obstructs the discussion of the event horizon considerably. As we have
seen, light rays penetrate the horizon without difficulty. This suggests to
use light rays as coordinate lines. Therefore we introduce in- and outgoing
Eddington-Finkelstein coordinates. By combining both, we arrive at
Kruskal-Szekeres coordinates, which provide a regular coordinate system for
the whole Schwarz\-schild spacetime.}

\subsubsection*{Eddington-Finkelstein coordinates}

\begin{figure}
\begin{center}
\bigskip
\includegraphics[height=6.2cm]{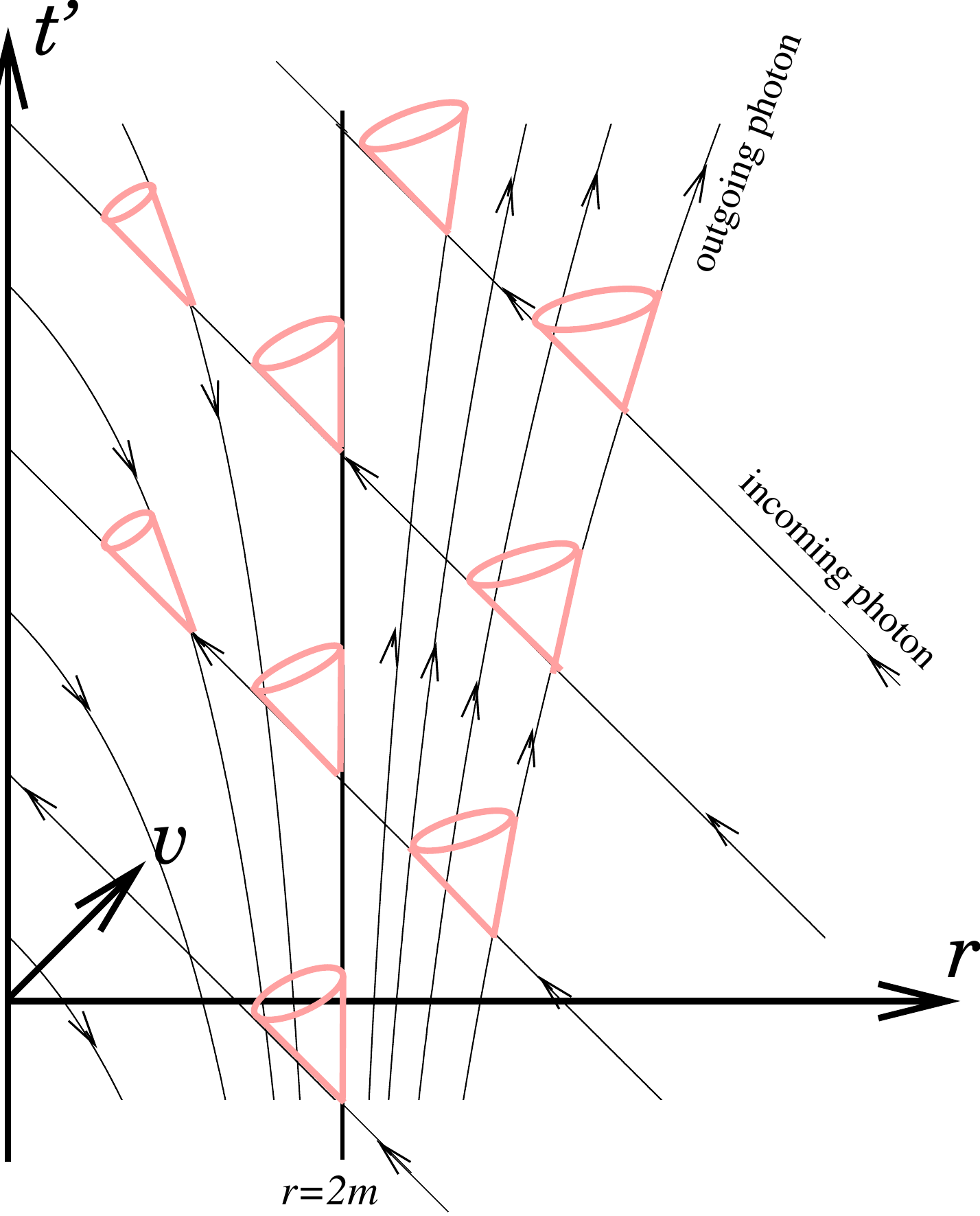}
\includegraphics[height=6cm]{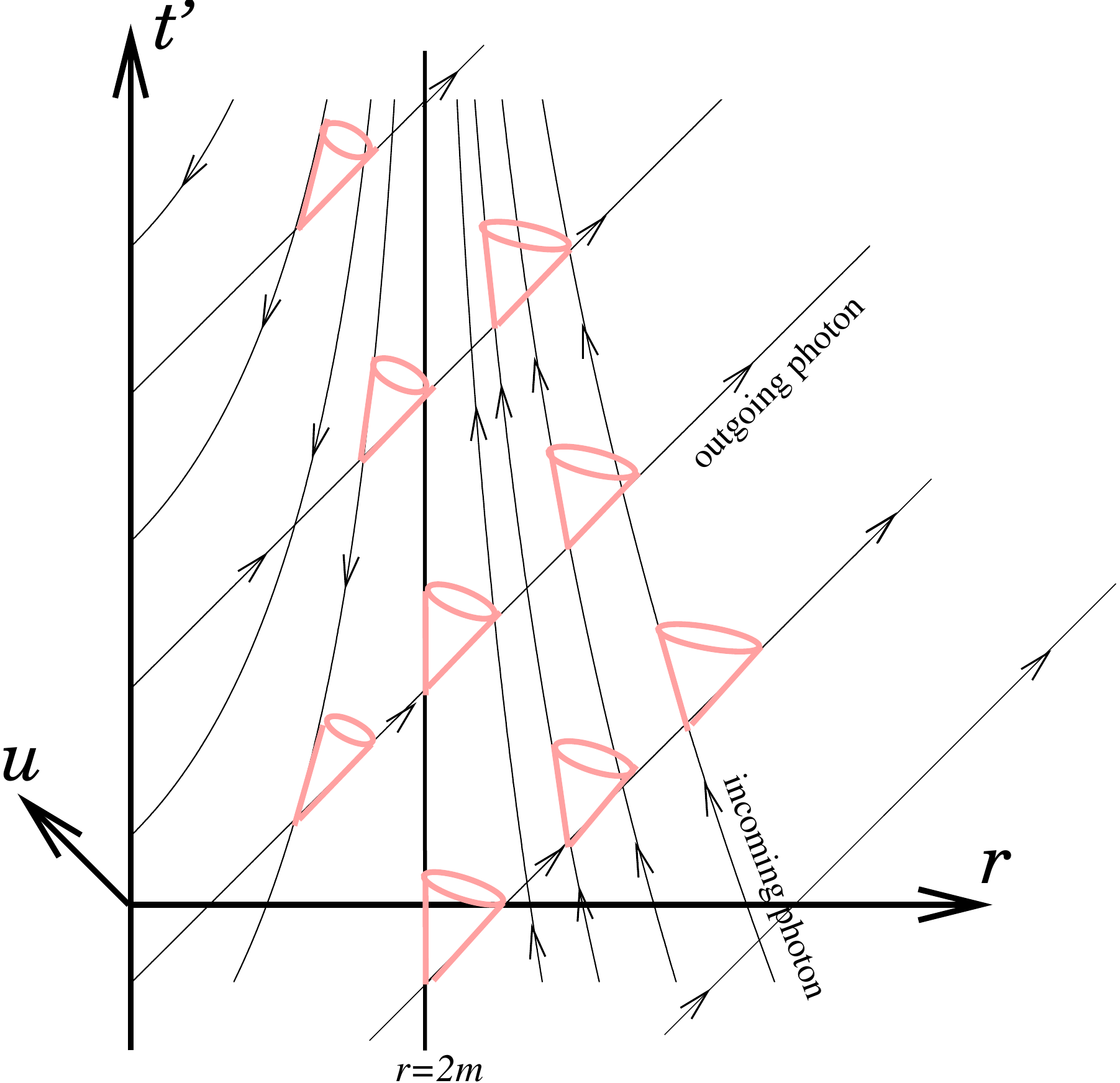}
\end{center}
\caption{In- and outgoing Eddington-Finkelstein coordinates (where we
  introduce $t'$ with $v=t'+r$, $u=t'-r$). The arrows indicate the
  direction of the original Schwarzschild coordinate time (and thereby
  the direction of the Killing vector $\partial_t$). The {\it left}
  figure illustrates a {\it black hole}: All incoming photons
  traverse the event horizon and terminate in the singularity. The
  {\sl right} figure illustrates a {\sl white hole:} All outgoing
  photons emerge {}from the singularity, cross the horizon, and
  propagate out to infinity.\label{chap1:eddi}}
\end{figure}

In relativity, light rays, the quasi-classical trajectories of photons, are
null geodesics. In special relativity, this is quite obvious, since in
Minkowski space the geodesics are straight lines and ``null''  just 
means $v=c$. A more rigorous argument involves the solution of
the Maxwell equations for the vacuum and the subsequent determination of the
normals to the wave surface (rays) which turn out to be null geodesics. 
This remains 
valid in general relativity. Null geodesics can be easily obtained by
integrating the equation $0=ds$. We find for the Schwarzschild metric,
specializing to radial light rays with $d\phi=0=d\theta$, 
\begin{equation}\label{lightinschw}
t=\pm \left( r + 2m \, \ln \left|\frac{r}{2m}-1\right|\right) + const\,.
\end{equation}
If we denote with $r_0$ the solution of the equation $r + 2m \ln \left|
\frac{r}{2m}-1\right| = 0$, we have for the $t$-coordinate of the light ray
$t(r_0)=:v$. Hence, if $r=r_0$,  we can use $v$ to label light rays. 
In view of this, we introduce $v$ and $u$ 
\begin{eqnarray}\label{chap1:vv}
v &:=& t+r+2m \ln\left|\frac{r}{2m}-1\right| \,, \\ \label{chap1:uu}
u &:=& t-r-2m \ln \left|\frac{r}{2m}-1\right| \,.
\end{eqnarray}
Then ingoing null geodesics are described by $v=const$, outgoing ones 
by $u=const$, see \fref{chap1:eddi}. 
We define {\em ingoing Eddington-Finkelstein
coordinates} by replacing the ``Schwarzschild time'' $t$ by $v$.
In these coordinates $(v,r,\theta,\phi)$, the metric becomes
\begin{equation}\label{chap1:metoutedd}
ds^2 = - \left( 1-\frac{2m}{r}\right) \, dv^2 + 2 dv \, dr + r^2 \,
d\Omega^2 \,.
\end{equation}
For radial null geodesics $ds^2=d\theta=d\phi=0$, we find two solutions
of (\ref{chap1:metoutedd}), namely $v=const$ and $v=4m \, \ln
\left|r/2m-1\right|+2r +const$. 
The first one describes infalling photons, i.e., $t$ increases if $r$ 
approaches $0$. At $r=2m$, there is no singular behavior any longer for 
incoming photons.
Ingoing Eddington-Finkelstein coordinates are particular useful in order to
describe the gravitational collapse.
Analogously, for outgoing null
geodesics take $(u,r,\theta,\phi)$ as new coordinates. 
In these {\em outgoing Eddington-Finkelstein coordinates} the metric reads 
\begin{equation}
ds^2 = - \left( 1-\frac{2m}{r}\right) \, du^2 -2 \, du \, dr + r^2 \,
d\Omega^2 \,.
\end{equation}
Outgoing light rays are now described by $u=const$, ingoing light rays by 
$u=-(4m \, \ln\left|r/2m -1\right| + 2r) + const$. 
In these coordinates, the hypersurface $r=2m$ (the ``horizon'') can be
recognized as a null hypersurface (its normal is null or lightlike) and as a
semi-permeable membrane.

%%%%%%%%%%%%%%%%%%%%%%%%%%%%%%%%%%%%%%%%%%%%%%%%%%%%%%%%%%%%%%%%%%%%%%%%%%%%%
\subsubsection*{Kruskal-Szekeres coordinates}
%%%%%%%%%%%%%%%%%%%%%%%%%%%%%%%%%%%%%%%%%%%%%%%%%%%%%%%%%%%%%%%%%%%%%%%%%%%%%

\begin{figure}[h]
\medskip
\includegraphics[width=12cm]{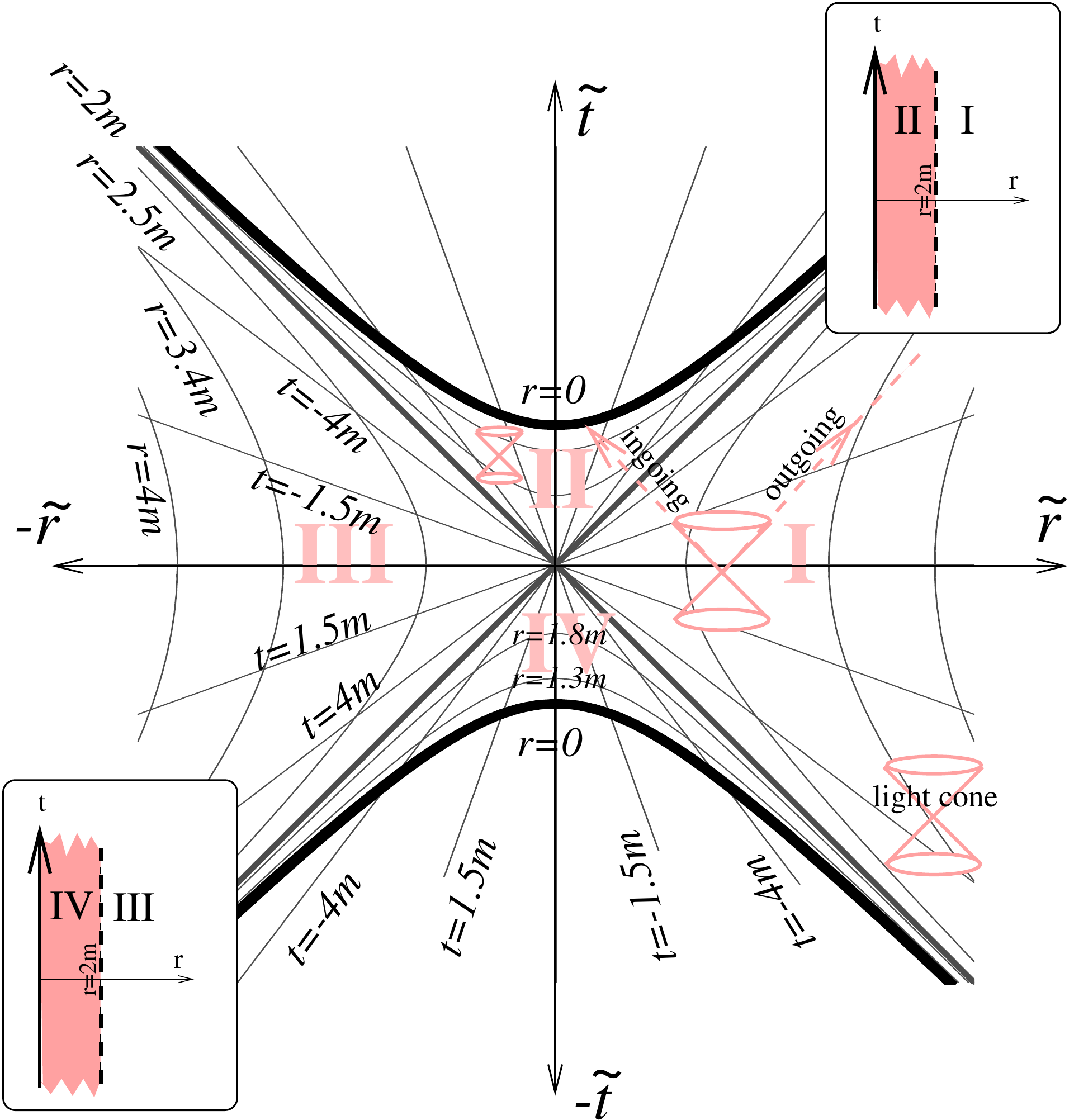}
\caption{Kruskal-Szekeres diagram of the Schwarzschild
  spacetime.\label{KruskalFig}}
\end{figure}

  Next we try to
combine the advantages of in- and outgoing Eddington-Finkelstein
coordinates in the hope to obtain a fully regular coordinate system of
the Schwarzschild spacetime. Therefore we assume coordinates
$(u,v,\theta,\phi)$. Some (computer) algebra yields the corresponding
representation of the metric:
\begin{equation} 
ds^2 = -\left(1-\frac{2m}{r(u,v)} \right) \, du\,dv + r^2(u,v) \, d\Omega^2
\,.
\end{equation}
Unfortunately, we still have a coordinate singularity at $r=2m$. 
We can get rid of it by reparametrizing the surfaces $u=const$ and
$v=const$ via
\begin{equation}\label{chap1:uvtilde}
\tilde v = \exp \left(\frac{v}{4m} \right)\,, \qquad 
\tilde u = - \exp \left(-\frac{u}{4m}  \right)\,.
\end{equation}
In these coordinates, the metric reads 
($r=r(\tilde u,\tilde v)$ is implicitly given by (\ref{chap1:uvtilde})
and (\ref{chap1:uu}), (\ref{chap1:vv}), $r_{\rm S}=2m$)
\begin{equation}
ds^2 = - \frac{4r_{\rm S}^3}{r(\tilde u,\tilde v)} \, \exp \left(
-\frac{r(\tilde u,\tilde v)}{2m}\right) \, d\tilde v \, d\tilde u 
+ r^2(\tilde u,\tilde v) \, d\Omega^2 \,.
\end{equation}
Again, we go back {}from $\tilde{u}$ and $\tilde{v}$ to time- and space-like
coordinates:
\begin{equation}\label{oldcoo}
\tilde{t} := \frac{1}{2} \, \left( \tilde{v} + \tilde{u}\right)\,,
\qquad
\tilde{r} := \frac{1}{2} \, \left( \tilde{v} - \tilde{u} \right) \,.
\end{equation}
In terms of the original Schwarzschild coordinates we have
\begin{eqnarray}\label{chap1:cookrusez1}
\tilde{r} &=& \sqrt{\left|\frac{r}{2m}-1\right|} \,\, 
\exp\left(\frac{r}{4m}\right) \, \cosh\frac{t}{4m}\,, \vspace{5pt}\\
\label{chap1:cookrusez2} \tilde{t} &=& \sqrt{\left| \frac{r}{2m}-1\right|} 
\,\, \exp\left(\frac{r}{4m}\right) \,  \sinh\frac{t}{4m}\,.
\end{eqnarray}
The Schwarzschild metric 
\begin{equation}\label{chap1:metkrusez}
ds^2 = \frac{4r_{\rm S}^3}{r} \, \exp\left(-\frac{r}{2m}\right) \, 
\left(-d\tilde{t}\,{}^2 + d\tilde{r}^2 \right) + r^2 \, d\Omega^2 \,,
\end{equation}
in these {\em Kruskal-Szekeres} coordinates
$(\tilde{t},\tilde{r},\theta, \phi)$, behaves regularly at the
gravitational radius $r=2m$.  If we substitute (\ref{chap1:metkrusez})
into the Einstein equation (via computer algebra), then we see that it
is a solution for all $r>0$.  Eqs.(\ref{chap1:cookrusez1}) and
(\ref{chap1:cookrusez2}) yield
\begin{equation}
\tilde{r}^2 - \tilde{t}\,{}^2 = 
\left|\frac{r}{2m}-1\right| \, \exp\left(\frac{r}{2m}\right) \,.
\end{equation}
Thus, the transformation is valid only for 
regions with $\left|\tilde{r}\right| > \tilde{t}$. However, we can find a
set of transformations which cover the entire ($\tilde t$, $\tilde
r$)-space. They are valid in different
domains, indicated here by I, II, III, and IV, to be explained below:

%{\color{red} \bf @@@ Check signs, see Wei-Tou}
\begin{eqnarray}
 {\rm (I)} &\Bigg\{ &
\begin{array}{rcl}
\tilde{t} &=& \sqrt{\frac{r}{2m}-1} \,\, \exp\left(\frac{r}{4m}\right) 
\, \sinh \frac{t}{4m} \vspace{4pt}\\
\tilde{r} &=& \sqrt{\frac{r}{2m}-1} \,\, \exp\left(\frac{r}{4m}\right) \, 
\cosh\frac{t}{4m}
\end{array}\label{Kruskal1}\\ 
{\rm (II)} &\Bigg\{ &
\begin{array}{rcl}
\tilde{t} &=& \sqrt{1-\frac{r}{2m}} \,\, \exp\left(\frac{r}{4m}\right) \, 
\cosh \frac{t}{4m} \vspace{4pt} \\ 
\tilde{r} &=& \sqrt{1-\frac{r}{2m}} \,\, \exp\left(\frac{r}{4m}\right) \, 
\sinh\frac{t}{4m}
\end{array}\label{Kruskal2}\\ 
{\rm (III)} &\Bigg\{ &
\begin{array}{rcl}
\tilde{t} &=& -\sqrt{\frac{r}{2m}-1} \,\, \exp\left(\frac{r}{4m}\right) \, 
\sinh \frac{t}{4m} \vspace{4pt} \\
\tilde{r} &=& -\sqrt{\frac{r}{2m}-1} \,\, \exp\left(\frac{r}{4m}\right) \, 
\cosh\frac{t}{4m} 
\end{array}\label{Kruskal3}\\
{\rm (IV)} &\Bigg\{ &
\begin{array}{rcl}
\tilde{t} &=& -\sqrt{1-\frac{r}{2m}} \,\, \exp\left(\frac{r}{4m}\right) \, 
\cosh \frac{t}{4m} \vspace{4pt} \\
\tilde{r} &=& -\sqrt{1-\frac{r}{2m}} \,\, \exp\left(\frac{r}{4m}\right) \, 
\sinh\frac{t}{4m}
\end{array}
\end{eqnarray}
The inverse transformation is given by
\begin{eqnarray}
&&\left(\frac{r}{2m}-1 \right) \,\, \exp\left(\frac{r}{2m}\right)
=\tilde{r}^2-\tilde{t}\,{}^2\,,\label{Kruskal5} \\
&&\frac{t}{4m} = \left\{\begin{array}{ll}
{\rm Artanh} \, \tilde{t}/ \tilde{r}\,, & \text{ for (I) and
(III)}\,,\vspace{5pt}\\
{\rm Artanh} \, \tilde{r}/\tilde{t}\,, & \text{ for (II) and (IV)\,.}
\end{array}\right.\,.\label{Kruskal6}
\end{eqnarray}
The Kruskal-Szekeres coordinates $(\tilde{t},\tilde{r},\theta,\phi)$
cover the entire spacetime, see \fref{KruskalFig}. 
By means of the transformation equations
we recognize that we need two Schwarzschild coordinate systems in
order to cover the same domain. Regions (I) and (III) both correspond
each to an asymptotically flat universe with $r>2m$. Regions (II) and
(IV) represent two regions with $r<2m$. Since $\tilde{t}$ is a time
coordinate, we see that the regions are time reversed with respect to
each other. Within these regions, real physical singularities
(corresponding to $r=0$) occur along the curves
$\tilde{t}\,{}^2-\tilde{r}^2=1$.  {}From the form of the metric we can
infer that radial light-like geodesics (and therewith the light cones
$ds=0$) are lines with slope $1$.  This makes the discussion of the
causal structure particularly simple.

%%%%%%%%%%%%%%%%%%%%%%%%%%%%%%%%%%%%%%%%%%%%%%%%%%%%%%%%%%%%%%%%%%%%%%%%%%%%
\subsection{Penrose-Kruskal diagram}
%%%%%%%%%%%%%%%%%%%%%%%%%%%%%%%%%%%%%%%%%%%%%%%%%%%%%%%%%%%%%%%%%%%%%%%%%%%%

{\em We represent the Schwarzschild spacetime in a
manner analogous to the Penrose diagram of the Minkowski spacetime. To this
end, we proceed along the same line as in the Minkowskian case.}

\begin{figure}
\begin{picture}(130,80)
\put(0,0){\includegraphics[width=12cm]{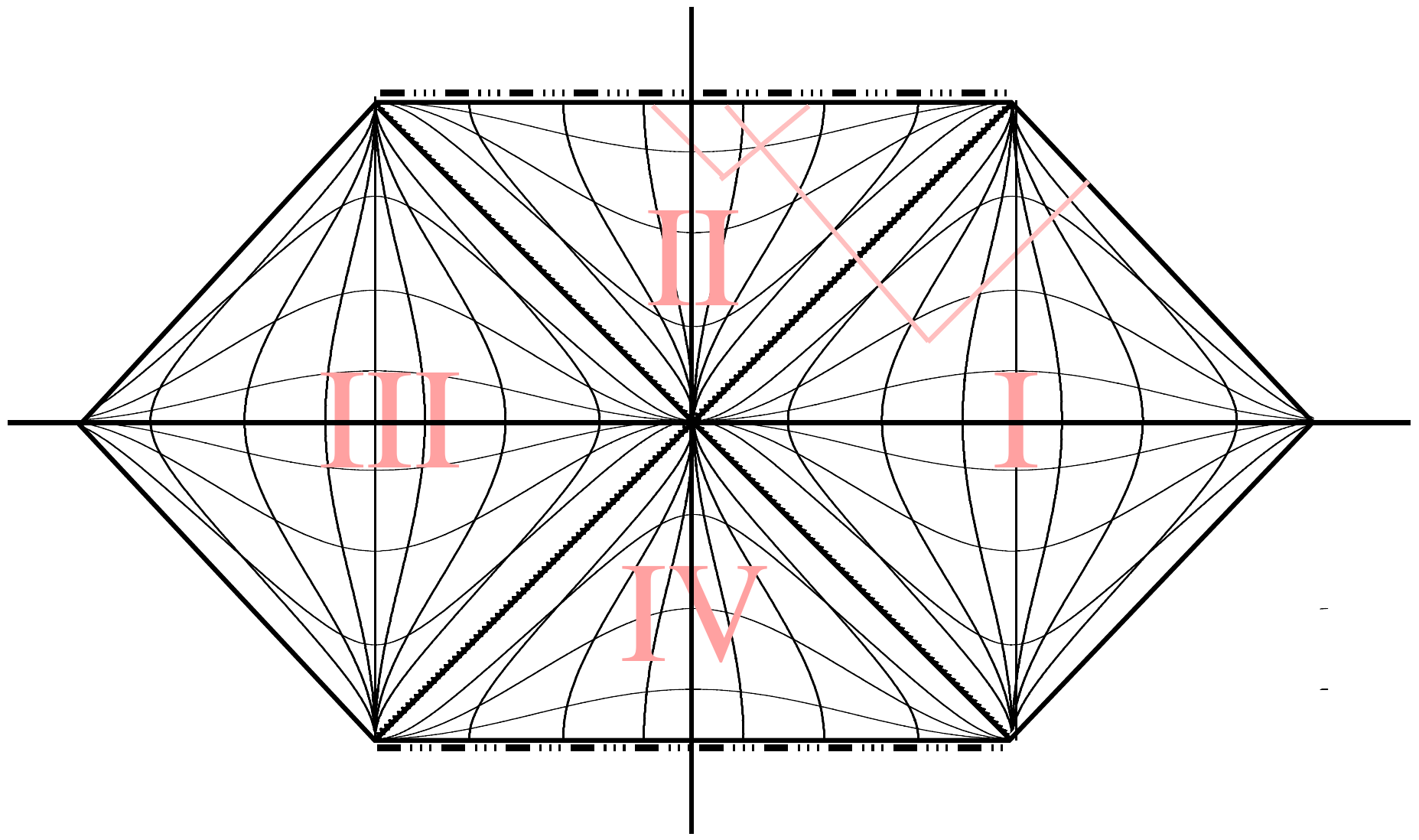}}
\put(2,37){\mbox{\Large $I_0$}}
\put(0,31){\mbox{\Large $-\pi$}}
\put(111,37){\mbox{\Large $I_0$}}
\put(111,31){\mbox{\Large $\pi$}}
\put(85,64){\mbox{\Large $I^+$}}
\put(30,64){\mbox{\Large $I^+$}}
\put(85,2){\mbox{\Large $I^-$}}
\put(30,2){\mbox{\Large $I^-$}}
\put(45,64.5){\mbox{\large singularity $r=0$}}
\put(45,4){\mbox{\large singularity $r=0$}}
\put(67,45){\mbox{\begin{rotate}{45} \Large$\bf r=2m$\end{rotate}}}
\put(10.5,49){\mbox{\Large $\cal I^+$}}
\put(100.5,49){\mbox{\Large $\cal I^+$}}
\put(10.5,17){\mbox{\Large $\cal I^-$}}
\put(100.5,17){\mbox{\Large $\cal I^-$}}
\end{picture}
\caption{ Penrose-Kruskal diagram of the Schwarzschild 
spacetime. Region II corresponds to a black hole, region IV to a white
hole. Regions I and III correspond to two universes.
\label{chap1:ssskru}}

\end{figure}
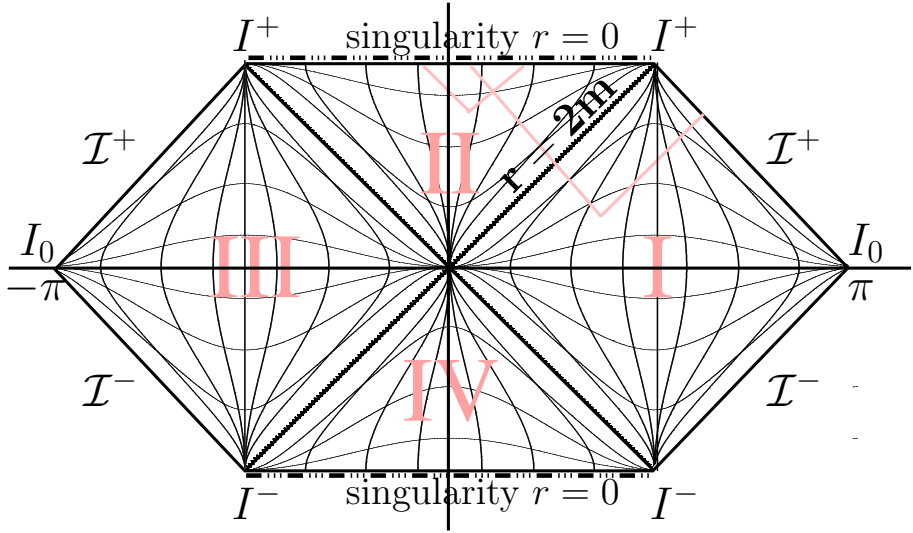

First, we switch again to null-coordinates $v'=\tilde{t}+\tilde{r}$ 
and $u'=\tilde{t}-\tilde{r}$ and perform
a conformal transformation which maps infinity into the finite (again, by
means of the tangent function). Finally we return to a time-like
coordinate $\hat t$ and a space-like coordinate $\hat r$. We perform these
transformations all in one according to
%\begin{eqnarray}
\[
\hspace{2.5cm}
\tilde{t}+\tilde{r} = \tan \frac{\hat t + \hat r}{2}\,, \qquad
\tilde{t}-\tilde{r} = \tan \frac{\hat t - \hat r}{2}\,.
\hspace{2.5cm}\mbox{(61,62)}
\]
%\]
\addtocounter{equation}{2}
%\end{eqnarray}
The Schwarzschild metric then reads\small
\begin{equation}
ds^2 = \frac{r_{\rm S}^3}{r(\hat r,\hat t)} \, 
\frac{
      \exp\left(-\frac{r(\hat r,\hat t)}{2m}\right) \, 
      \left(-d\hat t\,{}^2 + d \hat r^2 \right)
     }
     {
       \cos^2 \frac{\hat t + \hat r}{2} \, 
       \cos^2 \frac{\hat t - \hat r}{2} 
     }
+r^2(\hat t, \hat r) \, d\Omega^2 \,,
\end{equation}\normalsize
where the function $r(\hat t,\hat r)$ is implicitly given by\small
\begin{equation}
\left(\frac{r}{2m}-1\right) \, \exp\left(\frac{r}{2m}\right) =
\tan \frac{\hat t + \hat r}{2} \, \tan \frac{\hat t - \hat r}{2}\,.
\end{equation}\normalsize
The corresponding Penrose-Kruskal diagram is displayed in
\fref{chap1:ssskru}\,. The notations for the different infinities can be
extracted {}from Table 1. In contrast to Minkowski space, light rays and
particles may not escape to infinity but enter the black hole (II). 
Likewise, light rays and particles may not emerge {}from infinity but {}from the
white hole (IV).

\pagebreak

\subsection[The Reissner-Nordstr\"om metric]{Adding electric charge and 
the cosmological constant: Reissner Nordstr\"om}

As mentioned in the historical remarks, soon after Schwarzschild's solution,
the first generalizations, including electric charge and the
cosmological constant were published. We can be even quicker \dots
We already calculated the Ricci tensor for the Gullstrand-Painlev\'e
ansatz. If we use the well-known energy-momentum tensor for a point
charge\cite{Birkbook}, the field equation may be written
as\footnote{Einstein's gravitational constant is denoted by $\kappa$,
$\lambda_0=\sqrt{\frac{\epsilon_0}{\mu_0}}$ is the admittance of the vacuum.
With $c=1$ and $G=1$ we have $\kappa\lambda_0=2$.}
\begin{equation}
R_{\mu}{}^\nu - \frac{1}{2}\, R \,\delta_\mu^\nu + \Lambda \,\delta_\mu^\nu
= \kappa \, \lambda_0\,\frac{q^2}{2r^4}\,\mathrm{diag}(-1,-1,1,1)\,.
\end{equation}
Taking the trace, we find $R=4\Lambda$ and arrive at
\begin{equation}
R_{2}{}^2=R_3{}^3 = \frac{2\partial_r(r\psi)}{r^2} = \Lambda - 
\frac{q^2}{r^4}\,.
\end{equation}

This equation can be integrated elementarily,
\begin{equation}
2\psi = \frac{1}{3} \, \Lambda
r^2-\frac{q^2}{r^2}+\frac{2\alpha}{r}\,.
\end{equation} 
This function also solves the remaining two field equations. The
integration constant $\alpha$ is again the mass $m$. Substituted into
(\ref{gullstrand_ansatz}) and transformed to Schwarzschild coordinates
($f=1-2\psi$) the solution reads
\begin{equation}  
ds^2 =-f(r)\,dt^2+\frac{dr^2}{f(r)}+r^2\,d\Omega^2\,,
\end{equation}
with
\begin{equation}\label{reissner}
f(r) := 1 - \frac{2m}{r} + \frac{q^2}{r^2} -\frac{\Lambda}{3}\,
r^2\,.
\end{equation}
A detailed derivation using Schwarzschild coordinates and computer algebra
can be found in Puntigam et al.\cite{Puntigam}

A discussion of the Reissner-Nordstr\"om(-de Sitter) solution can be
found in Griffiths \& Podolsky\cite{Griffiths:2009}, for example. We
only remark, that we recover the Schwarzschild solution for $q=0$ and
$\Lambda=0$. The algebraic structure of the solution is identical to
the Schwarzschild case.  Thus, we find, in general, a singularity at
$r=0$. However, a pure cosmological solution, $m=0, q=0$ and $\Lambda
\ne 0$, possesses no singularity and no horizon! On the other hand, an
electrically charged black hole, $\Lambda=0$, exhibits {\em two}
horizons,
\begin{equation}
f(r)=0\qquad \Leftrightarrow \qquad r_\pm = m \pm \sqrt{m^2 -q^2}\,.
\end{equation}
In this respect, the charged black hole shows some similarities to a
rotating (Kerr) black hole. We will pick up this discussion in
Sec.\ref{kerrconcept}.

\color{black}

%%%%%%%%%%%%%%%%%%%%%%%%%%%%%%%%%%%%%%%%%%%%%%%%%%%%%%%%%%%%%%%%%%%%%%%%%%%%
\subsection{The interior Schwarzschild solution and the TOV equation}
%%%%%%%%%%%%%%%%%%%%%%%%%%%%%%%%%%%%%%%%%%%%%%%%%%%%%%%%%%%%%%%%%%%%%%%%%%%%

In the last section we investigated the gravitational field outside a
spherically symmetric mass-distribution. Now it is time to have a look
inside matter, see Adler, Bazin, and Schiffer \cite{adler}.  Of
course, in a first attempt, we have to make decisive simplifications
on the internal structure of a star. We will consider cold catalyzed
stellar material during the later phase of its evolution which can be
reasonably approximated by a perfect fluid. The typical mass densities
are in the range of $\approx 10^7 \, {\rm g}/{\rm cm^3}$ (white
dwarfs) or $\approx 10^{14} \, {\rm g}/{\rm cm^3}$ (neutron stars,
e.g., pulsars). In this context we assume vanishing angular momentum.

We start again {}from a static and spherically symmetric metric
\begin{equation}
ds^2 = - {\rm e}^{A(r)} \, c^2 \, dt^2 + {\rm e}^{B(r)} \, dr^2 + r^2 \,
d\Omega^2
\end{equation}
and the energy-momentum tensor
\begin{equation}
T_{\mu\nu} = \left(\rho+\frac{p}{c^2} \right) \, u_\mu \, u_\nu +
p \, g_{\mu\nu}\,,
\end{equation}
where $\rho=\rho(r)$ is the spherically symmetric mass density and $p=p(r)$
the pressure (isotropic stress). This has to be supplemented by the equation
of state which, for a simple fluid, has the form $p=p(\rho)$.

We compute the non-vanishing components of the field equation by means of 
computer algebra as (here $\kappa=8\pi G/c^4$ is Einstein's
gravitational constant and $()' = d/dr$)
\begin{eqnarray}\label{chap1:innfieequ1}
-{\rm e}^{B} \kappa r^2 c^2 \rho + {\rm e}^{B} + B'r -1 &=&0\,,\\
\label{chap1:innfieequ2}
-{\rm e}^{B} \kappa p r^2 -{\rm e}^{B} + A' r + 1 &=&0\,,\\
\label{chap1:innfieequ3}
-4{\rm e}^{B} \kappa p r + 2 A'' r + (A')^2r - A'B' r + 2A' - 2B' &=&0\,.
\end{eqnarray}
The $(\phi,\phi)$-component turns out to be equivalent to the
$(\theta,\theta)$-component. For convenience, we define a {\em mass
function} $m(r)$ according to
\begin{equation}\label{chap1:relmasfun}
{\rm e}^{-B} =: 1-\frac{2m(r)}{r}\,.
\end{equation}

We can differentiate (\ref{chap1:relmasfun}) with respect to $r$ and find, 
after substituting (\ref{chap1:innfieequ1}), a differential equation for 
$m(r)$ which can be integrated, provided $\rho(r)$ is assumed to be known
\begin{equation}\label{chap1:masfunsol}
m(r) = \int_0^r \frac{\kappa}{2} \,\rho c^2  \, \xi^2 \, d\xi\,.
\end{equation}
Differentiating (\ref{chap1:innfieequ2}) and using all three components of
the field equation, we obtain a differential equation for $A$:
\begin{equation}\label{chap1:a'1}
A'=-\frac{2p'}{p+\rho c^2}\,.
\end{equation}
We can derive an alternative representation of $A'$ by substituting 
(\ref{chap1:relmasfun}) into (\ref{chap1:innfieequ2}).
Then, together with (\ref{chap1:a'1}), we arrive at the 
{\em Tolman-Oppenheimer-Volkoff} (TOV) equation 
\begin{equation}\label{chap1:tov}
p'=-\frac{(\boldsymbol{\rho} c^2 +p)({\bf m} + \kappa \, pr^3/2)}{{\bf r}({\bf r}-2m)}
\,.
\end{equation}
Terms that survive in the Newtonian limit are emphasized by boldface
letters.  The system of equations consisting of
(\ref{chap1:masfunsol}), (\ref{chap1:a'1}), the TOV equation
(\ref{chap1:tov}), and the equation of state $p=p(\rho)$ forms a
complete set of equations for the unknown functions
$A(r),\,\rho(r),\,p(r),$ and $m(r)$, with
\begin{equation}
ds^2=-{\rm e}^{A(r)} \, c^2 \,dt^2 + \frac{dr^2}{1-\frac{2m(r)}{r}} \, 
+r^2\,d\Omega^2 \,.
\end{equation}
These differential equations have to be supplemented by initial
conditions. 

In the center of the star, there is, of course, no enclosed mass. 
Hence we demand $m(0)=0$. The density has to be finite at the
origin, i.e. $\rho(0)=\rho_c$, where $\rho_c$ is the density of the central
region. At the surface of the star, at $r=R_\odot$, we have to match 
matter with vacuum. In vacuum, there is no pressure which requires
$p(R_\odot)=0$. Moreover, the mass function should then yield the total mass
of the star, $m(R_\odot):={GM_\odot}/{c^2}$. 
Finally, we have to match the components of the 
metric. Therefore, we have to demand $\exp[A(r_0)] = 1-2m(R_\odot)/R_\odot$.

Equations (\ref{chap1:innfieequ1}), (\ref{chap1:innfieequ2}),
(\ref{chap1:innfieequ3}) and certain  regularity
conditions which generalize our boundary conditions, that is,
\begin{itemize}
\item regularity of the geometry at the origin,
\item finiteness of central pressure and density,
\item positivity of central pressure and density,
\item positivity of pressure and density,
\item monotonic decrease of pressure and density,
\end{itemize}        
impose conditions on the functions $\rho$ and $p$. Then, even without the 
explicit knowledge of the equation of state, the general
form of the metric can be determined. For recent work, see Rahman and 
Visser \cite{rahman} and the literature given there.

We can obtain a simple solution, if we assume a constant mass density
\begin{equation}
\rho=\rho(r) = const.
\end{equation}
One should mention here that $\rho$ is not the physically observable fluid
density,  which results {}from an appropriate projection of the
energy-momentum tensor into the reference frame of an observer.
Thus, this model is not as unphysical as it may look at the first.
However, there are serious but more subtle objections which we will not 
discuss further in this context. 

% ab eq.82, ab not discuss further in this context

When $\rho={\rm const.}$, we can explicitly write down the mass function
(\ref{chap1:masfunsol}),
\begin{equation}
m(r) = \frac{r^3}{2\widehat R^2}\,,\qquad \mbox{with }
\widehat R = \sqrt{\frac{3}{\kappa\rho c^2}}\,,
\qquad m_\odot := \frac{R_\odot^3}{2\hat R^2}\,.
\end{equation}
This allows immediately to determine one metric function
\begin{equation}
e^B = \frac{1}{1-\frac{r^2}{\widehat R^2}}\,.
\end{equation}
The TOV equation (\ref{chap1:tov}) factorizes according to
\begin{equation}
\frac{dp}{dr}= -\frac{1}{2}\, (\rho c^2 + p)(1+\kappa \widehat R^2 p)\,
\frac{r}{\widehat R^2 - r^2}\,.
\end{equation}
It can be elementarily solved by separation of variables,
\begin{equation}
p(r) = \rho c^2 \,
 \frac{ \sqrt{\widehat R^2-R_\odot^2} - \sqrt{\widehat R^2-r^2}}
{\sqrt{{\widehat R^2}-r^2} - 3\sqrt{{\widehat R^2}-R_\odot^2}}\,.
\end{equation}
Using (\ref{chap1:a'1}) as $A'=-2\left[\ln(p+\rho c^2)\right]'$ 
and continuous matching to the exterior, eventually
yields the {\em interior \& exterior Schwarzschild solution} 
for a spherically symmetric body \cite{schwarzschild2}
\begin{equation}\label{chap1:innssm}
  ds^2  = \left\{
\begin{array}{lc}
-\left(\frac{3}{2}\sqrt{1-\frac{R_\odot^2}{\widehat
          {R^2}}}-\frac{1}{2}\sqrt{1-\frac{r^2}{\widehat {R^2}}}
    \right)^2 c^2 \, dt^2 + \frac{1}{1-\frac{r^2}{\widehat
          {R^2}}}\,dr^2+r^2d\Omega^2\,, &\quad r\le R_\odot\,,
  \vspace{5pt}\\ 
-\left(1-\frac{2m_\odot}{r}\right)\, c^2 \, dt^2
    +\frac{1}{1-\frac{2m_\odot}{r}}\,dr^2+r^2\,d\Omega^2\,, & \quad r>R_\odot\,.
\end{array}\right.
\end{equation}
The solution is only defined for $R_\odot < \widehat R$.  For the
Sun\footnote{To ascertain the consistency of dimensions and units, we
  recollect the basic definitions:
\[ [G]=\frac{(\rm m/\rm s)^4}{\rm N}= \frac{\rm m^3}{\rm kg\,\rm
  s^2}\,, \qquad \kappa = \frac{8\pi G}{c^4}\,.
\]
The mass $M$ carries the unit kg, the mass {\em parameter} has the
dimension of a length:
\[
m := \frac{GM}{\rm c^2}\,,\quad [m] = \frac{\rm m^3\,\rm kg\,\rm
  s^2}{\rm kg\,\rm s^2\, \rm m^2}=\rm m\,.
\]
The definition of $m(r)$ in Eq.(\ref{chap1:masfunsol}) is
consistent. For $\rho=const$, we have
\[
m(r)=\frac{\kappa}{2}\,\rho c^2\,\frac{1}{3}\,r^3 =
\frac{G}{c^2}\,
\frac{4}{3}\,\pi\,r^3\,\rho
=\frac{GM(r)}{c^2}\,.
\]
Here $\rho$ denotes the physical mass density, $[\rho]=\rm kg/\rm
m^3$. Thus
\[
M(r) := \frac{4}{3}\,\pi\,r^3\,\rho
\]
is the physical mass with the unit kg.  } we have $M_\odot\approx2
\times 10^{30}\,{\rm kg}$, $R_\odot\approx 7\times 10^8\,{\rm m}$ and
accordingly $\rho_\odot \approx 1.4 \times 10^3 \, {\rm kg/m^3}$. This
leads to $\widehat R\approx 3 \times 10^{11}\,{\rm m}$, that is, the
radius of the star $R_\odot$ is much smaller than $\widehat{R}$:
$R_\odot <\widehat{R}$. Hence the square roots in (\ref{chap1:innssm})
remain real.

The condition $R_\odot < \widehat R$ suggests that a sufficiently massive object
cannot be stable since no static gravitational field seems possible. This  
conjecture can be further supported. Even before reaching $\widehat R$, the
central pressure becomes infinite,
\begin{equation}
p(0) \, \rightarrow\,\infty \mbox{ for } R_{\odot}\,\rightarrow
\sqrt{\frac{8}{9}}\, \widehat R\,,\qquad \mbox{or}\qquad
m_\odot \, \rightarrow \frac{4}{9}\,R_\odot\,.
\end{equation}   
If there is no static solution and the situation remains spherically
symmetric, we are forced to the conclusion that such a mass distribution
must radially collapse; either in an infinite time or to a single point in
space. With reasonable simplifications, it was first shown by Oppenheimer
and Snyder\cite{oppenheimer:1939} that the second alternative is true: A
very massive object collapses to a black hole. As various singularity
theorems show today, this behavior is indeed generic, see Chru\'sciel et
al.\cite{Chrusciel:2012jk} and Sec.3.10.
\color{black}
\vfill

%%%%%%%%%%%%%%%%%%%%%%%%%%%%%%%%%%%%%%%%%%%%%%%%%%%%%%%%%%%%%%%%%%%%%%%%%%
\newpage

\section{The Kerr metric (1963)} 

{\it After some historical reminiscences (Sec.3.1), we point out how
  one can arrive at the Kerr metric (Sec.3.2). For that purpose, we
  derive, in cylindrical coordinates, the four corresponding partial
  differential equations and explain how this procedure leads to the
  Kerr metric. In Sec.3.3, we display the Kerr metric in three
  classical coordinate systems. Thereafter we develop the concept of
  the Kerr black hole (Sec.3.4). In Secs.3.5 to 3.7, we depict and
  discuss the geometrical/kinematical properties of the Kerr metric.
  Subsequently, in Sec.3.8, we turn to the multipole moments of the
  mass and the angular momentum of the Kerr metric, stressing
  analogies to electromagnetism. In Sec.3.9, we present the
  Kerr-Newman solution with electric charge.  Eventually, in Sec.3.10,
  we wonder in which sense the Kerr black hole is distinguished {}from
  the other stationary axially symmetric vacuum spacetimes, and, in
  Sec.3.11, we mention the rotating disk metric of Neugebauer-Meinel
  as a relevant interior solution with matter. }

\bigskip

\null\hfill\begin{minipage}{10cm} \small{\it ....When I turned to
    Alfred Schild, who was still sitting in the armchair smoking
    away, and said ``Its rotating!'' he was even more excited than I
    was. I do not remember how we celebrated, but celebrate we
    did!}\vspace{-7pt}

  \null\hfill{ Roy P.\ Kerr (2009)}
\end{minipage}

\subsection{Historical remarks}

{\em The search for axially symmetric solutions of the Einstein
  equation started in 1917 with static and was extended in 1924 to
  stationary metrics. It culminated in 1963 with the discovery of the
  Kerr metric.}\bigskip

The Schwarzschild solution, as we have seen, describes the
gravitational field of a spherically symmetric body. Obviously, most
planets, moons, and stars rotate so that spherical symmetry is lost
and one spatial direction is distinguished by the 3-dimensional angular
momentum vector $\bm{J}$ of the body. Hence the next problem to attack
was to search for the gravitational field of a massive rotating body.

When one considers a {\it static} and axially symmetric
situation---this is the case if the body does not carry angular
momentum---then one can choose the rotation axis as the $z$-axis of a
cylindrical polar coordinate system: $x^1=z$, $x^2=\rho$ and
$x^3=\phi$. Then static axial symmetry means that the components of
the metric $g_{\mu\nu}=g_{\mu\nu}(z,\rho)$ do not depend on the time $t$ and
the azimuthal angle $\phi$ (we have here one timelike and one
spacelike Killing vector\footnote{{\it Remark on Killing vectors:}
  Consider a point $P$ of spacetime with coordinates $x^\a$. We
  specify a direction $\xi^\mu$ at $P$. If we have a flat Minkowski
  space, the components $g_{\mu\nu}$ of the metric, given in Cartesian
  coordinates, would not change under a motion in the
  $\xi$-direction. However, in a curved spacetime, the $g_{\mu\nu}$
  will change in general. If $\xi^\mu$ fulfills the Killing equations
  (see Stephani \cite{Stephani})
\begin{equation}\label{Killing}
  \nabla_\mu \xi_\nu +\nabla_\nu \xi_\mu =0\,,
\end{equation}
with $\nabla$ as covariant derivative operator, then $\xi^\mu$ is
called a {\it Killing vector}, and this vector specifies a direction
under which the metric does not change. The Schwarzschild metric is
static, that is, it has one timelike Killing vector along the time
coordinate. Furthermore, it is spherically symmetric and thus has
three additional spacelike Killing vectors. In the Weyl case, because
of the axial symmetry around the $z$-axis, two of those spacelike
Killing vectors get lost. Left over in the Weyl case are the two
Killing vectors, one timelike $^{(1)}\xi^t=\partial_t$ and one
spacelike $^{(2)}\xi^\phi=\partial_\phi$.}).

Already in 1917, Weyl\cite{Weyl:1917} started to investigate static
axially symmetric vacuum solutions of Einstein's field equation. He
took cylindrical coordinates and proposed the following ``canonical''
form of the static axisymmetric vacuum line element:\footnote{Weyl
  used $\rho\,\rightarrow\, r\,,\;\phi\,\rightarrow\,\vartheta$.}
\begin{equation}\label{Weyl1917}
ds^2\stackrel{\text{}}{=}fdt^2-\left\{h(dz^2+d\rho^2)
+\frac{\rho^2d\phi^2}{f}\right\}\,;
\end{equation}
here $f=f(z,\rho)$ and $h=h(z,\rho)$ and
$(x^0=t,\,x^1=z,\,x^2=\rho,\,x^3=\phi)$. Weyl was led, in analogy to
Newton's theory, to a Poisson equation and found thereby a family of
static cylindrically symmetric solutions that could be understood as
the exterior field of a line distribution of mass along the rotation
axis. Similar investigations were undertaken by
Levi-Civita\cite{Levi-Civita} (1917/19).

In the year 1918, Lense and Thirring\cite{Lense:1918zz} investigated a
rotating body. They specified the energy-momentum tensor of a slowly
rotating ball of matter of homogeneous density and integrated the
Einstein equation in lowest approximation. They found, for a ball
rotating around the $z$-axis of a spatial Cartesian coordinate system,
the linearized Schwarzschild solution in isotropic coordinates, see
Table 2, together with two new ``gravitomagnetic'' correction terms in
off-diagonal components of the metric ($\kappa$ is Einstein's
gravitational constant):
\begin{equation}\label{LenseThirring}
  ds^2=\underbrace{\left(1-\frac{2\kappa M}{r}\right)dt^2
  -\left(1+\frac{2\kappa
      M}{r}\right)(dx^2+dy^2+dz^2)}_{\text{linearized Schwarzschild}}
 \; \underbrace{-\,\frac{4\kappa
      J_z}{r^3}(xdy-ydx)dt}_{\text{gravitomagnetic\ term}}\,;
\end{equation}
this is valid for $\kappa M\ll r$ and $ \kappa J_z\ll r^2$.  This
gravitomagnetic effect (``the Lense-Thirring effect'') is typical for
GR: in Newton's theory a rotating rigid ball has the same
gravitational field as a non-rotating one. Gravitomagnetism is alien
to Newton's gravitational theory.

In the meantime, the Lense-Thirring effect has been experimentally
confirmed by the Gravity Probe B experiment, see Everitt et
al.\cite{everitt}. They took a gyroscope in a satellite falling freely
around the (rotating) Earth. The spin axis of the gyroscope pointed to
a fixed guide star. Because of the gravitomagnetic term in
(\ref{LenseThirring}), the gyroscope executed a (very small) {\it
  Lense-Thirring precession.}\footnote{For related experiments, see
  Ciufolini et al.\cite{ciufolini,Ciufolini:2004rq} and Iorio et
  al.\cite{Iorio:2008vm,Iorio:2010rk} A recent comprehensive review was
  given by Will\cite{will}. A textbook presentation may be found in
  Ohanian \& Ruffini\cite{Ohanian3rd}.} This can be understood as an
interaction of the spin of the gyroscope with the spin of the Earth
(spin-spin interaction). Since the gyroscope moves along a 4d geodesic
of a spacetime curved by the mass of the Earth, an additional {\it
  geodetic precession} occurs that has to be experimentally separated
{}from the Lense-Thirring term. The geodetic precession had already been
derived earlier by {\it de Sitter}\cite{deSitter:1916} in
1916.\footnote{De Sitter had applied it to the Earth-Moon system
  conceived as a gyroscope precessing around the Sun (the rotation of
  which can be neglected). This effect can nowadays be measured by
  Lunar Laser Ranging, see Will\cite{will}.}

In spherical polar coordinates we have $y dx - x dy = r^2\sin^2\!\theta\,
d\phi$. Thus, the gravitomagnetic cross-term in (\ref{LenseThirring})
may be rewritten as $({4\kappa J_z\sin^2\!\theta} /{r})\,dt\,d\phi$.
A comparison with (\ref{Weyl1917}) shows that the canonical Weyl form
of the static metric is too narrow for describing rotating
bodies. 

{}From 1919 on, there appeared further articles on axisymmetric
solutions.  Levi-Civita\footnote{See Ref.\cite{Levi-Civita}, note 8
  with the subtitle {\it ``Soluzioni binarie di Weyl''.}} (1919)
reacted to Weyl's article, and Bach\cite{Bach:1922} (1922) pushed the
Lense-Thirring line element to the second order in the approximation.

Then, in 1924, Lanczos\cite{Lanczos:1924} extending the Weyl ansatz,
started to investigate {\it stationary}\footnote{Stationary spacetimes
  are those that admit a time-like Killing vector. Static spacetimes
  are stationary spacetimes for which this Killing vector is
  hypersurface orthogonal; physically this implies time reversal
  invariance and thus the absence of rotation.}
solutions. He found an exact solution for uniformly rotating
dust. However, his work was apparently partially overlooked. Later,
Akeley\cite{Akeley:1931a,Akeley:1931b} (1931), Andress\cite{Andress}
(1930) and, in a more definite form, Lewis\cite{Lewis:1932} (1932)
generalized the static Weyl metric to a {stationary} one by taking
into account the gravitomagnetic term of Lense-Thirring. Lewis (1932)
wrote, in cylindrical polar coordinates ($x_1\rightsquigarrow \rho\,,
x_2\rightsquigarrow z$),
\begin{equation}\label{Le}
  ds^2 %\stackrel{\text{Le}}{=}
=fdt^2-(e^\mu dx_1{}^2+e^\nu dx_2{}^2
+ld\phi^2)-2mdt\,d\phi\,.
\end{equation}
He found some exact solutions, typically for rotating cylinders, but
not for rotating balls. It became definitely clear that, in the
axially symmetric case, we may have many different exact vacuum
solutions, in contrast to the case of spherical symmetry with,
according to the Birkhoff theorem, the Schwarzschild solution as being
unique.

Not much later, van Stockum\cite{vanStockum:1937} (1937) determined
the gravitational field of an infinite rotating cylinder of dust
particles, thereby recovering the Lanczos solution, inter alia. He
fitted one of the interior matter solutions of Lewis to an exterior
vacuum solution. Continuing on this line of research,
Papapetrou\cite{Papapetrou:1953} (1953) started {}from the Andress-Lewis
line element, putting it in a slightly different form, suitable for
all stationary axisymmetric vacuum
solutions:%\setcounter{footnote}{0}:
\begin{equation}\label{Pa}
  ds^2 =%\stackrel{\text{Pa}}{=}
  - e^\mu (d\rho^2 + dz^2) - l d\phi^2 
  - 2 m\, d\phi\, dt + fdt^2\,.
\end{equation}
The functions $\mu,l,m$, and $f$ depend only on $\rho$ and
$z$. Papapetrou integrated the field equations and found exact
stationary rotating vacuum solutions. However, his solution carried
either {\it mass and no angular momentum} or {\it angular momentum and
  no mass.} Thus\cite{Papapetrou:1953}, ``this solution is very
special and physically of little interest.''
% \begin{eqnarray}\label{Le}
%   ds^2 &\stackrel{\text{Le}}{=}& fdt^2-(e^\mu dx_1{}^2+e^\nu dx_2{}^2
% +ld\phi^2)-2mdt\,d\phi\,,\\ \label{Pa}
%   ds^2 &\stackrel{\text{Pa}}{=}& - e^\mu (d\rho^2 + dz^2) - l d\phi^2 
%   - 2 m d\phi dt + fdt^2\,.\\ \label{Bu}
%   ds^2 &\stackrel{\text{Bu}}{=}&-f^{-1}\left[e^{2\gamma}(dr^2+dz^2)
%     +r^2d\phi^2 \right]+f(dt-\omega d\phi)^2\,.
% \end{eqnarray}

A year later, a new result was published, which gave the problem of
finding solutions for a rotating ball a new
direction. Petrov\cite{Petrov:1954} (1954), {}from Kazan, classified
algebraically the Einstein vacuum field, that is, the Weyl curvature
tensor, according to its eigenvalues and eigenvectors. This
information reached the West, in the time of the Cold War, with some
delay. A bit later, Pirani\cite{Pirani:1956wr} (1957) developed a
related formalism. It was the Petrov classification and the picking of
a suitable class for the gravitational field of an isolated body
(Petrov class D, with two double principal null directions) that
finally led to the discovery of the Kerr solution during 1963, ten
years after the unphysical solutions of Papapetrou.

Accordingly, it turned out to be a formidable task to find an exact
solution for a rotating ball and it was only found nearly half a
century after the publication of Einstein's field equation, namely in
1963 by Roy Kerr \cite{Kerr:1963}, a New Zealander, who worked at
the time in Texas within the research group of Alfred Schild. It is a
2-parameter solution of Einstein's vacuum field equation with mass $M$
and rotation (or angular momentum) parameter $a:=J/M$.

The story of the discovery of the Kerr solution was told by Kerr
himself at a conference on the occasion of his 70th birthday
\cite{kerr2}. A decisive starting point of Kerr's investigations was,
as mentioned, the Petrov classification. Melia, in his popular
book\cite{Melia} ``Cracking the Einstein Code'', which does not
contain any mathematical formula---apart {}from those appearing in two
copies of Kerr's notes and on a blackboard in another figure---has
told this fascinating battle for solving Einstein's equation, see also
the Kerr story in Ferreira\cite{Ferreira}.

Dautcourt \cite{Dautcourt} discussed the work of people who were
involved in this search for axially symmetric solutions but who were
not so fortunate as Kerr. In particular, Dautcourt himself got this
problem handed over {}from Papapetrou in 1959 as a subject for
investigation. He used the results of Papapetrou (1953).  Dautcourt's
scholarly article is an interesting complement to Melia's book. In
particular, it becomes clear that the
(Lanczos-Akeley-Andress-Lewis-)Papapetrou line element (\ref{Pa}) was
the correct ansatz for the stationary axially symmetric metric and the
Kerr metric is a special case there{}from. The Papapetrou approach with
the line element (\ref{Pa}) was later, after Kerr's discovery, brought
to fruition by Ernst\cite{Ernst:1968} and by Kramer and
Neugebauer\cite{Kramer:1968}.

\subsection{Approaching the Kerr metric}

{\em We derive a 2nd order partial differential equation, the Ernst
  equation, that governs the stationary axially symmetric metrics in
  Einstein's theory. Subsequently, we sketch how the Kerr solution
  emerges as a simple case there{}from.}

\subsubsection*{Papapetrou line element and vacuum field equation}

In more modern literature, the Papapetrou line element (\ref{Pa}),
which describes some rotation around the axis with $\rho=0$, is
usually parametrized as follows\footnote{See Ernst \cite{Ernst:1968},
  Buchdahl\cite{Buchdahl}, de Felice \& Clarke\cite{felice}, Quevedo
  \cite{Quevedo:1990}, O'Neill\cite{O'Neill}, Stephani et
  al.\cite{S_K_2004}, Eq.(19.21), Griffiths \&
  Podolsk\'y\cite{Griffiths:2009}, and
  Sternberg\cite{SternbergDover}.},
\begin{eqnarray}\label{Bu}
  ds^2 %\stackrel{\text{Bu}}{=}
  &=&f(dt-\omega d\phi)^2-f^{-1}\left[e^{2\gamma}(d\rho^2+dz^2)
    +\rho^2d\phi^2 \right]\,,\\ \nonumber
  &&t\in(-\infty,\infty),\quad \rho\in [0,\infty),\quad z\in(-\infty,
  \infty),\quad\phi\in[0,2\pi)    \,;
\end{eqnarray}
we assume $f>0$. We compute the vacuum field equation of this
metric. Nowadays we can do this straightforwardly with the assistance
of a computer algebra system. During the 1960s, when this work was
mainly done, there were no computer algebra systems around. Hearn\cite{Hearn}
released the computer algebra system REDUCE in 1968. Back then,
one had to be in command of huge computer resources in order to bring
the underlying computer language LISP to work. Today, Reduce can run on
every laptop; for other computer algebra systems, see Grabmeier et
al.\cite{Johannes} and Wolfram\cite{Mathematica}.

Because of its efficiency, we will use Schr\"ufer's Reduce-package
EXCALC, which was built for manipulating expressions within the
calculus of exterior forms. For that purpose, we reformulate the
metric (\ref{Bu}) in terms of an orthonormal coframe of four 1-forms
$\vartheta^\a=e_i{}^\a dx^i$, with the unknown functions
$f=f(\rho,z)$, $\omega=\omega(\rho,z)$, and
$\gamma=\gamma(\rho,z)$, namely
\begin{eqnarray}\label{76}
  \vartheta^{\hat0} & = &\; {f}^{\frac{1}{2}}\;(d t 
  -\omega  d\,\phi)\quad =e_i{}^{\hat0}dx^i= {f}^{\frac{1}{2}}
  \;(dx^0-\omega dx^3)\,, \\
  \vartheta^{\hat1} & = &
  {f}^{-\frac{1}{2}}e^\gamma d\,\rho\hspace{35pt}
  =e_i{}^{\hat1}dx^i= {f}^{-\frac{1}{2}}e^\gamma dx^1\,,\\
  \vartheta^{\hat2} & = & {f}^{-\frac{1}{2}}e^\gamma d\,z\hspace{35pt} 
  =e_i{}^{\hat2}dx^i= {f}^{-\frac{1}{2}}e^\gamma dx^2\,,\\
  \vartheta^{\hat3} & = & {f}^{-\frac{1}{2}}\rho\, d\,\phi\hspace{37pt} 
  =e_i{}^{\hat3}dx^i= {f}^{-\frac{1}{2}}\,\rho\, dx^3\,.
\end{eqnarray}
Because of the orthonormality of the coframe $\vartheta^\a$, we have
\begin{equation}\label{80} 
ds^2\equiv
    g= +\,{\vartheta}^{\hat 0}\otimes
    {\vartheta}^{\hat 0} - {\vartheta}^{\hat 1}\otimes
    {\vartheta}^{\hat 1} - {\vartheta}^{\hat 2}\otimes
    {\vartheta}^{\hat 2} - {\vartheta}^{\hat 3}\otimes
    {\vartheta}^{\hat 3}\,.  
\end{equation}
Eqs.(\ref{76}) to (\ref{80}) are equivalent to (\ref{Bu}).
The corresponding computer code, as input for Reduce-Excalc, reads as
follows:\vspace{-10pt}

\begin{verbatim}
pform f=0, omega=0, gamma=0 $
fdomain f=f(rho,z), omega=omega(rho,z), gamma=gamma(rho,z);

coframe  o(0) = sqrt(f) * (d t - omega * d phi),
         o(1) = sqrt(f)**(-1) * exp(gamma) * d rho,
         o(2) = sqrt(f)**(-1) * exp(gamma) * d z,
         o(3) = sqrt(f)**(-1) * rho * d phi
    with signature (1,-1,-1,-1);
\end{verbatim}\vspace{-9pt}%$
\noindent Isn't that simple enough? {}From this data, the Einstein
equation is calculated, with the Einstein tensor $G^\mu{}_\nu$. The
complete, fairly trivial program is documented in the Appendix. Note
in particular that we used a \LaTeX\,interface allowing us to output the
expressions directly in \LaTeX. This computer output---without changing
anything of the formulas---after some post-editing for display
purposes, reads as follows:\small
%\newpage
\begin{eqnarray}
{G^{0}{}_{0}}&:=&
\left(4\cdot \partial _{\rho ,
                      \rho }f\cdot f\cdot \rho ^{2}
  -5\cdot \partial _{\rho }f^{2}\cdot \rho ^{2}
    +4\cdot \partial _{\rho }f\cdot f\cdot \rho% \nl 
    +4\cdot \partial _{z,
                       z}f\cdot f\cdot \rho ^{2}\right.\nonumber\\
  &&\left.  -5\cdot \partial _{z}f^{2}\cdot \rho ^{2} \nonumber
 -4\cdot \partial _{\rho ,
                       \rho }\gamma \cdot f^{2}\cdot \rho ^{2}
    -4\cdot \partial _{z,
                       z}\gamma \cdot f^{2}\cdot \rho ^{2}
    +3\cdot \partial _{\rho }\omega ^{2}\cdot f^{4}\right.\\
  &&\left.  +3\cdot \partial _{z}\omega ^{2}\cdot f^{4}
  \right)
  /\left(4\cdot e^{2\cdot \gamma }\cdot f\cdot \rho
    ^{2}\right)\,,\label{E00} \\
  %&& \nonumber\\
{G^{3}{}_{0}}&:=&
f\cdot 
        \left(2\cdot \partial_{\rho }f\cdot \partial_{\rho }
          \omega \cdot \rho 
          +2\cdot \partial_{z}f\cdot \partial_{z}\omega \cdot 
          \rho +\partial_{\rho,\rho }\omega \cdot f\cdot \rho 
          -\partial_{\rho }\omega \cdot f \right.\cr
       && \left.   +\partial_{z,z}\omega \cdot f\cdot \rho\right)
        /
       ( 2\cdot e^{2\cdot \gamma }\cdot \rho ^{2})\,,\label{E30}
\\ %&& \nonumber \\
{G^{1}{}_{1}}&:=&
(\partial _{\rho }f^{2}\cdot \rho ^{2}
        -\partial _{z}f^{2}\cdot \rho ^{2}
        -4\cdot \partial _{\rho }\gamma \cdot f^{2}\cdot \rho 
        -\partial _{\rho }\omega ^{2}\cdot f^{4}
        +\partial _{z}\omega ^{2}\cdot f^{4})\cr &&/(
        4\cdot e^{2\cdot \gamma }\cdot f\cdot \rho ^{2})\,,\label{E11}
\\ %&& \nonumber \\
{G^{1}{}_{2}}&:=&
(\partial _{\rho }f\cdot \partial _{z}f\cdot \rho ^{2}
        -2\cdot \partial _{z}\gamma \cdot f^{2}\cdot \rho 
        -\partial _{\rho }\omega \cdot \partial _{z}\omega 
        \cdot f^{4})/(
        2\cdot e^{2\cdot \gamma }\cdot f\cdot \rho ^{2})\,,\label{E12}
\\  %&&  \nonumber \\
G^{3}{}_{3}&:=&(-\partial_{\rho}f^{2}\cdot \rho ^{2}
        -\partial _{z}f^{2}\cdot \rho ^{2}
        -4\cdot \partial _{\rho,\rho}\gamma \cdot f^{2}\cdot \rho^{2}
        -4\cdot \partial_{z,z}\gamma \cdot f^{2}\cdot
        \rho^{2}\nonumber  \\ 
&&   -\partial _{\rho }\omega ^{2}\cdot f^{4}
        -\partial _{z}\omega ^{2}\cdot f^{4}) / (
        4\cdot e^{2\cdot \gamma }\cdot f\cdot \rho ^{2} )\,.\label{E33}
\end{eqnarray}\normalsize
This calculation of the Einstein tensor by machine did not require
more than about 15 minutes, including the programming and the typing
in; for sample programs, see Socorro et al.\cite{Socorro} and Stauffer
et al.\cite{reduce93}.

Inspecting these equations, it becomes immediately clear that the
numerator of (\ref{E30}) does not depend on $\gamma$. In order to get a
better overview, we abbreviate the partial derivatives of Reduce
$\partial_\rho f$ by subscripts, $f_\rho$, and drop the superfluous
multiplication dots of Reduce. We find
\begin{equation}\label{E30*}
  G^3{}_0=0\;\rightarrow\qquad 0= f(\omega_{\rho\rho}+\omega_{zz}
  -\frac{1}{\rho}\omega_\rho)+2(f_\rho \omega_\rho+f_z\omega_z)\,.
\end{equation}
Moreover, by subtracting (\ref{E33}) {}from  (\ref{E00}) we find another
equation free of $\gamma$:
\begin{equation}\label{E00-33}
 G^0{}_0-G^3{}_3=0\;\rightarrow\qquad
  0=f(f_{\rho\rho}+\frac{1}{\rho}f_\rho+f_{zz})-f_\rho^2
  -f_z^2+\frac{f^4}{\rho^2}
  (\omega_\rho^2 + \omega_z^2)\,.
\end{equation}
Left over are the equations (\ref{E11}) and (\ref{E12}), which can be
resolved with respect to the first derivatives of $\gamma$:
\begin{eqnarray}
  G^1{}_1=0\;\rightarrow\qquad  \gamma_\rho&=&\frac{\rho}{4f^2}(f_\rho^2-f_z^2)
  +\frac{f^2}{4\rho}(\omega_z^2-\omega_\rho^2)\,,\\
  G^1{}_2=0\;\rightarrow\qquad  \gamma_z&=& \frac{\rho}{2f^2}f_\rho f_z
  -\frac{f^2}{2\rho}\omega_\rho\omega_z\,.
\end{eqnarray}

Collected, we have these four equations determining the
  stationary axisymmetric vacuum metric:
\begin{eqnarray} \label{f-rho-rho}
  0&=&f(f_{\rho\rho}+\frac{1}{\rho}f_\rho+f_{zz})-f_\rho^2
  -f_z^2+\frac{f^4}{\rho^2}
  (\omega_\rho^2 + \omega_z^2)\,,\\ \label{omega-rho-rho}
  0&=& f(\omega_{\rho\rho}+\omega_{zz}
  -\frac{1}{\rho}\omega_\rho)+2(f_\rho \omega_\rho+f_z\omega_z)\,,\\
\label{gamma-rho}
  \gamma_\rho&=&\frac{\rho}{4f^2}(f_\rho^2-f_z^2)
  +\frac{f^2}{4\rho}(\omega_z^2-\omega_\rho^2)\,,\\ \label{gamma-z}
  \gamma_z&=& \frac{\rho}{2f^2}f_\rho f_z
  -\frac{f^2}{2\rho}\omega_\rho\omega_z\,.
\end{eqnarray}
Let us underline how effortless---under computer assistance---we
arrived at these four partial differential equations (PDEs) for
determining stationary axially symmetric solutions of Einstein's field
equation.

\subsubsection*{Ernst equation (1968)}

It is one step ahead, before we arrive at a still more convincing form of
these PDEs. After some attempts, one recognizes that
(\ref{omega-rho-rho}) can be written as
\begin{equation}\label{grad}
\left(\frac{f^2}{\rho}\omega_\rho\right)_\rho 
+\left(\frac{f^2}{\rho}\omega_z\right)_z=0\,.
\end{equation}
With the ansatz ($\Omega=\Omega(\rho,z)$),
\begin{equation}\label{ansatz}
\Omega_z=\frac{f^2}{\rho}\omega_\rho\,,\qquad
\Omega_\rho=-\frac{f^2}{\rho}\omega_z\,,
\end{equation}
Eq.(\ref{grad}) is identically fulfilled. We substitute (\ref{ansatz})
into (\ref{f-rho-rho}):
\begin{eqnarray}\label{Delta-f*}
  f(f_{\rho\rho}+\frac{1}{\rho}f_\rho+f_{zz})
  -f_\rho^2 -f_z^2+\Omega_\rho^2+\Omega_z^2=0\,.
\end{eqnarray}
Since (\ref{omega-rho-rho}) is already exploited, we can find $\Omega$
by differentiating the $\Omega$'s in (\ref{ansatz}) with respect to $z$
and $\rho$, respectively, and by adding the emergent expressions
($\omega_{\rho z}=\omega_{z \rho}$):
\begin{eqnarray}\label{Delta-omega*}
  f(\Omega_{\rho\rho}+\frac{1}{\rho}\Omega_\rho+\Omega_{zz})
  -2f_\rho\Omega_\rho-2f_z\Omega_z=0\,.
\end{eqnarray}

Eqs.(\ref{f-rho-rho}) and (\ref{Delta-omega*}) can be put
straightforwardly into a vector analytical form, if we recall that our
functions do not depend on the angle $\phi$:\footnote{In cylindrical
  coordinates, we have for a vector $\mathbf{V}$ and a scalar $s$ the
  following formulas, see Jackson\cite{Jackson}:
\renewcommand{\baselinestretch}{1.3}\footnotesize
\[
\begin{array}{rclrcl}
  \bm{\nabla} \cdot \mathbf{V}&=&\frac{1}{\rho}\,\partial_\rho(\rho V_1)
+\partial_z V_2  +\frac{1}{\rho}\partial_\phi V_3\,,&
  \nabla^2 s&\equiv&\Delta s=\frac{1}{\rho}\partial_\rho(\rho\,\partial_\rho s)
 +\partial^2_z s +\frac{1}{\rho^2}\partial^2_\phi s\,,\\
 \bm{\nabla}s&=&\mathbf{e_1}\partial_\rho s+\mathbf{e_2}\partial_z s
+\mathbf{e_3}\frac{1}{\rho}\partial_\phi s\,,&
 \bm{\nabla}s\cdot \bm{\nabla}s&=&(\partial_\rho s)^2+(\partial_z s)^2
+\frac{1}{\rho^2}(\partial_\phi s)^2\,.
\end{array}
\]
\renewcommand{\baselinestretch}{1}\small\normalsize}
\begin{eqnarray}\label{fDelf}
  f\Delta f- (\bm{\nabla}f)\cdot \bm{\nabla}f
  + (\bm{\nabla}\Omega)\cdot \bm{\nabla}\Omega&=&0\,,\\ \label{fDelOmega}
f\Delta\Omega-2(\bm{\nabla}f)\cdot\bm{\nabla}\Omega&=&0\,. 
\end{eqnarray}
The last equation can also be written as
$\bm{\nabla}\cdot(f^{-2}\bm{\nabla} \Omega)=0$. Eqs.(\ref{fDelf})
and (\ref{fDelOmega}) liberate ourselves {}from the cylindrical
coordinates, that is, this expression is now put in form independent
of the specific 3d coordinates.
With the potential ($i^2=-1$)
\begin{equation}\label{ErnstPot}
  \mathcal{E}:=f+i\Omega\,,
\end{equation}
which was found by Ernst\cite{Ernst:1968} and Kramer \&
Neugebauer\cite{Kramer:1968}, we find the Ernst
equation\cite{Ernst:1968}
\begin{equation}\label{Ernst}
  (\text{Re}\,\mathcal{E})\,\Delta\hspace{1pt}\mathcal{E}=\bm{\nabla}
  \mathcal{E}\bm{\cdot}  \bm{\nabla}\mathcal{E}\,,
\end{equation}
or, in components,
\begin{equation}
(\text{Re}\mathcal{E})\!\left[
\frac{\partial^2\mathcal{E}}{\partial
z^2}+\frac{1}{\rho}\,\frac{\partial}{\partial
\rho}\left(\rho\,\frac{\partial\mathcal{E}}{\partial\rho}\right)\right]
=
\left(\frac{\partial\mathcal{E}}{\partial z}\right)^{\!2}
+ \left(\frac{\partial\mathcal{E}}{\partial\rho}\right)^{\!2}\,.
\end{equation}
The ``Re'' denotes the real part of a complex quantity. Under stationary
axial symmetry---the corresponding metric is displayed in
(\ref{Bu})---the Ernst equation (\ref{Ernst}), together with
Eqs.(\ref{ErnstPot}, \ref{ansatz}, \ref{gamma-rho}, \ref{gamma-z}),
are equivalent to the vacuum Einstein field equation.

\subsubsection*{{}From Ernst back to Kerr}

This reduces the problem of axial symmetry to the solution of the
second order PDE (\ref{Ernst}). This method, which came along only
five years after Kerr's publication, led to many new exact solutions,
amongst them the Kerr solution (1963) as one of the simplest cases. We
are only going to sketch how one arrives at the Kerr solution
eventually. We follow here closely Buchdahl\cite{Buchdahl}.

One introduces a new complex potential $\xi$ by
\begin{equation}\label{xi}
\mathcal{E}=:\frac{\xi-1}{\xi+1}\,.
\end{equation}
Then the Ernst equation becomes
\begin{equation}\label{ernst*}
  (\xi\overline{\xi}-1)\Delta\xi=2\overline{\xi}\,\bm{\nabla}\xi\bm{\cdot}
    \bm{\nabla}  \xi\,,
\end{equation}
where the overline denotes complex conjugation. If one has a solution
of this equation, we can determine the functions $f$, $\omega$ and
$\gamma$ by
\begin{eqnarray}\label{Functionf}
f&=&\text{Re}\frac{\xi-1}{\xi+1}\,,\\
\omega_\rho&=&- 2\rho\,\frac{\text{Im}[(\overline{\xi}+1)^2\xi_z ]}
{(\xi\overline{\xi}-1)^2}\,,\qquad
\omega_z= 2\rho\,\frac{\text{Im}[(\overline{\xi}+1)^2\xi_\rho ]}
{(\xi\overline{\xi}-1)^2}\,,\\
\gamma_\rho&=&\rho\,\frac{\xi_\rho\overline{\xi}_\rho-\xi_z\overline{\xi}_z}
{(\xi\overline{\xi}-1)^2}\,,\hspace{40pt}
\gamma_z= 2\rho\,\frac{\text{Re}(\xi_\rho\overline{\xi}_z)}
{(\xi\overline{\xi}-1)^2}\,.
\end{eqnarray}

For rotating bodies, {\it spherical prolate coordinates} $x,y$, with a
constant $k$, are much more adapted:
\begin{equation}
\rho=k(x^2-1)^{\frac{1}{2}}(1-y^2)^{\frac{1}{2}}\,,\qquad z=kxy\,.
\end{equation}
It turns out that one simple potential solving the Ernst equation,
with the constants $p$ and $q$, is
\begin{equation}\label{KerrXi}
  \xi=px-iqy\qquad\text{with}\qquad p^2+q^2=1\,;
\end{equation}
it leads to the Kerr metric. For this purpose, one has to introduce
the redefined constants $m:=k/p$ (mass) and $a:=kq/p$ (angular
momentum per mass) and to execute subsequently the transformations
$px=(\tilde{\rho}/m)-1$ and $qy=(a/m)\cos\theta$ to the new
coordinates $\tilde{\rho}$ and $\theta$. Then one arrives at the Kerr
metric in Boyer-Lindquist coordinates, which is displayed in the table
on the next page. For more detail, compare, for instance, the books of
Buchdahl\cite{Buchdahl}, Islam\cite{Islam:1985},
Heusler\cite{Heusler:1996}, Meinel et al.\cite{MeinelEtAl2008}, or
Griffiths et al.\cite{Griffiths:2009}. By similar techniques, a Kerr
solution with a topological defect was found by Bergamini et
al.\cite{Bergamini:2003ch}.
%\newpage

Incidentally, in the context of the Ernst equation, Geroch made the
following interesting conjecture: A subset of all stationary axially
symmetric vacuum space-times, including all of its asymptotically flat
members, that is, in particular the Kerr solution, can be obtained
{}from Minkowski space by transformations generated by an
infinite-dimensional Lie group.  This conjecture was ``proved'' by
Hauser and Ernst\cite{Hauser:1981}, see also Ref.\cite{Hauser:1980}
However, the proof contained a mistake that was subsequently corrected
in Ref.\cite{Hauser:1987}

Starting from 4d ellipsoidal coordinates, Dadhich\cite{Dadhich} gave a
heuristic derivation of the Kerr metric by requiring, amongst other
things, that light propagation should be influenced by gravity.

\subsection{Three classical representations of the Kerr metric}

We collected these three classical versions of the Kerr metric in
Table 4, see also Visser\cite{VisserKerrFest}.%\vspace{-10pt} 
Three more coordinate systems should at least be mentioned:
\begin{itemize}
\item {\it Pretorius \& Israel}\cite{PretoriusIsrael} double null coordinates: 

very convenient to tackle the initial value problem
\item{\it Doran}\cite{Doran:1999gb} coordinates: 

Gullstrand-Painlev\'e like; useful in analog gravity

\item{\it Debever/Pleba\'nski/Demia\'nski}\cite{pd} coordinates: 

components of the metric
are rational polynomials; convenient for (computer assisted)
calculations
\end{itemize}

\newpage\small
\begin{tabular}{||c||}
\hline\hline\hline
\vbox{
\vspace{2mm}
\centerline{\bf\large Table 4. Kerr metric: the three classical representations}
\vspace{1mm}}
\\\hline\hline\hline
\vbox{\null\vspace{2mm}
{\bf Kerr-Schild}\hfill $(t,x,y,z)$ \hfill{\rm Cartesian background}
\vspace{0.5cm}
\begin{eqnarray}
ds^2 = &-& dt^2+dx^2+dy^2+dz^2\nonumber\\
&+&\frac{2mr^3}{r^4+a^2z^2}\left(dt + \frac{r(x\,dx+y\,dy)}{a^2+r^2}
+\frac{a(y\,dx-x\,dy)}{a^2+r^2}+\frac{z}{r}\,dz\right)^{\! 2}\nonumber
\end{eqnarray}}
\\\hline
\vbox{
\begin{equation}
x^2+y^2+z^2=r^2+a^2\left(1-\frac{z^2}{r^2}\right)\,,\qquad
r=r(x,y,z)\nonumber
\end{equation}}
\\\hline  
\hline  
\vbox{\begin{eqnarray*}
%t &=& u+r\\
x &=& (r\cos\phi + a \sin\phi)\sin\theta \\
y &=& (r\sin\phi - a \cos\phi)\sin\theta \\
z &=& r\cos\theta
\end{eqnarray*}}
\\\hline\hline  
\vbox{\null\vspace{2mm}
{\bf Boyer-Lindquist}\hfill$(t,r,\theta,\phi)$\hfill
{\rm Schwarzschild like}
\vspace{0.5cm}
\begin{eqnarray}\label{Kerrmetric}
  ds^2=&-&\left(1-\frac{2mr}{\rho^2}\right)dt^2-\frac{4mra\sin^2\!\theta}
{\rho^2}\,dt\,d\phi  \nonumber\\
  &+&\frac{\rho^2}{\Delta}\,dr^2+\rho^2d\theta^2+\left(r^2+a^2
+\frac{2mra^2\sin^2\!\theta}{\rho^2} \right) \sin^2\!\theta\, d\phi^2
\nonumber
\end{eqnarray} }
\\\hline
\vbox{
\begin{equation*}
\rho^2  :=  { r ^2 + a^2 \cos^2\theta}\qquad
\Delta  :=   r ^2  -2m r +a^2 =(r-r_+)(r-r_-)
\end{equation*}
}
\\\hline\hline
\vbox{\begin{eqnarray*}
dv &=& dt + \frac{r^2+a^2}{\Delta}\,dr\\
d\varphi &=& d\phi + \frac{a}{\Delta}\,dr
\end{eqnarray*}}
\\\hline\hline
\vbox{\null\vspace{2mm}
{\bf Kerr original} \hfill$(v,r,\theta,\varphi)$ \hfill 
{\rm Eddington-Finkelstein like}
\vspace{0.5cm}
\begin{eqnarray}
ds^2 = &-&\left(1-\frac{2mr}{\rho^2}\right) \left(d v - a\, \sin^2\theta 
\,d\varphi \right)^2\nonumber\\
&+ &2 \left(dv - a \sin^2\theta \, d \varphi \right)
\left(dr - a \sin^2\theta \, d \varphi \right)
+ \rho^2 \left(d\theta^2 + \sin^2\theta \, d\varphi^2\right)
\nonumber
\end{eqnarray}}
\\\hline\hline
\hline  
\vbox{\null\vspace{2mm}
\begin{equation*}
r_{{\text{E}_\pm}}:= m \pm \sqrt{m^2-a^2 \, \cos^2 \theta}\qquad
r _{\pm} := m \pm \sqrt{m^2-a^2}
\end{equation*}
}\\
\hline\hline
\end{tabular}\normalsize

As input for checking the {\it Kerr} solution, we use the {\it
  orthonormal coframe}\cite{SternbergDover}
\begin{eqnarray}\label{KerrCoframe}
\vartheta^{\hat 0} & := & \frac{\sqrt{{\color{magenta}\epsilon}\Delta}}{\rho
} \left(
  dt - a \, \sin^2\! \theta \, d\phi \right) \,,\\
\vartheta^{\hat 1} & := & \frac{\rho }{\sqrt{{\color{magenta}\eps}\Delta}}\,
d r \,,\\
\vartheta^{\hat 2} & := & \rho  \, d\theta\,, \\
\vartheta^{\hat 3} & := & \frac{\sin\theta}{\rho } \left[( r  ^2
                    + a^2 )\, d\phi -a  dt \right]\,.
\end{eqnarray}
We introduced the sign function, which is convenient for discussing
the different regions in the Penrose-Carter diagram:
\begin{equation}\label{KerrEpsi}
{\color{magenta}\eps}=\begin{cases}
+1\text{  for  }r>r_+ \text{ or  }r<r_-\,,,\\
-1\text{  for  }r_-<r<r_+ \,.\\
\end{cases}
\end{equation}
The metric can then be written in terms of the coframe as
\begin{equation}\label{anhmetric}
  ds^2\equiv g={\color{magenta}\eps}( -{\vartheta}^{\hat 0}\otimes
 {\vartheta}^{\hat 0}
  + {\vartheta}^{\hat 1}\otimes
  {\vartheta}^{\hat 1}) + {\vartheta}^{\hat 2}\otimes {\vartheta}^{\hat 2} +
  {\vartheta}^{\hat 3}\otimes {\vartheta}^{\hat 3}\,.
\end{equation}

% \Text...  % \\subsection{???}%Appendix subsection}
% \Text...
% \begin{equation}
% \mu(n, t) = ...
% \label{appen1}
% \end{equation}

{}From Table 4 it is not complicated to read off the Schwarzschild and
the Lense-Thirring metric as special cases.
In comparison to the Schwarzschild metric, the Kerr solution includes a new
parameter $a$ which will be related to the angular momentum. However, it   
should be noted that, by setting $a=0$, the Kerr metric reduces to the    
Schwarzschild metric, as it should be 
($\rho^2 \rightarrow r^2$ and $\Delta \rightarrow r^2 - 2mr$):
\begin{eqnarray}
  ds^2&=&-\left(1-\frac{2mr}{r^2}\right)dt^2
+ \frac{r^2}{r^2-2mr} \, dr^2+r^2d\theta^2+r^2 \sin^2\!\theta\,
d\phi^2\,.%\nonumber \\
\end{eqnarray}
By canceling $r^2$ in the $dr^2$-term, we immediately recognize the
Schwarzschild metric.

Considering the parameter $a$ we should note the following fact.
For small values of the parameter $a$, where we may neglect terms of the
order of $a^2$, we arrive at ($\rho^2 \rightarrow r^2$ and $\Delta
\rightarrow r^2 - 2mr$) and
\begin{eqnarray}
  ds^2=&-&\left(1-\frac{2m}{r}\right)dt^2
+ \frac{1}{1-\frac{2m}{r}}\,dr^2
+ r^2 \, \left(d\theta^2+\sin^2\!\theta\, d\phi^2\right)\nonumber \\
&-&\frac{4ma\sin^2\!\theta} {r}\,dt\,d\phi  
\,.
\end{eqnarray}
Since, in spherical coordinates we have
$y\, dx - x\, dy = r^2\,\sin^2\theta \, d\phi$, the crossterm
may be rewritten as $\frac{4ma\sin^2\!\theta} {r}\,dt\,d\phi =
\frac{4ma}{r^3}\, ( x \, dz - y\, dx)$.
Thus, in the limiting case $a^2 \ll 1$, the Kerr metric yields the
Lense-Thirring metric, provided we identify $ma=J_z$.

\vfill
\pagebreak

\subsection{The concept of a Kerr black hole}\label{kerrconcept}

{\it We come back to our \fref{chap2:lakes} with ``Schwarzschild''
versus ``Kerr''. The Kerr space\-time may be visualized by a vortex,
where the water of the lake spirals towards the center. Much of the
above said for the Schwarzschild case is still valid. However, one
important difference occurs. The stationary limit and the event
horizon separate, which will be illustrated by corresponding graphical
representations.}

\bigskip

In case of a vortex, the flow velocity of the in-spiraling water has
two components. The radial component which drags the boat towards the
center whereas the additional angular component forces the boat to
circle around the center. Again, the {\it stationary limit} is defined
by the distance at which the boat ultimately can withstand the radial
and circular drag of the water flow. Beyond the stationary limit the
situation is not as hopeless as in the Schwarzschild case. Using all
its power, the boat may brave the inward flow. But then it has not
enough power to overcome the angular drag and is forced to orbit the
center. By means of a clever spiral course the boat may even escape
beyond the stationary limit. The stationary limit is not necessary an
event horizon. At some distance, nearer to the center than the
stationary limit, also the pure radial flow of water will exceed the
power of the boat. There, inside the stationary limit, is the event
horizon.

In order to investigate the structure of the Kerr spacetime, we first look
at ``strange behavior'' of the metric components in Boyer-Lindquist
coordinates. The following cases can be distinguished: 

\bigskip

\renewcommand{\baselinestretch}{1.5}\small\normalsize
\begin{tabular}{ll}
$\Delta = 0$ & $g_{rr}$ becomes singular,\\
$\rho^2 = 2mr$ & $g_{tt}$ vanishes,\\
$\rho^2=0$ & $g_{rr}$ and $g_{\theta\theta}$ vanish,
  the other components are singular.
\end{tabular}
\renewcommand{\baselinestretch}{1}\small\normalsize

\bigskip

As we have extensively discussed in the previous section, singularities
of components of the metric {\em may} signify physical effects but, on the
other hand, may only be due to ``defective'' coordinates. Thus, we will
proceed along similar lines to investigate the nature of these
singularities.

We will not address the geodesics of the Kerr metric in detail. For an
elementary discussion the reader is referred to Frolov and
Novikov\cite{frolov} and to the more advanced discussion in Hackmann
et al.\cite{Hackmann:2014tga}.

\pagebreak

\subsubsection*{Depicting Kerr geometry}

We draw a picture of the spatial appearances
and relations of the various horizons and the singularity of the Kerr
metric. {}From outside to inside these are, explicitly,
\begin{equation*}\renewcommand{\baselinestretch}{1.3}\normalsize
\begin{array}{llccccl}
\mbox{outer} & \mbox{ergosurface\qquad} & \quad r_{E+}\quad
&:=&m&+&\sqrt{m^2-a^2\,\cos^2\theta}\\
&&&&&& \qquad\qquad\updownarrow\mbox{\tiny joined at polar axis}\\
\mbox{event} & \mbox{horizon} & r_{+}
&:=&m&+&\sqrt{m^2-a^2}\\
&&&&&& \qquad\updownarrow\mbox{\tiny merge for  $a\to m$}\\
\mbox{Cauchy} & \mbox{horizon} & r_{-}
&:=&m&-&\sqrt{m^2-a^2}\\
&&&&&& \qquad\qquad\updownarrow\mbox{\tiny joined at polar axis}\\
\mbox{inner} & \mbox{ergosurface} & r_{E-}
&:=&m&-&\sqrt{m^2-a^2\,\cos^2\theta}\\
&&&&&& \uparrow\mbox{\tiny lies on the rim for $\theta=\pi/2$}\\
&\mbox{singularity} & r & = & 0
\end{array}\renewcommand{\baselinestretch}{1}\small\normalsize
\end{equation*}\addtocounter{equation}{1}
For $a=0$, inner ergosurface and Cauchy horizon vanish, whereas outer
ergosurface and event horizon merge to the Schwarzschild horizon.
To visualize the various surfaces we use Kerr-Schild
quasi-Cartesian coordinates.
The radial coordinate $r$ of the Boyer-Lindquist coordinates is
related to the coordinates $x,y,z$ of the Kerr-Schild coordinates via, see
Table 4,
\begin{equation}\label{implicit_r}\small
x^2+y^2+\frac{r^2+a^2}{r^2}\,z^2 = r^2+a^2\,,\qquad
z = r \,\cos\theta\,.\normalsize
\end{equation}
Substituting $r=0$, $r=r_\pm$, $r=r_{E_\pm}$, and a little bit of algebra
yields:
\begin{itemize}
\item{\bf Singularity $r=0$}

Since $r=0$ leads to $z=0$, we get the equation of a circle of radius $a$ in
the equatorial plane,
\begin{equation}
x^2+y^2=a^2\,.
\end{equation}
For $a=0$, the ring collapses to the Schwarzschild singularity.

A closer inspection shows that the structure of the singularity is more
complex\cite{GarciaCompean:2012gt,Manko:2014qwa}.
\item{\bf Horizons $r=r_\pm$}

In this case we arrive at the equation for an oblate (for $a<m$) ellipsoid,
\small
\begin{equation}
\frac{x^2}{a_1^2}+\frac{y^2}{a_2^2}+\frac{z^2}{a_3^2}=1\,,
\end{equation}
\normalsize
where
%\begin{equation}
$
a^2_1=a^2_2=r_\pm^2+a^2 > a^2_3=\frac{1}{r_\pm^2}\,.
$
%\end{equation}
\end{itemize}
\pagebreak
%\begin{figure}[hp]\label{chap3:horizons}
\begin{center}
\includegraphics[width=12.5cm]{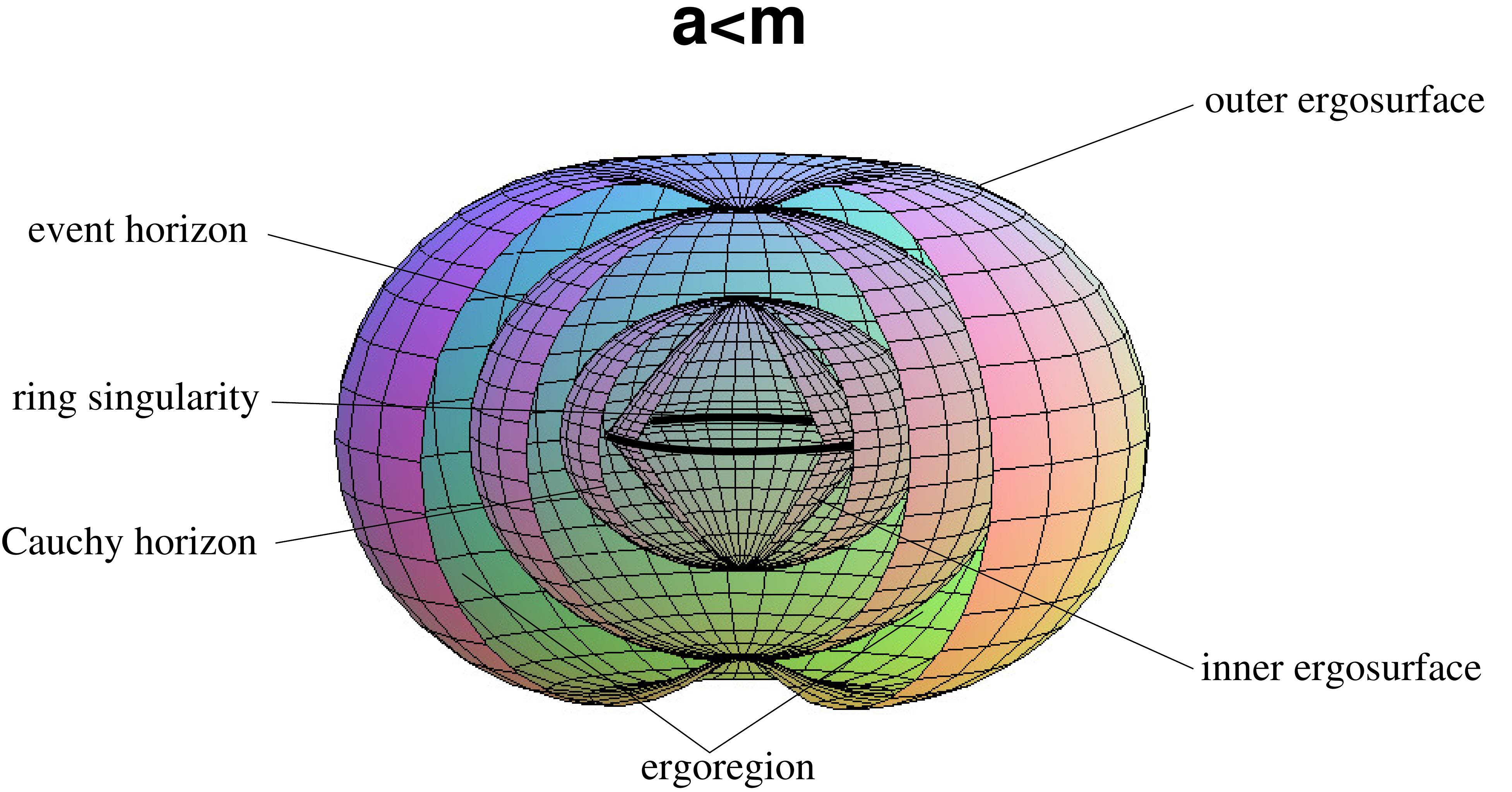}
\bigskip\bigskip
\includegraphics[width=12.5cm]{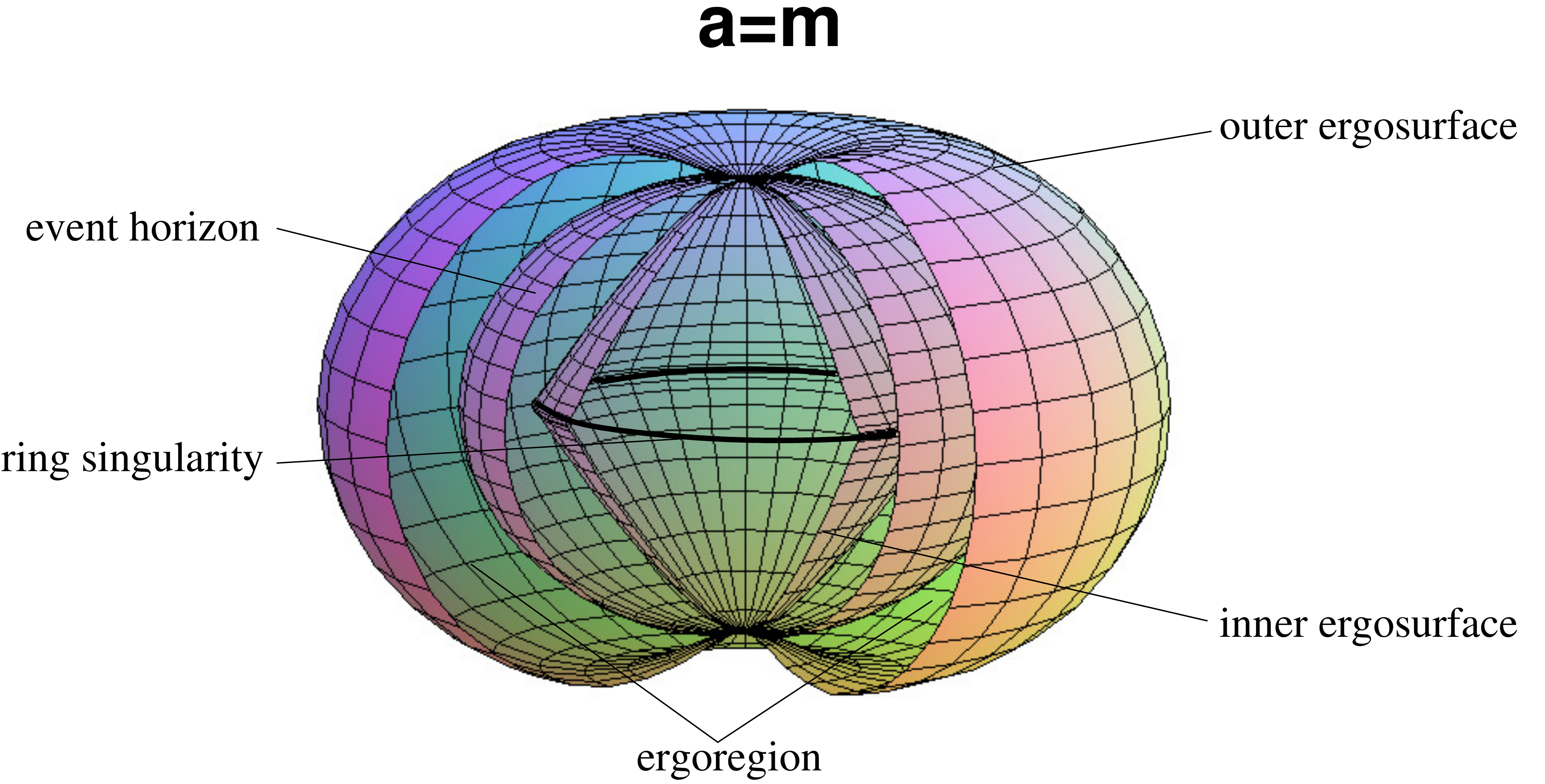}
\bigskip\bigskip
\includegraphics[width=13.5cm]{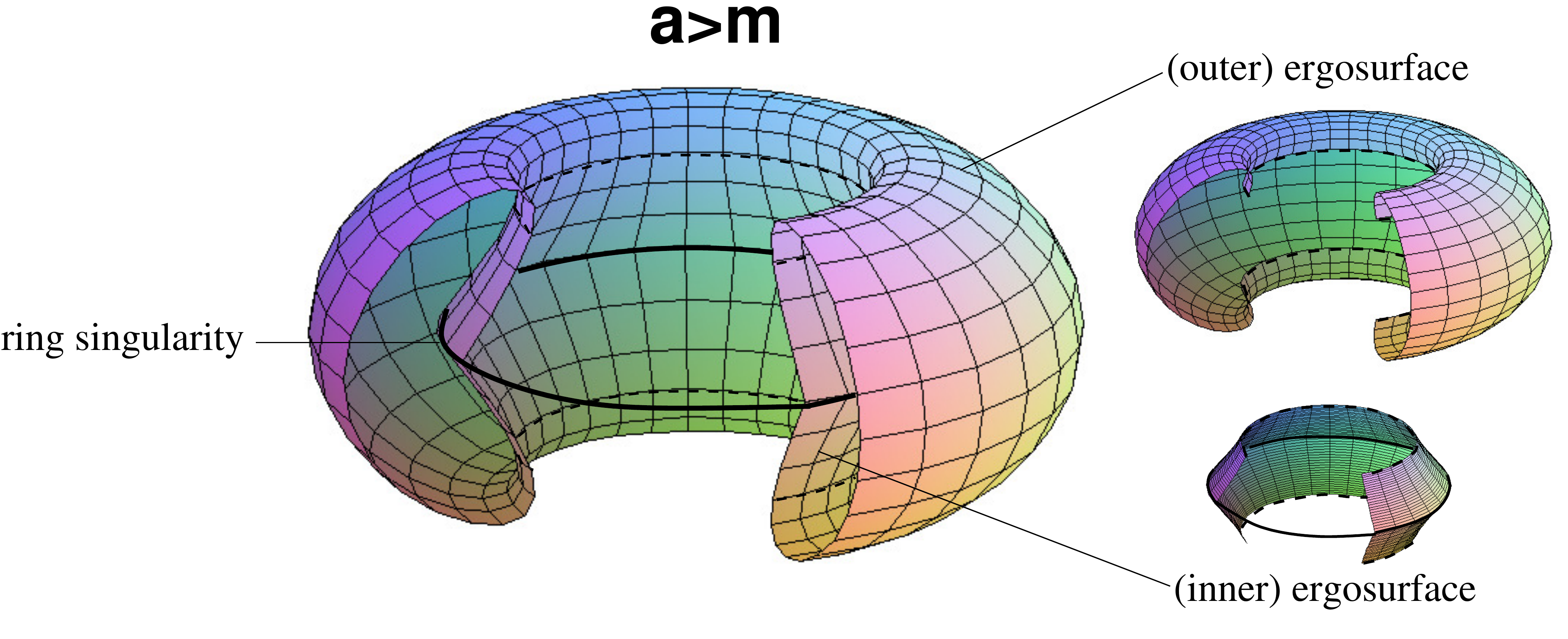}
\end{center} {Fig.~12: Ergosurfaces, horizons, and singularity for
  slow, extremal (`critical'), and fast Kerr black
  holes.\label{chap3:horizons}}
%\end{figure}
\addtocounter{figure}{1}
\newpage

\begin{itemize}
\item{\bf Ergosurfaces $r=r_{E_\pm}(\theta)$}

Things are a little bit more involved in this case because $r$ is not
constant. We can also derive a ``ellipsoid-like'' equation (for $a\leq m$),
\begin{equation}
\frac{x^2}{a_1^2(\theta)}+\frac{y^2}{a_2^2(\theta)}
+\frac{z^2}{a_3^2(\theta)}=1\,,
\end{equation}
now with 
\begin{equation}
a^2_1(\theta)=a^2_2(\theta)=r^2_{E_\pm}(\theta)+a^2\,,\quad 
a^2_3(\theta)=\frac{1}{r^2_{E_\pm}(\theta)}\,.
\end{equation}
The $\theta$-dependence will deform the ellipsoid. 
On the equatorial plane with $\theta=\pi$ we have $r^2_{E_\pm}=0$. Hence,
$a_1=a_2=a$ and $a_3$ diverges. This results in a non-regular rim on which
the ring singularity is located. 

For $a>m$, the $r_{E_\pm}$ is partly not defined, since the term under the
square-root changes sign, and becomes negative if
\begin{equation}
\cos(\theta) = \frac{m}{a}\,.
\end{equation}
This defines two rings with $\theta_1 = \arccos(m/a)$,
$\theta_2=\pi-\theta_1$ and $r=m$.  As a consequence, the outer
ergosurface only extends to these rings {}from the outside, and the
inner ergosurface up to the rings {}from the inside. The outcome is a
kind of torus. The center-facing side is constituted by a part of the
inner ergosurface (along with the ring singularity), whereas the
outside facing parts are given by a part of the outer ergosurface.

An extensive discussion of the embedding of the ergosurfaces into
Euclidean space, together with corresponding Mathematica-programs, can be
found in  Ref.\cite{Marsh:2007}.

\end{itemize}
\bigskip

The surfaces are visualized in Fig.~12.
We did not use a
faithful embedding but rescaled axes in order to achieve better visibility.

The presence of the term $\sqrt{m^2-a^2}$ requires the distinction of three
different cases dependent on the values of the mass parameter $m$ and the
angular momentum parameter $a$:
\begin{center}
$m>a \rightsquigarrow$ slow rotation
\hfill
$m=a\rightsquigarrow$ critical rotation
\hfill
$m<a\rightsquigarrow$ fast rotation
\end{center}
The slow rotating case shows the richest structure. Both ergosurfaces
and both horizons are present and distinct {}from each other. As $a$
approaches $m$, the event and the Cauchy horizon draw nearer and
nearer. At critical rotation, $a=m$, both horizons merge into one
single event horizon with $r=m$.  Eventually, for fast rotation $a>m$,
the event horizon disappears and reveals the naked ring singularity
which now is located at the inner edge of the now toroidal shaped
(outer) ergosurface.

\subsection{The ergoregion}

{\it We explore the region between the outer ergosurface and the event
  horizon. There it is not possible to stand still, anything has to
  rotate, even the event horizon. The compulsory rotation in the
  ergoregion allows one to extract energy {}from the black hole. This
  so-called Penrose process leads to black hole thermodynamics.}

\subsubsection*{Constrained rotation}

The outer ergosurface, $r=R_{E_+}$, is defined by the equation
$g_{00}(r_{E_+})=0$. Thus, it is a surface of infinite redshift and a
Killing horizon. 
For a third characterization of the ergo surfaces we have to deal
not only with radial but also with rotational motion.
Consider the Kerr metric in Boyer-Lindquist coordinates with $dr=d\theta=0$,
\[
ds^2 = g_{tt} dt^2 + 2\, g_{t\phi} \, dt \, d\phi + g_{\phi\phi} d\phi^2\,,
\]
or, after dividing by $dt^2$,
\[
\left(\frac{ds}{dt}\right)^2 = g_{tt} + 2 g_{t\phi}\,\frac{d\phi}{dt}+
g_{\phi\phi}\, \left(\frac{d\phi}{dt}\right)^2\,.
\]
The explicit form of the metric components is not needed here.
Note that $\Omega =\frac{d\phi}{dt}$ is the angular velocity with respect 
to a distant observer,
\begin{equation}\label{rotation}
\left(\frac{ds}{dt}\right)^2 = g_{tt} + 2 g_{t\phi}\,\Omega+g_{\phi\phi}
\,\Omega^2\,.
\end{equation}
The worldline of a particle has to be timelike, $ds^2<0$. Since the last
equation is quadratic in $\Omega$, this is only possible between the roots
$ds^2=0$,
\[
\Omega_{\text{min}/\text{max}} :=
-\frac{g_{t\phi}}{g_{\phi\phi}} \mp \sqrt{\left(
\frac{g_{t\phi}}{g_{\phi\phi}}\right)^2 - \frac{g_{tt}}{g_{\phi\phi}}}\,.
\]
What does
\[
\Omega_{\text{min}} < \Omega < \Omega_{\text{max}}
\]
mean?! In flat Minkowski spacetime (with Cartesian coordinates), 
$\Omega_{\text{min}/\text{max}}=\pm 1$ implies that a
particle, e.g., may freely circle around a point, restricted only by
the condition $|v|=|r\cdot \Omega| < c$. In the Kerr spacetime,
at $r=r_{E_\pm}$,
the smallest possible value of $\Omega$ becomes $0$. The particle may just
stay at rest, but can rotate only in one direction, namely in direction of
the angular momentum of the black hole.
Beyond $r_{E_+}$, $\Omega$ is forced to be larger than zero: The particle
must co-rotate with the black hole.

The preceding statement is only correct for radially infalling particles. In
general, the influence of the rotating black hole on the motion of particles
is more complex\cite{rindler}.

\subsubsection*{Rotation of the event horizon}

The behavior of $\Omega_\pm$ on the event horizon is quite remarkable. By
using the identities
\begin{equation}\label{somerelations}
2mr_\pm = r_\pm^2 + a^2\,,\qquad \rho^2-r^2 = a^2\,\cos^2\theta
= a^2-a^2\,\sin^2\theta\,,
\end{equation}
one finds for the event horizon ($r=r_+$),
\begin{equation}\label{eventomega}
\Omega_+=\Omega_-=\Omega_{H} := \frac{a}{2mr_+}\,.
\end{equation}
To interpret this result we use (\ref{rotation}) and write
\begin{equation}
g_{\mu\nu}\,l^\mu\,l^\nu |_{r=r_+} = 0 \,,\qquad
\mbox{with }l^\mu=(1,0,0,\Omega_H)\,.
\end{equation}
The integral lines of $l^\mu=\dot x^\mu$,
\begin{equation}
x^\mu = (t,r_+,\theta_0, \Omega_H \, t)\,,
\end{equation}
define a lightlike hypersurface rotating with a uniform angular velocity:
The event horizon of a Kerr black hole rotates ``rigidly'' with $\Omega_H$,
see in this context Frolov \& Frolov\cite{Frolov:2014dta}.
A consequence of this finding is discussed in the next paragraph.

\subsubsection*{Penrose process and black hole thermodynamics}

The (outer) ergosurface is a Killing horizon, not an event horizon. It
is possible for particles to pass {}from the inside to the outside. This
allows for a peculiar scenario: Since inside the Killing horizon the
particle is forced to spin around, it picks up an additional
rotational energy. This energy can be partly extracted by means of the
Penrose process. An infalling particle traverses the Killing horizon,
picks up rotational energy and subsequently decays into two parts. If
one part plunges into the event horizon, the other part, carrying away
some of the rotational energy, can return to the outside of the
Killing horizon. Thus, the region between Killing and event horizon is
justly labeled as ``ergoregion'' ({}from Greek ergon = work).

The observation that energy can also be extracted {}from the black hole
gave rise to black hole thermodynamics. The next question is then how
the parameters change if the black hole is infinitesimally
disturbed. It was Bekenstein\cite{bekenstein} who established a
relation between the variations of the mass, the angular momentum, and
the area of the event horizon.
Using the coframe (\ref{KerrCoframeLambda}), for $\lambda=0$, we find
for the area of the event horizon
\begin{equation}\label{kerrsurface}
  A=\int\limits_{\tiny\begin{array}{c}{r=r_+}\\ t={\rm
        const.}\end{array}} 
  \vartheta^{\hat 2} \wedge \vartheta^{\hat 3}
  = \int\limits_{\tiny\begin{array}{c}{r=r_+}\\ t={\rm const.}\end{array}}
  \sin\theta\,(r^2+a^2)\,d\theta \wedge d\phi
  = 4\pi\,\left(r_+^2+a^2\right)\,.
\end{equation}
We can rewrite (\ref{kerrsurface}), using (\ref{somerelations}) and $J=ma$,
\begin{equation}
A=8\pi m r_+ = 8\pi \,(m^2-\sqrt{m^4-J^2})\,.
\end{equation}
The differential of this equation is
\begin{equation}\label{chap3:da}
dA=\frac{\partial\,A}{\partial m} \,dm +\frac{\partial\,A}{\partial J}\,dJ
= \frac{8\pi}{\kappa}\,dm -\frac{8\pi}{\kappa}\,\Omega_H\,dJ\,,
\end{equation}
with
\begin{equation}
\kappa = \frac{1}{2m}\,\frac{\sqrt{m^4-J^2}}{m^2+\sqrt{m^4-J^2}}\,,
\qquad
\Omega_H = \kappa\,\frac{J}{\sqrt{m^4-J^2}}\,.
\end{equation}
The parameter $\Omega_H$ is
the angular velocity of the horizon (\ref{eventomega}). 
The parameter $\kappa$ is the surface gravity. 
Eq.(\ref{chap3:da}) can be rewritten as
\begin{equation}\label{chap3:firstlaw}
dm = \frac{\kappa}{8\pi}\, dA + {\Omega_H}\,dJ\,.
\end{equation}
The infinitesimal change of the mass, $dm$, is proportional to the the
infinitesimal change of the energy, $dE$. The term ${\Omega_H}\,dJ$
describes the infinitesimal change of the rotational energy. This
suggests the identification of (\ref{chap3:firstlaw}) with the first
law of thermodynamics. The analogy is still more compelling by
observing that, for a given black hole of initial (or irreducible)
mass $m$, the area of the horizon is always increasing. Even by
exercising a Penrose process, which extracts rotational energy {}from
the black hole, a fragment of the incoming particle will fall into the
black hole thereby increasing its mass and, in turn, the area of the
horizon. Accordingly, the area $A$ of the horizon behaves formally as
if it is proportional to an entropy $S$ and the surface gravity
$\kappa$ as if it is proportional to a temperature $T$. In fact, the
{\it Hawking temperature} and the {\it Bekenstein-Hawking entropy}
turn out to be
\begin{equation}\label{interprete}
  T=\frac{\hbar}{2\pi k_{\text{B}}}\,\kappa\,,\qquad S=\frac{1}{4G\hbar}\,A\,,
\end{equation}
with $k_{\text{B}}$ as the Boltzmann constant.
Eq.(\ref{chap3:firstlaw}) together with its thermodynamical
interpretation (\ref{interprete}) can be considerably generalized
thereby establishing the new discipline of ``black hole
thermodynamics", see Heusler\cite{Heusler:1996} and
Carlip.\cite{Carlip:2014}

%The inner horizons are much less easily accessible, the physics beyond them
%much more fragile. We will come back to this below.

\subsection{Beyond the horizons}

{\em In the Schwarzschild spacetime, event horizon and Killing horizon
  coincide.  In the Kerr spacetime, for $m>a$, there is an outer
  Killing horizon, an event horizon, an inner Killing horizon and an
  inner horizon.  So far, all the coordinate systems we used for the
  Kerr metric show singularities at the outer and inner horizons
  $r=r_\pm$. The construction of a regular coordinate system is
  possible along the same lines as for the Schwarzschild metric.  Of
  course, the corresponding calculations are much more involved for
  the Kerr case. Therefore, we will give more a kind of heuristic
  approach to motivate Kruskal-like coordinates for the Kerr metric.}

\subsubsection*{Using light rays as coordinate lines}

Our first task is to construct Eddington-Finkelstein like coordinates
for the Kerr metric by considering radial light rays.  We restrict
ourselves to the case $\theta=0=\phi$. The Kerr metric in
Boyer-Lindquist coordinates reduces to ($\theta=0\quad\rightarrow\quad
\rho^2=r^2+a^2=\Delta + 2mr$):
\[
ds^2=-\frac{\Delta}{\rho^2}\,dt^2+\frac{\rho^2}{\Delta}\,dr^2\,.
\]
Hence, for in-/out-going light rays, $ds^2=0$, we find
\[
dt = \pm\frac{\rho^2}{\Delta} \, dr 
= \pm\frac{r^2+a^2}{(r-r_+)(r-r_-)} \, dr
\]
or, explicitly,\small
\begin{equation}\label{eddikerrtime}
  \pm t =  \int dr\frac{r^2+a^2}{(r-r_-)(r-r_+)} 
  = r + \frac{r_+^2+a^2}{r_+-r_-}\ln \left|r-r_+\right|
  - \frac{r_-^2+a^2}{r_+-r_-}\ln \left|r-r_-\right|+ const. 
\end{equation}\normalsize
Unlike in the Schwarzschild spacetime, there form {\bf two} event
horizons, at $r=r_-$ and $r=r_+$, respectively. However, as 
$a\rightarrow 0$, $r_-$ goes to $0$, whereas $r_+$ approaches $2m$ and
the Schwarzschild situation is reproduced.

We next focus on the  (Boyer-Lindquist) coordinates $(t,r)$ and
how the horizons etc.\ will appear in terms of the new coordinates.
The other coordinates and the regularity of the metric is not
addressed. However, all the details can be found in the literature, see
Refs.\cite{Boyer:1966qh,Carter:1973,hawking}.
Using (\ref{eddikerrtime}) analogously to (\ref{lightinschw}), we introduce
Eddington-Finkelstein like coordinates for Kerr,
\begin{eqnarray}\label{vutildekerr}
v & := & t+r+\sigma_+ \, \ln|r-r_+| -  \sigma_-\, \ln |r-r_-|\,,\\
u & := & t-r-\sigma_+ \, \ln|r-r_+| +  \sigma_-\, \ln |r-r_-|\,,
\end{eqnarray}
where (according to the notation in Ref.\cite{Boyer:1966qh})
\begin{equation}
\sigma_\pm := \frac{r_\pm^2+a^2}{r_+-r_-}=
\frac{mr_\pm}{\sqrt{m^2-a^2}}\,.
\end{equation}
Again, we can get rid of the coordinate singularity by rescaling $u$
and $v$ analogously to (\ref{chap1:uvtilde}). Since we have two
horizons, $r=r_+$ and $r=r_-$, we have to decide with respect to which
singularity we rescale.  We firstly choose $r_+$ and define, see
(\ref{chap1:uvtilde}),
\begin{eqnarray}
\tilde v &:=& \exp\left(\frac{v}{2\sigma_{+}}\right)
=\frac{|r-r_+|^\frac{1}{2}}{|r-r_-|^\frac{\nu}{2}}
\,e^\frac{r+t}{2\sigma_{\!+}}\,,\\
\tilde u &:=& -\exp\left(-\frac{u}{2\sigma_+}\right)
=-\frac{|r-r_+|^\frac{1}{2}}{|r-r_-|^\frac{\nu}{2}}
\,e^\frac{r-t}{2\sigma_{\!+}}\,,
\end{eqnarray}
with
\begin{equation}
\nu := \frac{\sigma_-}{\sigma_+}= \frac{r_-}{r_+}>1\,.
\end{equation}
Again, we go back to time- and space-like coordinates, exactly like in
(\ref{oldcoo}),
\begin{equation}
\tilde t := \frac{1}{2}\left(\tilde v + \tilde u\right)\,,\qquad
\tilde r := \frac{1}{2}\left(\tilde v - \tilde u\right)\,.
\end{equation}
Then we work out the four coordinate patches exactly like
(\ref{Kruskal1}) to (\ref{Kruskal6}). We arrive at a Kruskal like coordinate
system.  However, there arises an important difference: The coordinate
system still is singular for $r=r_-$. 
%\pagebreak
This can be most easily seen
{}from the analog to (\ref{Kruskal5}), the inverse transformation to $r$, which
now reads
\begin{equation}
\tilde r^2 -\tilde t^2 = - \tilde v \tilde u 
= \frac{r-r_+}{(r-r_-)^\nu}\, e^\frac{r}{\sigma_{\!+}}\,.
\end{equation}
The horizon $r=r_+$ is regular in this coordinate system and is described by
$\tilde r=\pm \tilde t$. The transformation(s) are valid in the domain
$r_-<r<+\infty$

\renewcommand{\baselinestretch}{1.2}\small\normalsize
\begin{tabular}{lcll}
$r=r_+$ &:& $\tilde r=\pm \tilde t$ & as for Schwarzschild\\
$r\to+\infty$ &:& $\tilde r^2-\tilde t^2 \to +\infty$ & particularly $\tilde
r \to \pm \infty$ for $\tilde t =0$\\
$r\to r_-$ &:&  $\tilde r^2-\tilde t^2 \to -\infty$ & particularly $\tilde
t \to \pm \infty$ for $\tilde r =0$ \\
$r=r_{E_+}$ &:&  $\tilde r^2-\tilde t^2 = {\rm const.}>0$ & 
hyperbolas in I, II patches
\end{tabular}
\smallskip
\renewcommand{\baselinestretch}{1}\small\normalsize

\noindent
In contrast to the Schwarzschild case, the full upper and lower
halfplanes of the $(\tilde r,\tilde t)$ plane
is covered. It is not limited by the hyperbolas of the
Schwarzschild singularity $r=0$!

We can regularize with respect to $r_-$ by introducing
\begin{equation}
\tilde v := -\exp\left(-\frac{v}{2\sigma_-}\right)\,,\qquad
\tilde u :=  \exp\left(\frac{u}{2\sigma_-}\right)\,.
\end{equation}
Now we find
\begin{equation}
\tilde r^2 -\tilde t^2  
= \frac{r-r_-}{(r-r_+)^\frac{1}{\nu}}\, e^{-\frac{r}{\sigma_{\!-}}}\,.
\end{equation}
This coordinate system covers the domain $-\infty<r<r_+$. Like the first
coordinate system, it contains also the region between 
the horizons, $r_-<r<r_+$. This time, $r\geq r_+$ is excluded.

\renewcommand{\baselinestretch}{1.2}\small\normalsize
\begin{tabular}{lcll}
$r=r_-$ &:& $\tilde r = \pm \tilde t$ &as above\\
$r\to-\infty$ &:& $\tilde r^2-\tilde t^2 \to \infty$ & particularly $\tilde
t \to \pm \infty$ for $\tilde r =0$\\
$r\to r_+$ &:&  $\tilde r^2-\tilde t^2 \to -\infty$ & particularly $\tilde 
r \to \pm \infty$ for $\tilde t =0$\\
$r=r_{E_-}$ &:& $\tilde r^2-\tilde t^2 = {\rm const.}>0$ & hyperbola in
I*, II* patches\\
$r=0$ &:&  $\tilde r^2-\tilde t^2 = -\frac{r_-}{r_+}<0$& hyperbola in   
I*, II* patches                  
\end{tabular}
\smallskip
\renewcommand{\baselinestretch}{1}\small\normalsize

\noindent
Again, the whole $(\tilde r,\tilde t)$ plane is covered. Note that the
spacetime extends beyond the ring(!) singularity.
\subsection{Penrose-Carter diagram and Cauchy horizon}

{\it We compactify the Kruskal-like coordinate system for
Kerr, yielding conformal Penrose-Carter diagrams. We discuss the analytical
extension and the role of the inner horizon as Cauchy horizon.}

\medskip

\begin{figure}
\begin{center}
\includegraphics[width=9.0cm]{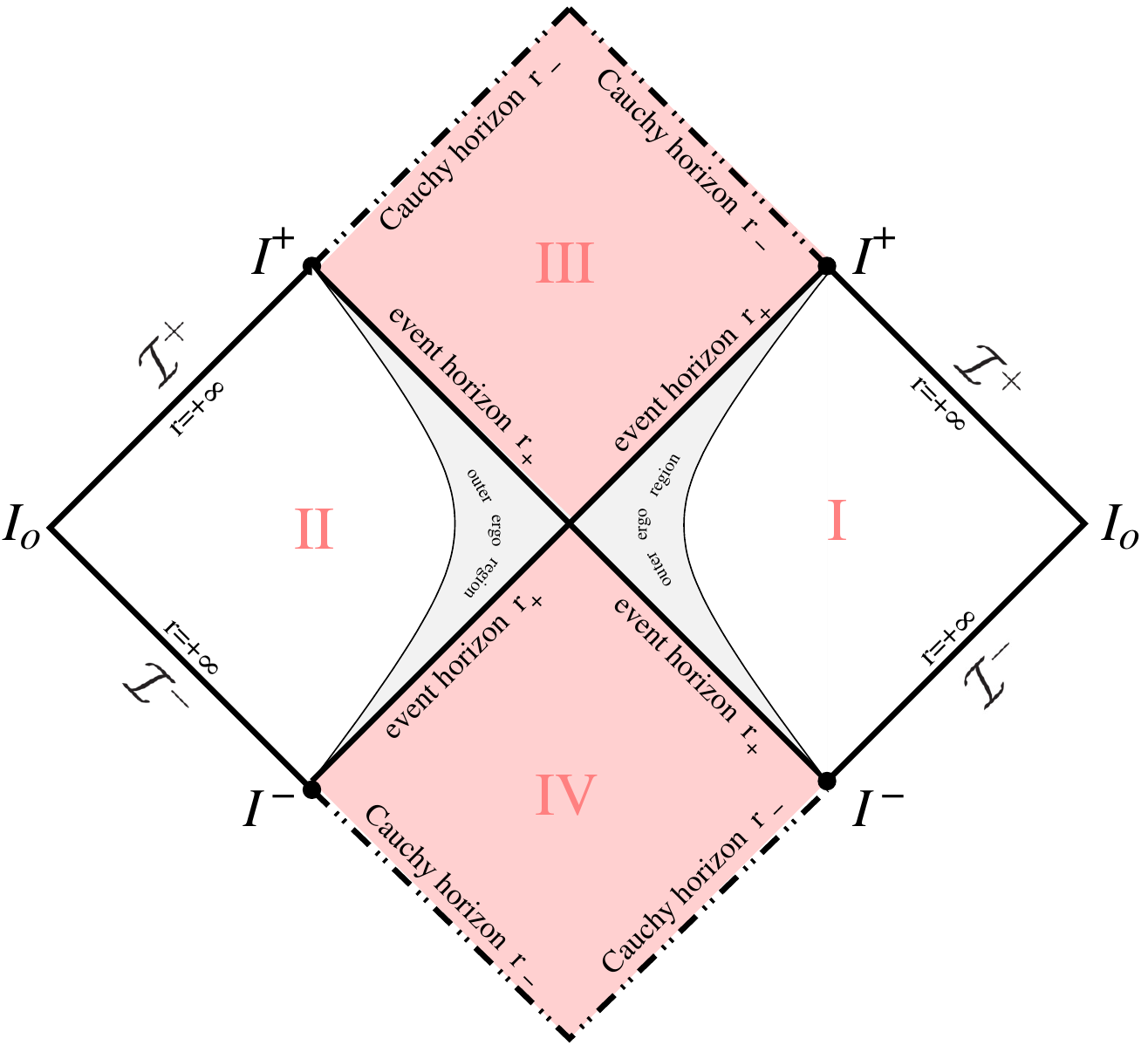}\qquad
\end{center}
\caption{The Penrose diagram for the Kerr spacetime for $r>r_-$. 
\label{pencart1}}
\end{figure}

\begin{figure}
\begin{center}
\includegraphics[width=9.0cm]{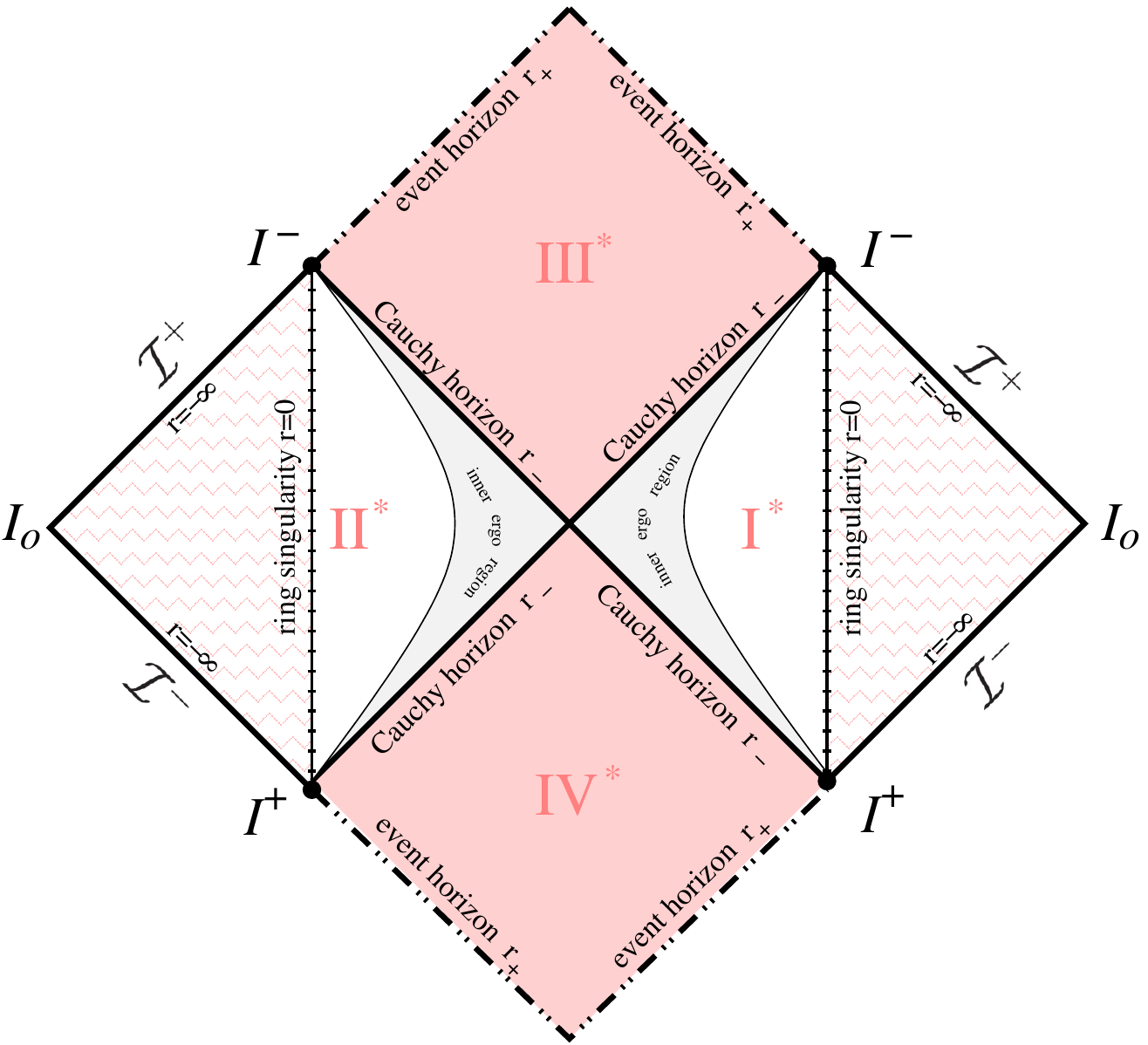}
\end{center}
\caption{The Penrose diagram for the Kerr spacetime for $r<r_-$. 
\label{pencart2}}
\end{figure}

In order to draw a {\it Penrose-Carter diagram} for the Kerr spacetime,
we compactify the coordinates via the tangent function like in
Sec.2.6. The result looks at first quite similar to Schwarzschild
in \fref{chap1:ssskru}. However, the cutoff at $r=0$ vanishes. The 
diagrams \fref{pencart1} and \fref{pencart2} both 
show the entire compactified $(\tilde r,\tilde t)$-space.
%\pagebreak
%\pagebreak

The two coordinate sets overlap in the region between the horizons. Thus,
the corresponding coordinate patches have to be identified. 
And we can even draw beyond that \dots
\begin {figure}
\begin{center}
\includegraphics[width=8.5cm]{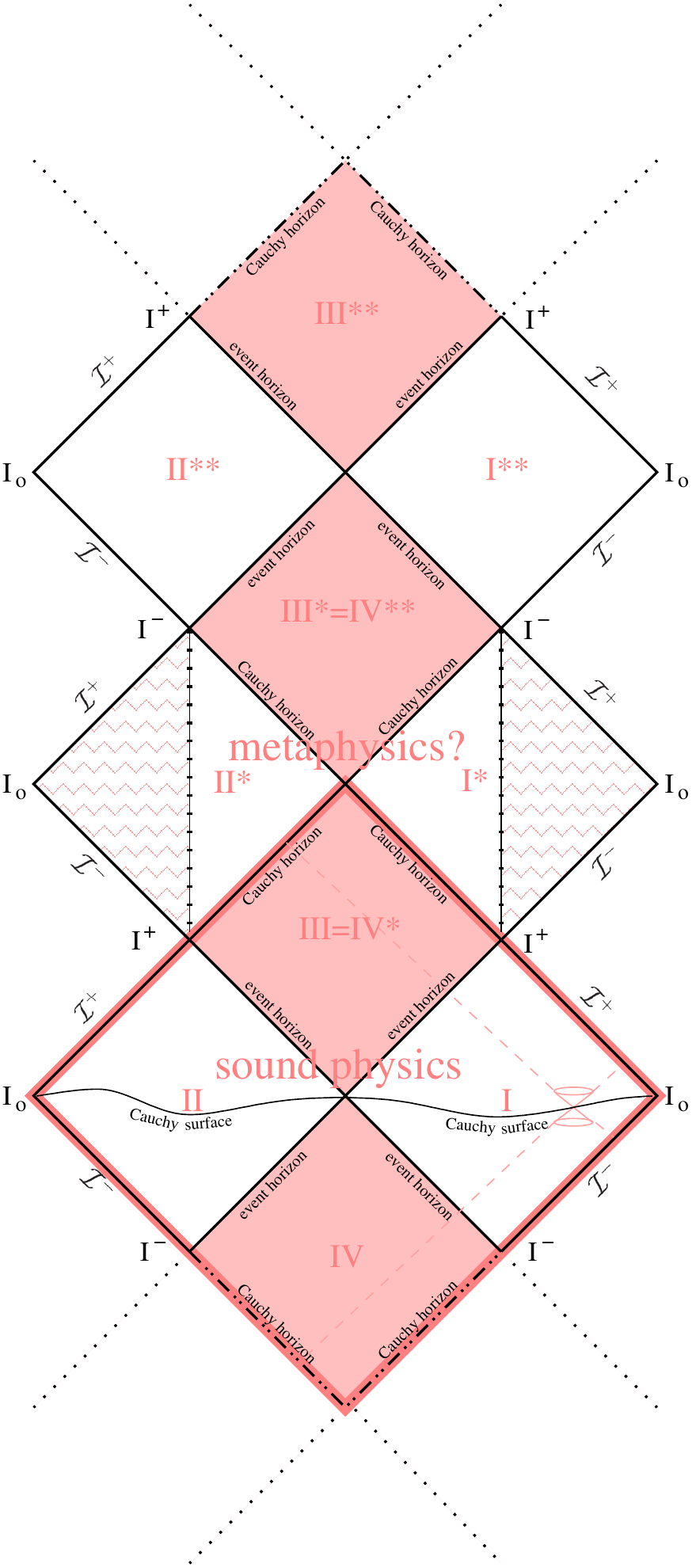}
\end{center}
\caption{Maximal analytic extension of the Kerr spacetime\label{teppich}}%\caption{Maximally extended Kerr spacetime. The douptfull regions are shaded
%\dots @@@IMPROVE}
\end{figure}
%\newpage
Patch II is identified with patch IV*, II* with another patch
IV**. And so on: We find an infinite sequence of coordinate systems.
Formally, this constitutes a {\em maximal analytic extension} of the
Kerr spacetime.  Alas, there are good reasons for not believing in
such vast an extension.

\begin{figure}[h]
\begin{center}
\includegraphics[width=8cm]{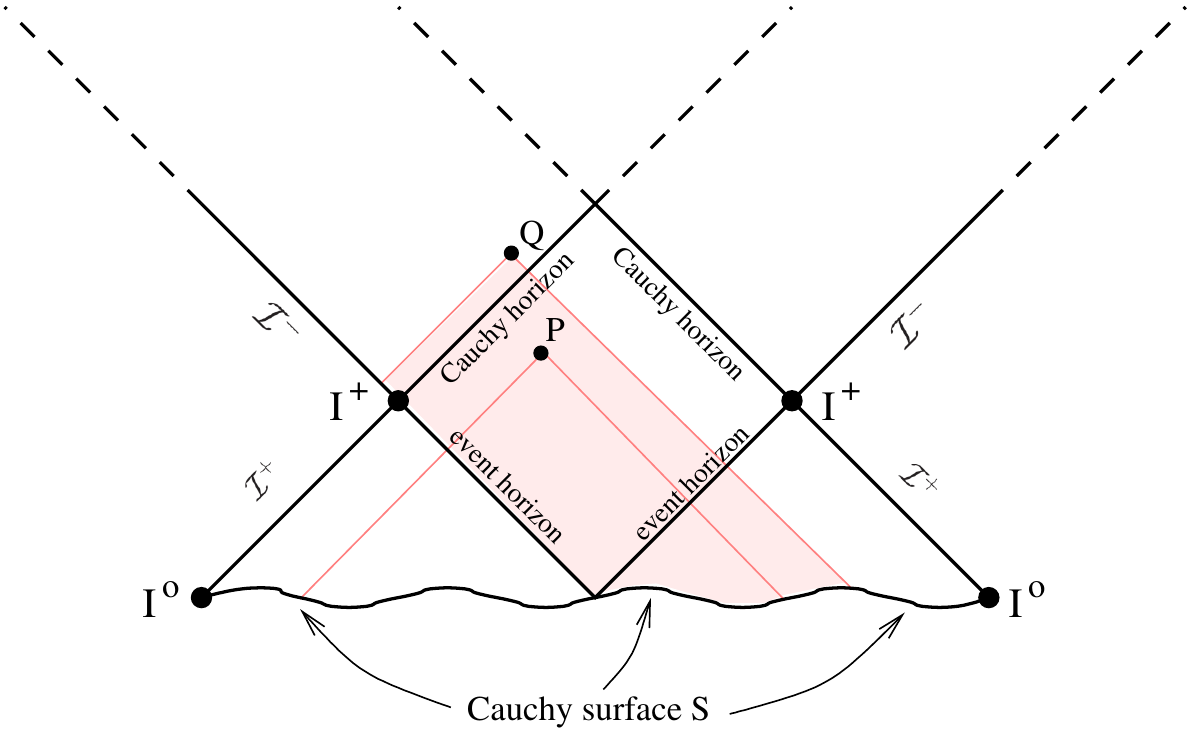}
\end{center}
\caption{Cauchy horizon: The causal past of a point P {\it outside} 
the {\it Cauchy horizon} of S is entirely determined by the information 
given on the
Cauchy surface S. A point Q {\it inside} the Cauchy horizon receives
also information {}from ${\cal I^-}$.  Evidently, initial data on S are
not sufficient to uniquely determine events at point Q. The surface
that separates the two regions ``causally determined by S'' and ``not
causally determined by S" is called Cauchy horizon.
\label{chap3:cauchyhorizon}}
\end{figure}

The Kerr metric is a vacuum solution of Einstein's field equation---it
describes a {\em totally empty} spacetime. To render it physically
meaningful, we should regard it as the spacetime structure generated
by a sensible physical source. One may ask then, why a single source
should produce an infinite number of spacetimes. And it is even
worse. The regions beyond the Cauchy horizon are exceptionally badly
behaved. Consider the Cauchy surface in regions I+II of
\fref{teppich}. All light rays and particle trajectories {}from the past
intersect this surface only once. Then the field equations will tell
us their future development, see Franzen\cite{Franzen}, for
example. In \fref{teppich} this is roughly indicated by the little
light cone. However, even {\em total} knowledge of the world in I+II
does not determine what might be going on in regions I*+II*. That is
why $r=r_-$ is called a Cauchy horizon, see
\fref{chap3:cauchyhorizon}. Thus, I* and II* are not only
beyond the Cauchy horizon but also beyond predictable, sound physics.
Moreover, the zigzagged region beyond the singularity is physically
doubtful.  In this region, the asymptotics is reversed, see the
permutation of $I^+$ and $I^-$. As a consequence, the asymptotic mass
in I* picks up a minus sign as compared to I. So the {\em same
  source} possesses a positive mass $+m$ in I and a negative mass
$-m$ in I*, which seems strange. Moreover, it turns out that these regions
are crowded with closed timelike curves.  The whole extension is not
globally hyperbolic. Thus one should restrict to the ``diamond of
sound physics'', I+II+III+IV. To do this consistently, one has to
devise a physical mechanism preventing traveling beyond the Cauchy
horizon, that is, the Cauchy horizon should become singular in some
sense (cosmic censorship, see Penrose\cite{Penrose:2005}).

\pagebreak
\subsection{Gravitoelectromagnetism, multipole moments}

{\em The curvature tensor of the Kerr metric is calculated. By
  squaring it suitably, we find the two quadratic curvature invariants.
  Subsequently, we determine the gravitoelectric
  and the gravitomagnetic multipole moments of the Kerr metric, and we
  mention the Simon-Mars tensor the vanishing of which leads to the
  Kerr metric.}\bigskip

The analogy between gravity and electrostatics became apparent when
the Coulomb law was discovered in 1785. The gravitational and the
electrostatic forces both obeyed an inverse-square law, with the
difference that the mass can only be positive whereas the electric
charge exists with both signs. Equal electric charges repel, opposite
ones attract; in contrast, gravity is always attractive.

In 1820 electromagnetism was discovered by Oersted, and the emerging
unified theory, called ``electrodynamics'' by Amp\`ere, eventually
found its expression in the Maxwell equations of 1864. Besides the
electric field $\bm{E}$ related to charge, we have the magnetic field
$\bm{B}$ related to {\it moving} charge. These fields, together with
the electric and magnetic excitations $\bm{D}$ and $\bm{H}$,
respectively, obey the Maxwell equations.

Newton's gravitational theory was only superseded in 1915/16 by
Einstein's gravitational theory, general relativity. However, already
in the 1870s physicists began to speculate whether, besides Newton's
``gravitoelectric'' field, related to mass at rest, there may also
exist a new ``gravitomagnetic'' field, accompanying moving mass; for
more details and references see Mashhoon\cite{Mashhoon:2003ax}. As we
saw above, these speculations became a solid basis in general
relativity. In (\ref{LenseThirring}), the gravitomagnetic
Lense-Thirring term surfaced, which found solid experimental
verification in the meantime. Thus, we can speak with justification of
gravitoelectromagnetism\cite{Mashhoon:2003ax} (GEM), a notion which
can guide our intuition, see in this context also Ni and
Zimmermann\cite{Ni:1978zz}.

\subsubsection*{Gravitoelectromagnetic field strength}

Electrodynamics is a linear theory, GR a nonlinear one. Still, if we
take a linearized version of GR, there are those strong analogies
between electrodynamics and gravitodynamics, as worked out, for
instance, nicely in Rindler's\cite{rindler} book.  However, the
analogies go even further, as pointed out particularly by
Mashhoon\cite{Mashhoon:2003ax}. Even in an arbitrary gravitational
field, if referred to a Fermi propagated reference frame with
coordinates $(T,\mathbf{X})$, GEM is a useful concept. If we apply the
geodesic deviation equation (\ref{chap1:geodev}) to such a frame, the
gravitoelectromagnetic field strength, representing the tidal forces,
turns out to be\cite{Mashhoon:2003ax}\footnote{Alternatively, we could
  generalize the Newtonian tidal force matrix of (\ref{TidalForce}) to
  the gravitoelectric and gravitomagnetic tidal force matrices, ${\cal
    E}_{ij}=R_{i0j0}$ and ${\cal B}_{ij}= \epsilon_{ikl}R_{klj0}$,
  respectively, see Scheel \& Thorne\cite{Scheel:2014hfa}. Both
  matrices are symmetric and trace-free.  Note that
  $^{\text{GEM}}F_{\alpha\beta}$ is an antisymmetric $4\times 4$
  matrix and ${\cal E}$ and ${\cal B}$ are both symmetric trace-free
  $3\times 3$ matrices.}
\begin{equation}\label{GEMF}
^{\text{GEM}}\!F_{\a\b}=-R_{\a\b0i}(T)\,X^i\,.
\end{equation}
If we develop  (\ref{chap1:geodev}) up to the order linear in the
velocity $\mathbf{V}:=d\mathbf{X}/dT$, we find
\begin{equation}\label{GEMF*}
\frac{d^2 X^i}{dT^2}=-R_{0i0j}X^j+2R_{ik0j}X^j V^k
=-\,^{\text{GEM}}\!F_{i0}  -2 \, ^{\text{GEM}}\!F_{ki}\,V^k   \,.   
\end{equation}
Now we recall that in electrodynamics the electric and the
magnetic fields $\mathbf{E}$ and $\mathbf{B}$, respectively, are
accommodated in the 4d electromagnetic field strength tensor according
to
\begin{align}\label{Faraday}
\left( {F}_{\a\b} \right) =  \left(
	\begin{array}{cccc}
		0 &-E_1  & -E_2 & -E_3 \\
		\diamond & 0 &B_3 & -B_2  \\
		\diamond & \diamond & 0 & B_1 \\
		\diamond & \diamond &\diamond  & 0\\
	\end{array} \right)= -\left( {F}_{\b\a} \right)\,. 
\end{align}
The diamond symbol $\diamond$ denotes matrix elements already known
because of the antisymmetry of the matrix involved. The corresponding
2-form reads $F=\frac{1}{2} F_{\a\b}\,dx^\a\wedge dx^\b$.  Keeping
(\ref{Faraday}) in mind, Eq.(\ref{GEMF*}) can be rewritten as a vector
equation
\begin{equation}\label{GEMF**}
\frac{d^2\mathbf{X}}{dT^2}=-\mathbf{^{\text{gr}}E}-2\mathbf{V}{\times}
\mathbf{^{\text{gr}}B}\,.
\end{equation}
In accordance with the equivalence principle, this equation of motion
is {\it independent} of the mass. The analogy with electromagnetism
requires that the gravito{\it electric} charge, in terms of the mass
$m$, is $-1$ and the gravito{\it magnetic} charge $-2$. In
electrodynamics, both quantities are $+1$. The difference comes {}from
the vector nature of the electromagnetic potential $A_\a$ as compared
to the tensor nature of the gravitational potential $g_{\a\b}$, that
is, helicity $1$ as compared to helicity $2$. The relation between the
gravitomagnetic to the gravitoelectic charge, that is, the {\it
  gyrogravitomagnetic ratio}, is two: $^{\text{gr}}\gamma=2$. 
Note that in Gravity Probe-B the authors specify the gyrogravitomagnetic
ratio as $1$. However, their gyros carried only orbital angular momentum
rather than spin angular momentum. Hence this is to be expected; for
more detailed discussions on this difference, see
Ref.\cite{Schucking,Ni:2009fg}.

It has been pointed out by Ni\cite{Ni:2011cp} that the ``measurement
of the gyrogravitational ratio of [a] particle would be a further step
\cite{Ni:2009fg} towards probing the microscopic origin of
gravity. GP-B serves as a starting point for the measurement of the
gyrogravitational factor of particles.''

\subsubsection*{Quadratic invariants}

In electrodynamics, we have two quadratic invariants \cite{Birkbook}:
\begin{equation}\label{ElmInvariants}
  \frac 12 F_{\a\b}F^{\a\b}=  \star(\star F\wedge F)=B^2-E^2\,,\qquad 
  \frac 12
  F_{\a\b}{F}^{*\a\b}= \star(F\wedge F)=2\mathbf{E}\mathbb{\cdot}\mathbf{B}\,,
\end{equation}
where we used for the tensor dual the notation ${F}^{*\a\b}:=\frac
12\varepsilon^{\a\b\gamma\d} F_{\gamma\d}$. We also employed the very
concise notation of exterior calculus with the Hodge star
operator.\footnote{The Hodge star $\star\omega$ of a p-form
  $\omega=(1/p!)\omega_{\mu_{1}\cdots\mu_{p}} dx^{\mu_1}\wedge
  \cdots\wedge dx^{\mu_p}$ is an $(n-p)$-form $\star\omega$, with the
  components
  $(\star\omega)_{\mu_1\cdots\mu_{n-p}}=(1/p!)\varepsilon^{\nu_1\cdots\nu_p}
  {}_{\mu_1\cdots\mu_{n-p}}\omega_{\nu_1\cdots\nu_p}$, where
  $\varepsilon$ is the totally antisymmetric unit tensor and $n$ the
  dimension of the space, see Eq.(C.2.90) in Ref.\cite{Birkbook}}

The first invariant is proportional to the Maxwell vacuum Lagrangian
and is an ordinary scalar, whereas the second one corresponds to a
surface term and is a pseudoscalar (negative parity).

Turn now directly to the Kerr metric and list for this example the
tidal gravitational forces, which are represented by the curvature
tensor. With its 20 independent components, it can be represented by a
trace-free symmetric $6\times 6$ matrix, see (\ref{symmetries*}). The
collective indices $A,B,..=1,...,6$ are defined as follows:
$\{\hat{t}\hat{r}, \hat{t}\hat{\theta}, \hat{t}\hat{\phi};
\hat{\theta}\hat{\phi}, \hat{\phi}\hat{r}, \hat{r}\hat{\theta}
\}\longrightarrow$ $\{1,2,3;4,5,6 \}$.
We throw the orthonormal Kerr coframe (\ref{KerrCoframe}) to
(\ref{anhmetric}) into our computer and out pops the $6\times 6$
curvature matrix,
\begin{align}
\left( {R}_{AB} \right)  =  \left(
	\begin{array}{cccccc}
		-2\mathbb{E} & 0 & 0 & 2\mathbb{B} & 0 & 0 \\
		\circ & \mathbb{E} & 0 & 0 & -\mathbb{B} & 0 \\
		\circ & \circ & \mathbb{E} & 0 & 0 & -\mathbb{B} \\
		\circ & \circ &\circ  & 2\mathbb{E} & 0 & 0 \\
	\circ	& \circ &\circ  &\circ  & -\mathbb{E} & 0 \\
		\circ	& \circ &\circ  &\circ  &\circ &  -\mathbb{E} \\
	\end{array} \right)=\left( {R}_{BA} \right)\,, 
\end{align}%\newpage
with
\begin{equation}\label{EBdef}
  \mathbb{E} := mr\,\frac{r^2 - 3a^2\cos^2\theta}
  {( r ^2 + a^2 \cos^2\theta)^3}\,,\qquad 
  \mathbb{B} :=ma\cos\theta\, \frac{ 3r^2 
- a^2\cos^2\theta}{( r ^2 + a^2 \cos^2\theta)^3}\,.
\end{equation}
It is straightforward to identify $\mathbb{E}$ as the gravitoelectric
and $\mathbb{B}$ as the gravitomagnetic component of the
curvature. This is in accordance with (\ref{GEMF}).

It is obvious how we should continue. Our gravitoelectromagnetic
invariants will be\footnote{In exterior calculus, we have the Euler
  4-form $E:=R_{\a\b}\wedge \star R^{\a\b}$, with $K= \star
  E$. Analogously, we have the Chern-Pontryagin 4-form
  $P:=-R^\a{}_\b\wedge R^\b{}_\a$, which is an exact form, with ${\cal
    P}:=\star P$, cf.\ Obukhov et al.\cite{Obukhov:1995eq}.}
\begin{eqnarray}\label{GEMinvariants}
  K&:=&   \frac{1}{2}R_{\a\b\gamma\d} R^{\a\b\gamma\d}=
  -\star\!\left(\star R_{\a\b}\wedge R^{\a\b}\right)\,,\\
  {\cal P}&:=&
  \frac{1}{4}\varepsilon^{\gamma\d\mu\nu}R_{\a\b\gamma\d} 
  R^{\a\b}{}_{\mu\nu}=\star\!\left( R_{\a\b} \wedge R^{\a\b}\right)\,.
\end{eqnarray}
Again, our program determines the {\it Kretschmann}\footnote{Usually in
  the literature\cite{ciufolini2,deFelice:1990}, the Kretschmann
  scalar is defined as $R_{\a\b\gamma\d}R^{\a\b\gamma\d}$, even though
  the electrodynamics analogy would suggest to include the factor
  $1/2$.}  scalar $K$ and the {\it Chern-Pontryagin} pseudoscalar
${\cal P}$ to be\cite{Boos}
\begin{align}\label{Jens}
  K = -24(\mathbb{B}^2-\mathbb{E}^2)\,, \qquad  \mathcal {P}
  =-48\,\mathbb{E}\,\mathbb{B}\,.
\end{align}
The similarity to (\ref{ElmInvariants}) is impressive. The GEM analogy
quite apparently applies to the full nonlinear theory. The results in
(\ref{Jens}), partly in more involved representations, can be found
in the literature, see, for instance, the books of de Felice \&
Clarke\cite{deFelice:1990} and of Ciufolini \&
Wheeler\cite{ciufolini2}, but compare also de Felice \&
Bradley\cite{deFelice:1988}, Henry\cite{Henry}, and Cherubini et
al.\cite{Cherubini:2002}.

Thus, the quadratic invariants $K$ and $\cal P$ confirm that the Kerr
metric is the exterior field of a rotating mass distribution. In order
to get more information about this distribution, we proceed, like in
electrodynamics, and look into the gravitoelectromagnetic multipole
moments of this rotating mass.

\subsubsection*{Gravitomagnetic clock effect of Mashhoon, Cohen, et
  al.}

According to the results of Lense-Thirring, the rotation of the Sun
changes the spacetime around it by inducing gravitomagnetic
effects. As we saw above, in a similar way the temporal structure
around a Kerr metric is affected by the angular momentum of the Kerr
source. Thus, a gravitomagnetic clock effect should
emerge,\footnote{This was first predicted by Cohen and
  Mashhoon\cite{Cohen:1993} and worked out in greater detail by
  Mashhoon et al.\cite{Mashhoon:1999,Mashhoon:1999nr}, see also Bonnor
  \& Steadman\cite{Bonnor} and the review papers in the workshop of
  L\"ammerzahl et al.\cite{Lammer}. In a similar way, there emerges
  also a gravitomagnetic {\it time delay,} see Ciufolini et
  al.\cite{Ciufolini:2002iq}.} the measurability of which requires very
accurate clocks. The effect can be demonstrated by two clocks that move
on equatorial orbits, one in prograde and the other in retrograde
orbit around the Kerr metric. It turns out\cite{Mashhoon:1999nr} that
the {\it prograde} equatorial clock is {\it slower} than the
retrograde one. This is not necessarily what our intuition would tell
us. It is connected with the fact that the dragging of frames in a
Kerr metric can sometimes turn out to be an ``antidragging'', thus
making this notion less intuitive,\cite{rindler} as we already recognized
in Sec.3.5.

Generalizations of this clock effect were studied, for example, by
Hackmann \& L\"ammerzahl\cite{Hackmann:2014aga}. The recent discussion
of the {\it Clocks around Sgr\,A$^{\star}$,} by Ang\'elil \&
Saha\cite{Angelil:2014yea} is, in effect, just one more manifestation
of the gravitomagnetic clock effect.

\subsubsection*{Multipole moments: gravitoelectric and gravitomagnetic
  ones}

In Newton's theory, one gets a good idea about a mass distribution and
its gravitational field by determining the multipole moments of the
mass distribution $M$. In GR, because of the existence
gravitomagnetism, we have to expect a new type of multipole moments,
namely the moments $J$ of the angular momentum distribution.

If a stationary axially symmetric line element of the form (\ref{Bu})
is asymptotically flat, then it is possible\cite{S_K_2004} to define
two sets of multipole moments, the gravitoelectric moments $M_s$
(``mass multipole moments'') and the gravitomagnetic moments $J_s$
(``angular momentum multipole moments''), for $s=0,1,2,...$. These
moments were found by Geroch\cite{Geroch:1970cd} for the static and by
Hansen\cite{Hansen:1974} for the stationary case. They were reviewed
by Quevedo\cite{Quevedo:1990} and used for constructing new exact
solutions by Quevedo \&
Mashhoon\cite{Quevedo:1989,Quevedo:1991zz}. Hansen computed the
multipole moments for the Kerr solution and found
\begin{align}\label{moments1}
s&=0    & M_0&=-m    & J_1&=\hspace{8pt} ma\\
s&=1    & M_2&= \hspace{8pt}ma^2  & J_3&=-ma^3\\
s&=2    & M_4&=-ma^4 & J_5&= \hspace{8pt}ma^5\\
s&=3 \dots   & M_6&= \hspace{8pt} ma^6 \dots & J_7&=-ma^7\dots
\end{align}
More compactly, we have
\begin{align}\label{moments2}
  M_{2s}&=(-1)^{s+1} m\, a^{2s}\,,&\qquad M_{2s+1}&=0\,;\\
  J_{2s}&=0\,,&\qquad J_{2s+1}&=(-1)^{s} m\, a^{2s+1}\,.  \label{moments3}
\end{align}

It is possible to introduce normalized multipole moments, see Meinel
et al.\cite{MeinelEtAl2008}, such that for Kerr we have
$\widetilde{M}_s + i \widetilde{J}_s = m(ia)^s$. Then the mass
monopole $\widetilde{M}_0=m$ is positive. Apparently, the Kerr metric
has a simple multipolar structure or, formulated differently, only
very specific matter distributions can represent the interior of the
Kerr metric.

Quevedo\cite{Quevedo:1990} compiled a number of theorems which
illustrate the use of the multipole moments:
\begin{enumerate}
\item[1)] A stationary spacetime is {\it static} if and only if all its
  gravito{\it magnetic} multipole moments vanish (Xanthopoulos 1979).

\item[2)] A static metric is flat if and only if all its gravito{\it
    electric} multipole moments vanish (Xanthopoulos 1979).

\item[3)] A stationary metric is axisymmetric if and only if all its
  multipole moments are axisymmetric (G\"ursel 1983).

\item[4)] Two metrics with the same multipole moments have the same
  geometry at large distances {}from the source (Beig \& Simon 1981;
  Kundu 1981; Van den Bergh \& Wils 1985).

\item[5)] Any stationary, axisymmetric, asymptotically flat solution of
  Einstein's vacuum equation approaches the Kerr solution
  asymptotically (Beig \& Simon 1980).

\item[6)] Any static, axisymmetric, asymptotically flat vacuum solution
  approaches the Schwarzschild solution asymptotically (Beig 1980).
\end{enumerate} \vspace{-5pt}
\noindent In the formulation of Stephani, Kramer, et al.\cite{S_K_2004}:
\begin{itemize} \vspace{-5pt}
\item[7)] A given asymptotically flat stationary vacuum spacetime is
  uniquely characterized by its multiple moments.
\end{itemize}  
We recognize that the knowledge of the multipole moments provides a
lot of insight into the physical properties of an exact solution. 

{}From the point of view of the Kerr solution, theorem 5), see Beig \&
Simon\cite{Beig:1980be}, is perhaps the most interesting one. It
underlines the central importance of the Kerr solution. The
considerations in the context of theorem 5) were further developed by
Simon\cite{Simon:1984a,Simon:1984b}. On the 3-dimensional spatial
slices of a stationary axially symmetric metric, he defined the 3d
{\it ``Simon tensor'',}\cite{Bini:2004} a kind of complexified
generalized Cotton-Bach tensor\cite{cotton}. The vanishing of the
Simon tensor then leads to the multipole moments of the Kerr
solution. Later, Mars\cite{Mars:2000gb}, see also
Mars\cite{Mars:1999yn} and Mars \& Senovilla\cite{Mars:2013qja},
generalized this approach and was led to the 4d {\it ``Simon-Mars
  tensor''}. In Ionescu \& Klainerman\cite{Ionescu}, one can find a
more extended discussion of the Simon-Mars tensor, see also
Wong\cite{Wong}. More recently, B\"ackdahl \& Valiente
Kroon\cite{Backdahl:2011np} have proposed replacing the Simon-Mars
tensor by another measure of ``non-Kerrness'', namely a scalar
parameter.

\subsection[The Kerr-Newman metric]{Adding electric charge and
the  cosmological constant: Kerr-Newman metric}

Enriching the Kerr metric by an electric charge is straightforwardly
possible. We start {}from the metric (\ref{anhmetric}) with coframe
(\ref{KerrCoframe}) to (\ref{KerrEpsi}). 
This coframe can accommodate the Kerr, the Schwarzschild, and the
Reissner-Nordstr\"om solutions. The different forms of the
function $\Delta$ suggest how a charged Kerr solution
should look like \dots

\medskip
\renewcommand{\baselinestretch}{1.3}\small
\begin{tabular}{llllll}
Schwarzschild &($m$)  & $\rho=r^2 $ & $\Delta =  r^2-2mr$ \\
Reissner-Nord.\ &($m$, $q$) &$\rho=r^2 $ & $\Delta =  r^2-2mr$&$+q^2$\\
Kerr &($m$, $a$) & $\rho=r^2+a^2\cos^2\theta $ 
& $\Delta = r^2-2mr$&&$+a^2$ \\ 
\hline
Kerr-Newman &($m$, $a$, $q$) & $\rho=r^2+a^2\cos^2\theta $ 
& $\Delta = r^2 - 2mr $&$+ q^2$ & $ + a^2$ \\ 
\end{tabular}  
\renewcommand{\baselinestretch}{1}\small\normalsize
\medskip

Charging the Schwarzschild solution 
is achieved by adding $q^2$ to the function $\Delta$.
Since the charged Kerr solution should encompass the
Reissner-Nordstr\"om solution, we tentatively keep the term $q^2$ for
the case $a\neq 0$. Now, we can indeed find a potential, 
\begin{equation}
A = - \frac{qr}{\rho^2}\,(dt- a \sin^2\theta\,d\phi )\,,
\end{equation}
such that the Einstein-Maxwell equations are fulfilled. 
The potential describes a line-like charge distribution at $\rho=0$, that
is, on the ring singularity of the Kerr spacetime, 
which is quite satisfying\cite{NewmanJanis}.
This charged Kerr solution was first worked out by 
Newman, Couch, Chinnapared, Exton,
Prakash, and R.~Torrence\cite{Newman:65} (1965), using ``methods which
transcend logic'', as Ernst\cite{Ernst:1968b} puts it. He, in turn, proceeded
{}from (120). Replacing\footnote{Here, q is {\it not} the charge but a
complex parameter in the solution of the Ernst equation.} 
$\xi$ by $\sqrt{1-qq^*}\,\xi$ generates a solution
of the Einstein-Maxwell equations with potential $A_t+iA_\phi =q/(\xi+1)$.

The Kerr and the Kerr-Newman solution behave quite similarly. We can
adopt most of the discussion of the Kerr metric by substituting
$a^2+q^2$ for $a^2$.

\begin{figure}[h]
\begin{center}
\includegraphics[width=12cm]{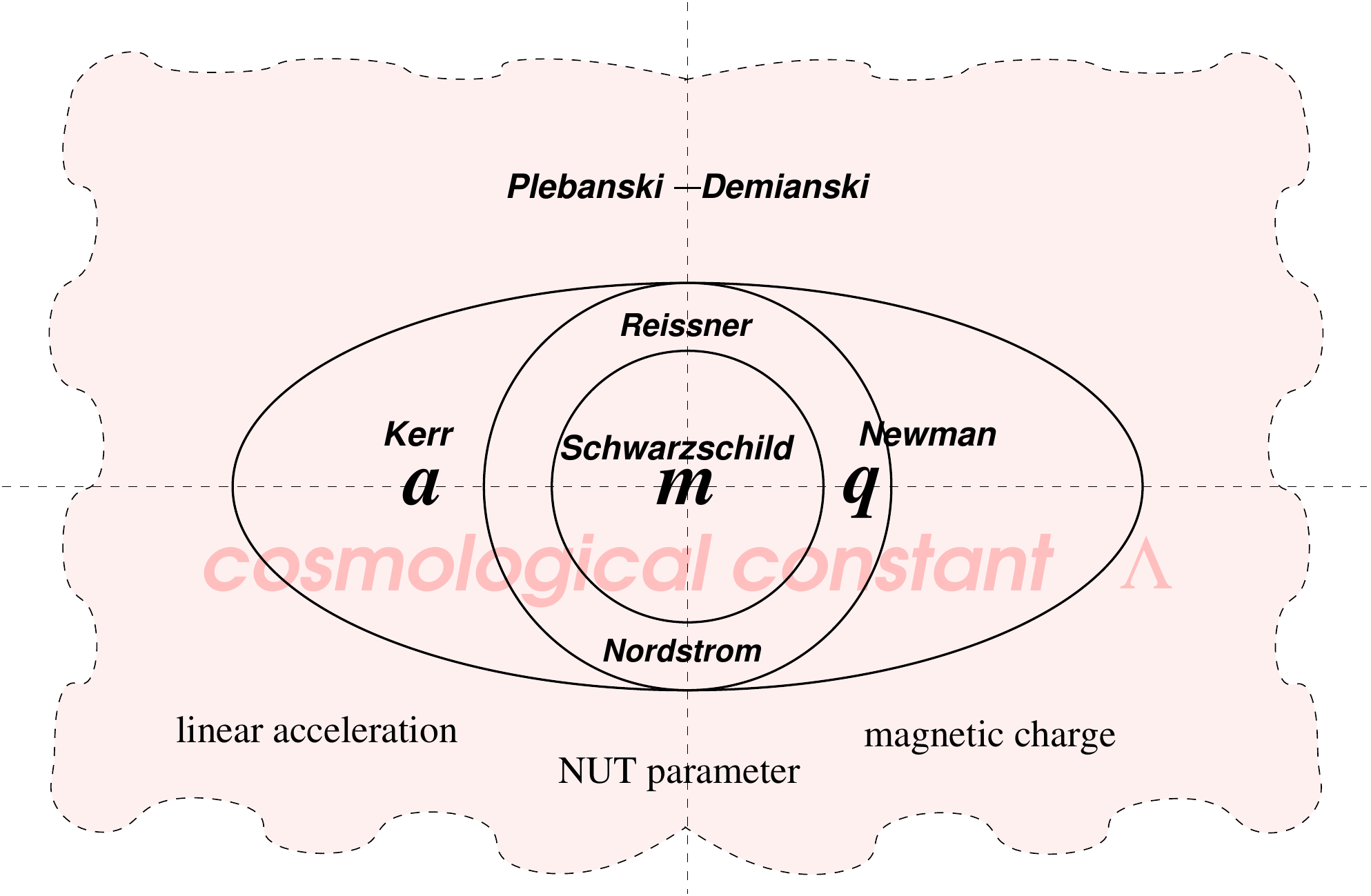}
\caption{Schematics of Petrov D solutions: The spherically symmetric
Schwarzschild solution with mass parameter $m$ is located in the center.
Adding an electric charge $q$ brings us to the Reissner-Nordstr\"om solution.
It is still spherically symmetric but adds a second horizon. The distance
between the horizons increases with the charge $q$. Setting the black hole
into rotation, the angular momentum parameter emerges, $a\neq 0$, and
reduces the spherical symmetry to an axial one. 
An oblate ergosurface (two, actually) forms. Event
horizon and ergosurface meet at the polar axis, the equatorial distance
increases with $a$. All these solutions can be {\em deSitter}\/ed, that
is, a cosmological constant $\Lambda$ is added. 
All presented solutions are subcases of the Plebanski-Demianski
solution, which adds three more parameters\cite{Griffiths:2009}.
\label{plebtoschw}}
\end{center}
\end{figure}

We can further generalize the Kerr-Newman metric to include also a
cosmological constant, see Sec.4.1., and even more parameters, see
\fref{plebtoschw}.

%\vspace{200pt}
\pagebreak
\subsection{On the uniqueness of the Kerr black hole}

{\it The Kerr black hole, up to some technical assumptions, is the
  unique solution for the stationary, axially symmetric case. We point
  to some of the literature where these results can be found.}

Because of the Birkhoff theorem, the Schwarzschild solution (mass
parameter $m$) represents the general spherically symmetric solution
of the Einstein vacuum field equation. The analogous is true in the
Einstein-Maxwell case for the 2 parameter Reissner-Nordstr\"om
solution (mass and charge parameters $m$ and $q$, respectively). Thus,
for spherical symmetry, we have a fairly simple situation.

In contrast, in the axially symmetric case, there does {\it not} exist
a generalized Birkhoff theorem. The 2-parameter Kerr solution (mass
and rotation parameters $m$ and $a$, respectively), is just a
particular solution for the axially symmetric case. As we saw in
Sec.3.8, the Kerr solution has very simple gravitoelectric and
gravitomagnetic multipole moments (\ref{moments2},\ref{moments3}).
Numerous solutions are known that represent the exterior of matter
distributions with different multipole moments. The analogous is valid
for the 3 parameter Kerr-Newman solution (parameters $m,a,q$), see
Stephani et al.\cite{S_K_2004} and Griffiths \&
Podolsk\'y\cite{Griffiths:2009}.

However, one can show under quite general conditions that the
Kerr-Newman metric represents the most general asymptotically flat,
stationary electro-vacuum {\it black hole solution} (``no-hair
theorem''), see Meinel's short review\cite{Meinel:2013cra}.  Important
contributions to the subject of black hole uniqueness were originally
made by Israel\cite{Israel:1967,Israel:1968},
Carter\cite{Carter:1973,CarterKerrFest}, Hawking \&
Ellis\cite{hawking}, Robinson\cite{Robinson:1975bv,RobinsonKerrFest},
and Mazur\cite{Mazur:1982db} (1967-1982), for details see the recent
review of Chru\'sciel et al.\cite{Chrusciel:2012jk}.

More recently Neugebauer and
Meinel\cite{Neugebauer:2000,Neugebauer:2003qe} found a constructive
method for proving the uniqueness theorem for the Kerr black hole
metric. This was extended to the Kerr-Newman case by
Meinel.\cite{Meinel:2011wu} By inverse scattering techniques, they
showed how one can construct the Ernst potential of the Kerr(-Newman)
solution amongst the asymptotically flat, stationary, and axially
symmetric (electro-)vacuum spacetimes surrounding a connected Killing
horizon.\vspace{4pt}

\noindent Let us then eventually pose the following
questions\cite{Carlip}:\vspace{-8pt}
\begin{itemize}
\item[(i)] Are axially symmetric, stationary vacuum solutions outside some
matter distribution ``Kerr''?  The answer is ``certainly not'', and it
makes sense to figure out ways to characterize the Kerr metric, see
Sec.3.8.
\item[(ii)] Is the Kerr solution the unique axially symmetric, stationary
vacuum black hole? The answer is essentially ``yes'' (modulo some
technical issues)---see, for example Mazur\cite{Mazur:1982db}.
\end{itemize}

The general tendency in the recent development of the subject is to
use additional scalar or other matter fields. They weaken the
uniqueness theorems, which is probably not too surprising.

Let us conclude with a quotation that may make you curious to learn
still more about the beauty of the Kerr metric: We have many different
axially symmetric solutions. The Kerr solution is characterized by
{\it ``stationary, axially symmetric, asymptotically flat, Petrov type
  D vacuum solution of the vanishing of the Simon tensor, admitting a
  rank-2 Killing-St\"ackel (KS) tensor of Segre type [(11)(11)]
  constructed {}from a (non-degenerate) rank-2 Killing-Yano (KY)
  tensor''}, see Hinoui et al.\cite{Hinoue:2014zta}.
\pagebreak

\subsection{On interior solutions with material sources}

{\it To match the Kerr (vacuum) metric to a material source
  consistently is one of the big unsolved problems. Only the rotating
  disc solution of Neugebauer \& Meinel provides some hope.}\bigskip

This section is added in order to draw your attention to an
unsolved problem, to the solution of which you might want to
contribute. Find a realistic material source for the Kerr metric in
the sense of an exact solution. Many unsuccessful attempts have been
made, see the early review of Krasi\'nski\cite{Krasinski:1978} of
1978. More recently, in 2006, Krasi\'nski\cite{PlebanskiKrasinski}
concludes {\it``that a bright new idea is needed, as opposed to
  routine standard tricks tested so far.''} This statement was not
made lightheartedly, Krasi\'nski knows what he is talking
about.

Many axially symmetric vacuum solutions were constructed. Quevedo \&
Mashhoon\cite{Quevedo:1991zz}, for example, deformed the multipole
moments of the Kerr(-Newman) metric and constructed appropriate
solutions of the Einstein(-Maxwell) equation that describe the
exterior gravitational field of a (charged) rotating mass.  It is
always the hope that somebody may find a suitable matter distribution
with the multipole moments of the Kerr solution---but this did not
happen so far; for another approach see Marsh\cite{marsh}.

We are only aware of one exact solution that fits into this general
context: It is the {\it infinitesimally thin and rigidly rotating
  dust} solution of Neugebauer \&
Meinel\cite{NeugebauerMeinel,Neugebauer:1995pm} (1993). It is an exact
analytical solution of the Einstein equation {\it with} matter. It
depends on 2 independent parameters, the radius $\rho_0$ of the disk
and its angular velocity $\Omega$. Petroff \&
Meinel\cite{Petroff:2001ub} developed, by means of an iterative
procedure, a post-Newtonian approximation of the solution that helps to
understand the Newtonian limit.

We recall that in electrostatics in flat space, for example, we
prescribe an electric charge distribution and we are used to solve the
corresponding boundary value problem within Maxwell's
theory. Similarly, Neugebauer \& Meinel specified a very thin rotating
disk of dust and solved the boundary value problem within GR. This is
a well-defined procedure. The problem is, however, that within a
non-linear theory, such as GR, it is extremely hard to
implement. Remarkably, for certain parameter values, the gravitational
field of the disk approach the extremal Kerr case. Accordingly, there
exists a certain relation to the Kerr problem. The desideratum would
be to find a rotating matter distribution the external field of which
coincides with the {\it complete} Kerr field.

Driven by the fact that the electrically charged Kerr solution, the
Kerr-Newman solution, has a g-factor of $2$, exactly like the electron
(see also Pfister \& King\cite{Pfister}),
Burinskii\cite{burinskii,Burinskii:2014xja} speculated that a soliton
like solution of the Dirac equation may be the source of the Kerr
metric, see also Burinskii \& Kerr\cite{Burinskii:1995fb}. Is that the
``bright new idea'' Krasi\'nski was talking about? We do not know but
a hard check of the Burinskii ansatz seems worthwhile.

\bigskip

\section{Kerr beyond Einstein}

{\it In generalizations of Einstein's theory of gravity, the
  Riemannian geometry of spacetime is often extended to a more general
  geometrical framework. We describe two such examples in which the
  Kerr metric still plays a vital role.}%\medskip

\subsection{Kerr metric accompanied by a propagating linear
  connection}

{\it We display the Kerr metric with cosmological constant that,
  together with an explicitly specified torsion, represents an exact
  vacuum solution of the two field equations of the Poincar\'e gauge
  theory of gravity with quadratic Lagrangian. }\medskip

In gauge theories of gravitation, see Blagojevi\'c et
al.\cite{Blagojevic:2013xpa}, the linear connection becomes a field
that is at least partially independent {}from the metric. It can be
either metric-compatible, then it is a connection with values in the
Lie-algebra of the Lorentz group $SO(1,3)$ and the geometry is called
a Riemann-Cartan geometry, or it can be totally independent, then it
resides in a so-called metric-affine space and the connection is
$GL(4,R)$-valued. For simplicity, we concentrate here on the former case, the
Poincar\'e gauge theory of gravity, but the latter
case is also treated in the literature\cite{Vlachynsky:1996zh,Baekler:2006de}.

Let us shortly sketch the theory. Gauging the Poincar\'e group leads
to a spacetime with torsion $T^\a$ and curvature $R^{\a\b}$
(Riemann-Cartan geometry\footnote{Experimental limits of a possible
  torsion of spacetime were recently specified in a remarkable paper
  by Obukhov et al.,\cite{Obukhov:2014fta} see also the literature
  given there.}):
\begin{eqnarray}\label{TR}
  T^\a &:= & D\vt^\a=d\vt^\a+\Gamma_\b{}^\a\wedge\vt^\b
  =\frac 12 T_{ij}{}^\a dx^i\wedge dx^j\,,\\
  R^{\a\b} &:=&d\Gamma^{\a\b}-\Gamma^{\a\gamma}\wedge
  \Gamma_\gamma{}^\b
=-R^{\b\a}=\frac 12 R_{ij}{}^{\a\b} dx^i\wedge dx^j\,.
\end{eqnarray}
Besides the coframe 1-form $\vt^\a$, the Lorentz connection 1-form
$\Gamma^{\a\b}=\Gamma_i{}^{\a\b}dx^i=-\Gamma^{\b\a}$ is a second field
variable of the gauge theory. For a Riemannian space, torsion $T^\a=0$
and $\Gamma^{\a\b}$ becomes the Levi-Civita connection. %see (\ref{Christ}).

We choose a model Lagrangian quadratic in torsion and
curvature, in actual fact (for $\hbar=1,c=1$),
\begin{equation}\label{V}
  V=-\frac{1}{2\kappa}(T^\a\wedge \vt^\b)\wedge \,\star(T_\b\wedge \vt_\a)
  -\frac{1}{2\varrho}R^{\a\b}\wedge \,\star R_{\a\b}\,,
\end{equation}
with Einstein's gravitational constant $\kappa$ (dimension
length-squared) and a dimensionless {\it strong} gravity coupling
constant $\varrho$. One can calculate the two vacuum field equations
by varying with respect to $\vt^\a$ and $\Gamma^{\a\b}$. In 1988, for
these two field equations, a Kerr metric with torsion\cite{mckerr} was
found as an exact solution.

We display here the orthonormal coframe and the torsion: The coframe
${\vartheta}^{\alpha}$, in terms of Boyer-Lindquist coordinates
$(t,r,\theta,\phi)$, reads (in the conventions used in Ref.\cite{mckerr})
\vspace{-8pt}
\begin{eqnarray}\label{KerrCoframeLambda}
\vartheta^{\hat 0} & := & \frac{\sqrt{\Delta}}{\rho
} \left(
  dt + a \, \sin^2\! \theta \, d\phi \right) \,,\\
\vartheta^{\hat 1} & := & \frac{\rho }{\sqrt{\Delta}}\,
d r \,,\\
\vartheta^{\hat 2} & := & \frac{\rho}{\sqrt{F}}  \, d\theta\,, \\
\vartheta^{\hat 3} & := & \frac{\sqrt{F}\sin\theta}{\rho } \left[a  dt+( r^2
                    + a^2 )\, d\phi  \right]\,.
\end{eqnarray}
As before, we have $\rho^2:= r^2 + a^2{\cos^{2} \theta}$. However, the
other structure functions pick up a cosmological constant $\lambda$:
\begin{eqnarray}\label{DSF}
F  := 1 + \frac{1}{3}{\lambda}a^2{\cos^2 \theta}\, ,\qquad
{\Delta}  :=  r^2+a^2-2Mr-\frac{1}{3}{\lambda}r^2(r^2+a^2)\,
.\label{DSF'}
\end{eqnarray}
The corresponding metric is called a Kerr-deSitter metric. The coframe
is orthonormal. Then the metric reads
\begin{equation}%\label{anhmetric}
  g = -{\vartheta}^{\hat 0}\otimes {\vartheta}^{\hat 0} + {\vartheta}^{\hat 1}\otimes
  {\vartheta}^{\hat 1} + {\vartheta}^{\hat 2}\otimes {\vartheta}^{\hat 2} +
  {\vartheta}^{\hat 3}\otimes {\vartheta}^{\hat 3}\,.
\end{equation}
It is a characteristic feature of these exact solutions that even
though the Lagrangian (\ref{V}) does {\it not} carry a cosmological
constant, in the coframe and the metric there emerges such a constant,
namely $\lambda:=-{3\varrho}/{(4\kappa)}$. This could be of potential
importance for cosmology.
\pagebreak

The torsion $T^{\alpha}$ of this stationary axially symmetric solution
of the Poincar\'e gauge theory reads
($\vt^{\a\b}:=\vt^\a\wedge\vt^\b$),
\begin{eqnarray}
  T^{\hat 0} & = & {\frac{\rho}{\sqrt{\Delta}}}\left[
    -v_{ 1}{\vartheta}^{\hat 0\hat 1} + {\frac{\rho}{\sqrt{\Delta}}}
    \left[v_{2}({\vartheta}^{\hat 0\hat 2}
      -{\vartheta}^{\hat 1\hat 2})+ v_{3}({\vartheta}^{\hat 0\hat 3}-{\vartheta}^{\hat 1\hat 3})\right]
      - 2v_{4}{\vartheta}^{\hat 2\hat 3}\right]\,, \cr% & & \cr
    T^{\hat 1} & = & T^{\hat 0}\,,%\cr & &
    \cr T^{\hat 2} & = & {\frac{\rho}{\sqrt{\Delta}}} \left[
      v_{5}({\vartheta}^{\hat 0\hat 2}-{\vartheta}^{\hat 1\hat 2})+
      v_{4}({\vartheta}^{\hat 0\hat 3}-{\vartheta}^{\hat 1\hat 3})\right]\,,\cr %& & \cr
    T^{\hat 3} & = & {\frac{\rho}{\sqrt{\Delta}}}\left[
      -v_{4}({\vartheta}^{\hat 0\hat 2}-{\vartheta}^{\hat 1\hat 2})+
      v_{5}({\vartheta}^{\hat 0\hat 3}-{\vartheta}^{\hat 1\hat 3})\right]\,,
\label{TORSION-U4}
\end{eqnarray}
with the following gravitoelectric and gravitomagnetic functions:\small
\begin{eqnarray}
 v_{1} & = & \frac{M}{{\rho^4}}(r^2-a^2{\cos^{2}\theta})\,,\quad 
v_{5}  =  \frac{Mr^2}{{\rho^4}}\,;\\
 v_{2} & = & -\frac{Ma^{2}r{\sin\theta}{\cos\theta}}{{\rho^5}}
 \sqrt{F}\,,\quad v_{3}  = \frac{Mar^2{\sin\theta}}{{\rho^5}}\sqrt{F}\,,\quad
 v_{4}  =  \frac{Mar\,{\cos\theta}}{{\rho}^4}\,.
 \label{torsionsubs}
\end{eqnarray}\normalsize
Metric and torsion of this exact solution are closely
interwoven. Note, in particular, that the leading gravitoelectric part
in the torsion, for small $a$, is $\sim M/r^2$, a definitive Coulombic
behavior proportional to the mass. For $a=0$, we find a
Schwarzschild-deSitter solution with torsion.

One may legitimately ask, why is it that the Lagrangian (\ref{V})
yields an exact solution with a Kerr-deSitter metric? The answer is
simple: The Lagrangian was devised such that the torsion square-piece,
in lowest order in $\kappa$, encompasses a Newtonian
approximation. This is already sufficient in order to enable the
existence of a Kerr-deSitter metric. One could even add another
torsion-square piece to $V$ for getting an Einsteinian approximation,
but this is not even necessary. Thus, only a Newtonian limit of some
kind seems necessary for the emergence of the Kerr structure.

\subsection{Kerr metric in higher dimensions and in string theory}

{\it There exist also Schwarzschild and Kerr metrics in higher
  dimensional spacetimes. These investigations are mainly motivated by
  supergravity and string theory.}\medskip

Tangherlini\cite{Tangherlini:1963bw} (1963) started to investigate
higher dimensional Schwarzschild solutions, with $n-1$ spatial
dimensions. He studied the (``planetary'') orbits in an
$n$-dimensional Schwarzschild field (``Sun'') and found that only for
$n=4$ we have stable orbits, see also Ortin\cite{Ortin}. According to
Tangherlini, this is then the only case that is interesting for
physics. Nowadays, however, many physicists hypothesize that higher
dimensions do exist because string theory suggests it.

Somewhat later, Myers and Perry\cite{Myers:1986un} (1986) generalized
these considerations to higher-dimensional Kerr metrics. In the
meantime a plethora of such higher-dimensional objects have been
found, see Allahverdizadeh et al.\cite{Allahverdizadeh:2010hi} and
Frolov \& Zelnikov\cite{frolov}. Recently Keeler et
al.\cite{Keeler:2012mq} investigated, in the context of string theory,
the separability of Klein-Gordon or Dirac fields on top of a
higher-dimensional Kerr type solutions. Lately Brihaye et
al.\cite{Brihaye:2014nba}, for example, discussed the exact solution
of a 5d Myers-Perry black holes as coupled to a to a massive scalar
field. The physical interpretations of these results remain to be
seen.

%%%%%%%%%%%%%%%%%%%%%%%%%%%%%%%%%%%%%%%%%%%%%%%%%%%%%%%%%%%%%%%%%%%%%%%%%%
%\newpage

%\section{Discussion and Outlook}

%\newpage
\section*{Acknowledgments}
We are grateful to Wei-Tou Ni (Hsin-chu) for inviting us to contribute
to his Einstein volume. Georg Dautcourt (Berlin), Anne Franzen
(Utrecht), Bahram Mashhoon (Columbia, Missouri), Reinhard Meinel
(Jena), Gernot Neugebauer (Jena), Hernando Quevedo (Mexico City), and
Walter Simon (Vienna) helped us to understand gravitoelectromagnetism,
the Kerr solution, and more. We would like to thank Jens Boos
(Cologne), Alberto Favaro (Oldenburg/Treviso), Eva Hackmann (Bremen),
Gerald Marsh (Chicago), and Herbert Pfister (T\"ubingen) for valuable
remarks. Our three referees, Steve Carlip (Davis), James Nester
(Chung-li), and Dah-Wei Chiou (Taipei) helped us to avoid a number of
slips. Our sincere thanks for their numerous suggestions.  We thank
Mikael R{\aa}gstedt (Djursholm), librarian of the Mittag-Leffler
Institute, for sending us a copy of Gullstand's original
paper\cite{Gullstrand} and for providing information about Gullstrand.

\bigskip

\pagebreak
%\end{document}
\begin{appendix}%[Appendix]
%%%%%%%%%%%%%%%%%%%%%%%%%%%%%%%%%%%%%%%%%%%%%%%%%%%%%%%%%%%%%%%%%%%%%%%%%%%%%
\noindent{\bf\Large Appendix}\vspace{-10pt}

%%%%%%%%%%%%%%%%%%%%%%%%%%%%%%%%%%%%%%%%%%%%%%%%%%%%%%%%%%%%%%
\section{ Exterior calculus and computer algebra}

We want to use as input the Papapetrou metric (\ref{Bu}). We take the
equivalent representation in the form of the orthonormal coframe of
the Eqs.(\ref{76}) to (\ref{80}). How such a Reduce-Excalc program can
be set up, is demonstrated in Stauffer et al.\cite{reduce93} and in
Socorro et al.\cite{Socorro}, for the Einstein 3-form, see
Heinicke\cite{heinicke}:
\begin{verbatim}
%*****************************************************
%  Coframe of Andress-Lewis-Papapetrou-Buchdahl metric
%*****************************************************
% file Buchdahl03.exi, 29 July 2014, fwh & chh
% in "Buchdahl03.exi";

load_package excalc;
off exp$
pform f=0, omega=0, gamma=0 $
fdomain f=f(rho,z), omega=omega(rho,z), gamma=gamma(rho,z) ;

coframe  o(0) = sqrt(f) * (d t - omega * d phi),
         o(1) = sqrt(f)**(-1) * exp(gamma) * d rho,
         o(2) = sqrt(f)**(-1) * exp(gamma) * d z,
         o(3) = sqrt(f)**(-1) * rho * d phi
    with signature (1,-1,-1,-1);

displayframe;
frame e$

%*****************************************************
%     Connection, curvature, and Einstein forms
%*****************************************************
pform conn1(a,b)=1, curv2(a,b)=2$
antisymmetric conn1, curv2$
factor o(0), o(1), o(2), o(3)$

conn1(-a,-b) := (1/2)*( e(-a)_|d o(-b) - e(-b)_|d o(-a) 
                     - (e(-a)_|(e(-b)_|d o(-c))) * o(c))$
curv2(-a,b) := d conn1(-a,b) - conn1(-a,c) ^ conn1(-c,b)$

% Einstein tensor = Einstein 0-form
pform einstein3(a)=3, einstein0(a,b)=0$
symmetric einstein0$

einstein3(-a)  := -(1/2) * curv2(b,-c) ^ # (o(-a) ^ o(-b) ^ o(c))$
einstein0(a,-b):= #( o(a) ^ einstein3(-b))$

on exp, gcd$
factor ^$
on nero;

einstein0(a,-b):= #( o(a) ^ einstein3(-b));

off nero;

% by inspection, we find 
einstein0(1,-1) + einstein0(2,-2); % equals 0
einstein0(0,-0) - einstein0(3,-3); % eliminates gamma

out "Buchdahl03.exo";

load_package tri;
on tex;
on TeXBreak;
einstein0(a,-b):=einstein0(a,-b);
off tex;
einstein0(a,-b):=einstein0(a,-b);
omega:=0;
einstein0(a,-b):=einstein0(a,-b);

shut "Buchdahl03.exo";
;end;
\end{verbatim}

 \end{appendix}

%%%%%%%%%%%%%%%%%%%%%%%%%%%%%%%%%%%%%%%%%%%%%%%%%%%%%%%%%%%%%%%%%%%%%%%%%%%

\end{document}